\documentclass[11pt,a4paper]{article}
\usepackage[francais]{babel}  
\usepackage{epsfig}
\usepackage[T1]{fontenc}    
\usepackage{graphics}
\usepackage{graphicx}
\usepackage{pstricks,pst-coil,pst-fill,pst-plot}
\usepackage[fleqn]{amsmath}    
\usepackage{amssymb}    
\usepackage{amsfonts}   
\usepackage{verbatim}   
\usepackage{mathrsfs}   
\usepackage{dsfont}
\usepackage{euscript}
\usepackage{yfonts}
\usepackage{enumerate}     
\usepackage{amsthm}         
\usepackage{txfonts}
\usepackage{marvosym}
\usepackage{vmargin}        
\usepackage{wasysym}		

\setmarginsrb{1.8cm}{2cm}{1.8cm}{2cm}{1cm}{1cm}{1cm}{1.6cm}
 \makeatletter
 \@addtoreset{equation}{section}
 \makeatother

\providecommand{\bysame}{\leavevmode\hbox to3em{\hrulefill}\thinspace}
\providecommand{\MR}{\relax\ifhmode\unskip\space\fi MR }

\providecommand{\href}[2]{#2}

\let\ua=\uparrow

\let\tend=\rightarrow

\long\def\symbolfootnote[#1]#2{\begingroup%
\def\thefootnote{\fnsymbol{footnote}}\footnote[#1]{#2}\endgroup}

\newtheorem{theorem}{Theorem}[section]
\newtheorem{prop}[theorem]{Proposition}
\newtheorem*{theorem*}{Theorem}
\newtheorem{cor}[theorem]{Corollary}
\newtheorem{defin}[theorem]{Definition}

\newtheorem{lemme}{Lemma}[section]

\def\Proof{\medskip\noindent {\it Proof --- \ }}

\def\qed{\hfill\rule{2mm}{2mm}}

\newcommand\beq{\begin{equation}}
\newcommand\enq{\end{equation}}
\newcommand\bem{\begin{multline}}
\newcommand\enm{\end{multline}}

\def\beqa{\begin{eqnarray}}
\def\eeqa{\end{eqnarray}}
\def\ba{\begin{array}}
\def\ea{\end{array}}

\newcommand{\f}[2]{{\ensuremath{%
    \mathchoice%
    {\dfrac{#1}{#2}}
    {\dfrac{#1}{#2}}
    {\frac{#1}{#2}}
    {\frac{#1}{#2}}
}}}
\newcommand{\tf}[2]{\ensuremath{#1/#2}}
\newcommand{\pa}[1]{\ensuremath{\left(#1\right)}}

\def\a{\alpha}

\def\ga{\gamma}
\def\Ga{\Gamma}

\def\de{\delta}

\def\De{\Delta}
\def\eps{\epsilon}
\def\veps{\varepsilon}
\def\la{\lambda}
\def\La{\Lambda}

\def\sg{\sigma}

\def\th{\theta}
\def\Th{\Theta}
\def\vth{\vartheta}

\def\vp{\varphi}

\newcommand{\mc}[1]{\ensuremath{\mathcal{#1}}}
\newcommand{\mf}[1]{\ensuremath{\mathfrak{#1}}}
\newcommand{\msc}[1]{\ensuremath{\mathscr{#1}}}

\newcommand{\bs}[1]{\ensuremath{\boldsymbol{#1}}}

\DeclareFontFamily{OT1}{pzc}{}
\DeclareFontShape{OT1}{pzc}{m}{it}{<-> s * [1.10] pzcmi7t}{}
\DeclareMathAlphabet{\mathpzc}{OT1}{pzc}{m}{it}

\def \i{ \mathrm i}

\newcommand{\ov}[1]{\ensuremath{\overline{#1}}}
\newcommand{\wt}[1]{\ensuremath{\widetilde{#1}}}
\newcommand{\wh}[1]{\ensuremath{\widehat{#1}}}

\newcommand{\Int}[2]{\ensuremath{\int\limits_{#1}^{#2}}}
\newcommand{\Oint}[2]{\ensuremath{\oint\limits_{#1}^{#2}}}

\newcommand{\sul}[2]{\ensuremath{\sum\limits_{#1}^{#2}}}
\newcommand{\pl}[2]{\ensuremath{\prod\limits_{#1}^{#2}}}

\newcommand{\R}{\ensuremath{\mathbb{R}}}
\newcommand{\Cx}{\ensuremath{\mathbb{C}}}

\newcommand{\Dp}[1]{\ensuremath{\partial_{#1}}}

\newcommand{\limit}[2]{\ensuremath{\underset{#1 \tend #2}{\longrightarrow} }}

\newcommand{\ex}[1]{\ensuremath{\e{e}^{#1}}}

\newcommand{\op}[1]{ \boldsymbol{ \texttt{#1} } }

\newcommand{\abs}[1]{\ensuremath{\left| #1 \right|}}
\newcommand{\Norm}[1]{\ensuremath{\abs{\abs{#1}} }}

\newcommand{\norm}[1]{\ensuremath{|| #1 ||}}

\newcommand{\dd}{\mathrm{d}}
\newcommand{\e}[1]{\ensuremath{\mathrm{#1}}}

\newcommand{\intff}[2]{\ensuremath{ [  #1 \,; #2 ] }}
\newcommand{\intfo}[2]{\ensuremath{ [  #1 \,; #2 [ }}
\newcommand{\intof}[2]{\ensuremath{ ]  #1 \,; #2 ] }}
\newcommand{\intoo}[2]{\ensuremath{ ]  #1 \,; #2 [ }}

\newcommand{\intn}[2]{\ensuremath{[\![ \, #1 \,;\, #2 \,]\!]}}

\begin{document}




\begin{center}
\begin{LARGE}
{\bf On condensation properties of Bethe roots associated with the XXZ chain}
\end{LARGE}

\vspace{1cm}

\vspace{4mm}
{\large Karol K. Kozlowski \footnote{e-mail: karol.kozlowski@ens-lyon.fr}}%
\\[1ex]
Univ Lyon, Ens de Lyon, Univ Claude Bernard, CNRS, Laboratoire de Physique, F-69342 Lyon, France. \\[2.5ex]

\vspace{4mm} {\bf MSC}: 30E15, 45G10, 45M05

\par

\vspace{40pt}

\centerline{\bf Abstract} \vspace{1cm}
\parbox{12cm}{\small I prove that the Bethe roots describing either the ground state or a certain class of "particle-hole" excited states of the XXZ spin-$1/2$ chain in any sector with magnetisation $\mf{m} \in \intff{0}{\tf{1}{2}}$
exist, are uniquely defined, and form, in the infinite volume limit, a dense distribution on a subinterval of $\R$. 
The results holds for any value of the anisotropy $\De \geq -1 $.
In fact, I establish an even stronger result, namely the existence of an all order asymptotic expansion of the counting function associated with such roots. 
As a corollary, these results allow one to prove the existence and form of the infinite volume limit of various observables attached to the model -the excitation energy,  momentum, 
the zero temperature correlation functions, so as to name a few- that were argued earlier in the literature.}

\end{center}

\vspace{40pt}

\section*{Introduction}

The XXZ spin-$\tf{1}{2}$ chain refers to a system of interacting spins in one dimension described by the Hamiltonian 
\beq
\op{H}_{\De} \, = \, J \sum_{a=1}^{L} \Big\{ \sigma^x_a \,\sigma^x_{a+1} +
  \sigma^y_a\,\sigma^y_{a+1} + \De  \,\sigma^z_a\,\sigma^z_{a+1}\Big\} \;.
\label{ecriture hamiltonien XXZ}
\enq
$\op{H}_{\De}$ is an operator on the Hilbert space of the model $\mf{h}_{XXZ}=\otimes_{a=1}^{L}\mf{h}_a$ with $\mf{h}_a \simeq \Cx^2$. 
The matrices $\sg^{w}$, $w=x,y,z$ are the Pauli matrices and $\sg_a^{w}$ stands for the operator on $\mf{h}_{XXZ}$ which acts as the Pauli matrix $\sg^{w}$
on $\mf{h}_a$ and as the identity on all other spaces appearing in the tensor product defining $\mf{h}_{XXZ}$. 
The Hamiltonian depends on two coupling constants: $J>0$ which represents the so-called exchange interaction and $\De$
which takes into account the anisotropy in the coupling between the spins in the longitudinal and transverse directions. 
Finally, the chain consists of an even number of sites $L\in 2 \mathbb{N}$. 

The Hamiltonian $\op{H}_{\De}$ commutes with the total spin operator 
\beq
\op{S}^z \, = \,  \sul{a=1}{L}  \sg_a^z \;. 
\enq
The Hilbert space of the model $\mf{h}_{XXZ}$ decomposes into the direct sum
\beq
\mf{h}_{XXZ} \; = \; \bigoplus_{N=0}^{L} \mf{h}_{XXZ}^{(N)} \qquad \e{with} \qquad \mf{h}_{XXZ}^{(N)} \; = \; 
\Big\{   \bs{v}  \in \mf{h}_{XXZ} \; : \;  \op{S}^z \cdot \bs{v} = (L-2N) \cdot \bs{v}   \Big\}  \;,  
\enq
what turns $\op{H}_{\De}$ into a block diagonal operator relatively to the above decomposition. I denote by  $\op{H}_{\De}^{(N)}$ the restriction of $\op{H}_{\De}$ to every subspace $ \mf{h}^{(N)}_{XXZ}$.

The spectrum and eigenvectors of the isotropic limit $\De=1$ of the XXZ chain have been first studied by Bethe \cite{BetheSolutionToXXX} 
in 1931 by means of the celebrated Bethe Ansatz. Then, in 1958, Orbach extended the approach to the case of the 
XXZ chain \cite{OrbachXXZCBASolution} with a general coupling $\De$. Within Bethe's Ansatz, the eigenvectors and eigenvalues
of $\op{H}_{\De}^{(N)}$ are parametrised by a collection of $N$ numbers $\{\la_a\}_1^{N}$, the so-called Bethe roots. The Bethe roots are 
constrained to satisfy a system of transcendental equations called the Bethe equations. For $-1 < \De  <1$ when 
the anisotropy can be parametrised as $\De = \cos(\zeta)$ with $\zeta \in \intoo{0}{\pi}$, these take the form 
\beq
\pl{b=1 }{ N  } \Bigg\{ \f{ \sinh(\la_a -\la_b + \i \zeta ) }{ \sinh(\la_b -\la_a + \i \zeta )  } \Bigg\} \, \cdot \, \bigg( \f{ \sinh( \tf{\i \zeta }{2} - \la_a  ) }{ \sinh( \tf{\i \zeta }{2} +\la_a  )  } \bigg)^L \; = \; (-1)^{N+1} 
\; , \quad a=1,\dots,N \;.
\label{ecriture BAE De entre moins 1 et 1}
\enq
When $\De=1$, the Bethe equations degenerate into the rational form 
\beq
\pl{b=1 }{ N } \Bigg\{  \f{ \la_a -\la_b + \i  }{ \la_b -\la_a + \i   } \Bigg\}  \, \cdot \, \bigg( \f{  \tf{\i  }{2} - \la_a  }{ \tf{\i  }{2} +\la_a    } \bigg)^L \; = \; (-1)^{N+1} 
\; , \quad a=1,\dots,N \;.
\label{ecriture BAE De egal 1}
\enq
Finally, for $ \De > 1$, the anisotropy can be parametrised as $\De = \cosh(\zeta)$ with $\zeta \in \R^+$ and the Bethe equations take the form 
\beq
\pl{b=1 }{ N } \Bigg\{ \f{ \sin(\la_a -\la_b + \i \zeta ) }{ \sin(\la_b -\la_a + \i \zeta )  } \Bigg\}  \, \cdot \, \bigg( \f{ \sin( \tf{\i \zeta }{2} - \la_a  ) }{ \sin( \tf{\i \zeta }{2} + \la_a  )  } \bigg)^L \; = \;  (-1)^{N+1}
\; , \quad a=1,\dots,N \;.
\label{ecriture BAE De plus grand que 1}
\enq
The question of the completeness of the Bethe Ansatz, namely whether the set of solutions to \eqref{ecriture BAE De entre moins 1 et 1}, \eqref{ecriture BAE De egal 1} or 
\eqref{ecriture BAE De plus grand que 1}, is in one-to-one correspondence with the set of eigenvectors of $\op{H}_{\De}^{(N)}$ is tricky and remained open for 
quite a long time. 
A positive answer has been given, for the XXX chain, by Mukhin, Tarasov and Varchenko \cite{MukhinTarasovVarchenkoCompletenessandSimplicitySpectBA}
provided that one agrees to slightly change the perspective and to characterise states in terms of polynomial solutions to an appropriate $\op{T}-\op{Q}$
equations. Also, completeness was established for certain generic inhomogeneous variants of the XXZ chain \cite{TarasovVarchenkoCompletenessBAXXZ}. 

Independently of completeness issues, there are numerous other questions related with the study of the equations
\eqref{ecriture BAE De entre moins 1 et 1}-\eqref{ecriture BAE De plus grand que 1}. One relates to identifying the solution of \eqref{ecriture BAE De entre moins 1 et 1}, \eqref{ecriture BAE De egal 1} 
or \eqref{ecriture BAE De plus grand que 1}, depending on the value of $\De$, giving rise to $\op{H}^{(N)}_{\De}$'s ground state, \textit{viz}. the eigenvector associated with 
the lowest eigenvalue. 
The answer has been obtained  by Yang-Yang in \cite{Yang-YangXXZproofofBetheHypothesis}. 
By applying a variant of the Perron-Frobenius theorem, Yang and Yang showed that $\op{H}^{(N)}_{\De}$ admits a unique ground state. In order to 
identify the roots describing the ground state, it is convenient to rewrite the Bethe equations in their logarithmic form
\beq
\f{1}{2\pi} \mf{p}(\la_a) \, - \,  \f{1}{2\pi L } \sul{a=1}{N} \th(\la_a-\la_b)  \, + \, \f{N+1}{2L} \; = \; \f{ \ell_a }{ L }\qquad \e{with}\quad  \ell_a \in \mathbb{Z} \;,
\label{ecriture Log BAE cas general tout delta}
\enq
where for $-1 < \De=\cos(\zeta) < 1$, with $\zeta \in \intoo{0}{\pi}$, 
\beq
\mf{p}(\la) \; = \; \i \ln \bigg( \f{ \sinh(\i \tf{\zeta}{2}+\la) }{  \sinh(\i \tf{\zeta}{2}-\la) } \bigg) \quad , \qquad 
\th(\la) \; = \; \i \ln \bigg( \f{ \sinh(\i \zeta+\la) }{  \sinh(\i \zeta-\la) } \bigg) \; ,
\label{ecriture bare phase and momentum De entre 1 et -1} 
\enq
for $\De=1$
\beq
\mf{p}(\la) \; = \; \i \ln \bigg( \f{ \tf{\i}{2}+\la }{ \tf{\i }{2}-\la } \bigg) \quad , \qquad 
\th(\la) \; = \; \i \ln \bigg( \f{ \i + \la }{  \i - \la } \bigg) \;, 
\label{ecriture bare phase and momentum De egal 1} 
\enq
and, for $\De=\cosh(\zeta)>1$,  $\zeta \in \R^+$,
\beq
\mf{p}(\la) = \vth\big( \la , \tf{\zeta}{2} \big) \quad , \qquad \th(\la) = \vth\big( \la , \zeta \big) \qquad \e{with} \qquad 
\vth\big( \la , \eta \big)\;= \; \Int{0}{\la} \f{ \sinh(2\eta)  \dd \mu  }{ \sin(\mu+\i \eta)  \sin(\mu  - \i \eta)  } \;. 
\label{ecriture bare phase and momentum De depasse 1} 
\enq
Yang and Yang proved that the ground state for $\De \geq -1$ is obtained from a real valued solution to \eqref{ecriture Log BAE cas general tout delta} corresponding to the specific choice of integers $\ell_a=a$. 
More precisely, Yang and Yang were analysing a reparametrised version of the Bethe equations in terms of $k_a$ variables, $\la_a=f_{\De}(k_a)$ for some explicit $f_{\De}$. 
It was, in fact, the equation in terms of the variables $k_a$ that has been initially obtained by Orbach \cite{OrbachXXZCBASolution}. 
Yang and Yang showed that the transformed counterpart of \eqref{ecriture Log BAE cas general tout delta} admits a unique real valued solution when $-1 < \De  \leq 0$. 
Furthermore, they showed that the transformed counterpart of \eqref{ecriture Log BAE cas general tout delta} when $\ell_a=a$ admits, in fact, solutions for any $\De \geq -1$ and that, among these, 
there exists one such that $\De\mapsto k_a(\De)$ is continuous in $\De>-1$. Yang and Yang established that it is precisely this particular solution 
that gives rise to the Bethe roots parametrising the ground state of $\op{H}_{\De}^{(N)}$. 
I will refer, henceforth, to this solution as the ground state Bethe roots. Yang and Yang, however, did not prove the uniqueness of solutions to \eqref{ecriture Log BAE cas general tout delta} for $\De > 0$
and $\ell_a=a$.

On top of the one parametrising the ground state of $\op{H}_{\De}^{(N)}$, the logarithmic Bethe Ansatz equations admit many other solutions upon varying the choices of integers $\ell_a$. For instance, when the choice of the ground state integers 
is slightly perturbed, as 
\beq
\ell_a = a \quad \e{for} \quad a \in \intn{1}{N}\setminus \{h_a\}_1^n \qquad \ell_{ h_{a} } \, = \, p_a \qquad \e{with} \quad 
\left\{ \ba{cc} h_a \in \intn{1}{N}   \\ 
		p_a \in \intn{-M_{\De} }{ M_{\De} +N } \setminus \intn{1}{N} \ea \right. 
\enq
with $M_{\De}$ an integer depending on $\De$, then the real-valued solution to the logarithmic Bethe Ansatz equations -if they exist- define so-called particle-hole excited states. 
One can also have complex valued solutions to the Bethe Ansatz equations, as already observed by Bethe \cite{BetheSolutionToXXX}. See \cite{CauxHagemansDeformedStringsInXXX}
for an extensive numerical analysis thereof in the case of small length XXX chains. In the present paper, I will \textit{not} discuss the complex valued solutions.

\vspace{3mm}

From the point of view of practical applications one is mostly interested in the behaviour of observables attached to the model in the so-called thermodynamic 
limit $L\tend + \infty$. Such observables can be the energy or momentum of an eigenstate or some correlation function. 
In practical situations  this amounts to computing either the limit or the first few terms in the large-$L$ asymptotic expansion of sums of the type
\beq
\f{1}{L} \sul{a=1}{N} f(\la_j)
\label{ecriture somme limite  a calculer}
\enq
where $f$ is some sufficiently regular function,  $\{ \la_a \}_1^N$ are the Bethe roots describing the ground state or some exited state "close" to it and the integer $N$
labelling the spin sector to which the Bethe vector belongs is $L$ dependent and grows with $L$ in such a way that $\tf{N}{L} \tend D\in \intff{0}{\tf{1}{2}}$.

In \cite{Yang-YangXXZStructureofGS}, Yang and Yang affirmed that the limit exists in the case of the Bethe roots for the ground state and that, for any sufficiently regular function $f$, it holds
\beq
\lim_{L\tend + \infty} \bigg\{ \f{1}{L} \sul{a=1}{N} f(\la_j) \bigg\} \; = \; \Int{-q}{q}\!  f(\la) \, \rho(\la\mid q) \cdot \dd \la \;. 
\label{ecriture convergence somme sur racines de Bethe limit thermo}
\enq
The pair $(q,\rho(\la\mid q))$ appearing above corresponds to the unique solution to the system of equations  for the unknowns $\big( Q , \rho(\la\mid Q) \big)$:
\beq
\rho(\la\mid Q) + \Int{-Q}{Q} \! K(\la-\mu) \rho(\mu \mid Q) \cdot \dd \mu \; =\;  \f{1}{2\pi}\mf{p}^{\prime}(\la) \qquad \e{and} \qquad D \; = \; \Int{-Q}{Q}\! \rho(\la \mid Q) \cdot \dd \la 
\label{ecriture probleme integal pour q rho de q}
\enq
where $D=\lim_{L\tend+\infty} \big(\tf{N}{L} \big)$  and $K(\la)=\tf{ \th^{\prime}(\la) }{ (2\pi) }$, \textit{viz}.
\beq
K(\la) \; =  \;   \left\{ \ba{c c c }   \f{ \sinh(2\zeta)    }{ 2\pi \sin(\mu+\i \zeta)  \sin(\mu  - \i \zeta)  } \hspace{1cm} & \e{for} & \;\; \De >1  \vspace{2mm} \\
				    \f{ 1    }{ \pi (1+\la^2)  } & \e{for} & \;\; \De =1  \vspace{2mm} \\  
					      \f{ \sin(2\zeta)   }{ 2\pi \sinh(\mu+\i \zeta)  \sinh(\mu  - \i \zeta)  } & \e{for} & \;\; -1<\De <1 \ea \right. \;.  
\label{ecriture noyau integral K tout delta}
\enq

Yang and Yang did prove that the system of equations on the pair of unknowns  $\big( Q , \rho(\la\mid Q) \big)$ does indeed admit a unique solution.
They however did not prove the statement relative to the existence of the limit in the \textit{lhs} of \eqref{ecriture convergence somme sur racines de Bethe limit thermo}
nor its value given by the \textit{rhs} of \eqref{ecriture convergence somme sur racines de Bethe limit thermo}.

\vspace{3mm}

Numerous works, starting from the pioneering handlings of H\'{u}lten \cite{HultenGSandEnergyForXXX}, did rely on the assumption that the limit of sums as in \eqref{ecriture somme limite  a calculer} exists and takes the form  \eqref{ecriture convergence somme sur racines de Bethe limit thermo},  
 be it when in the \textit{lhs} there appear ground state Bethe roots or those describing a certain class of excited states above the ground state. 
In particular, such properties were used in the exact (but not rigorous in the sense introduced by Baxter \cite{BaxterExactlySolvableLatticeModels})
calculations leading to characterising the ground state energy and spectrum of excitations of the infinite volume XXZ chain 
\cite{DescloizeauxGaudinExcitationsXXZ+Gap,DescloizeauxPearsonExcitationsXXX,HultenGSandEnergyForXXX,KlumperZittartzEigenvaluesofXYZPReciseStudyI,KlumperZittartzEigenvaluesofXYZPReciseStudyII},
testing the conformal structure of the spectrum of the XXZ chain \cite{BogoluibovIzerginReshetikhinFiniteSizeCorrectionsEnergyAndCriticalExponentsforXXZBoseFiniteH,JohnsonKrinskyMcCoyCorrelationLength8VModelAndXYZ,
KarowskiConformalPropertiesAndFiniteSizeCorrections,KlumperWehnerZittartzConformalSpectrumofXXZCritExp6Vertex,EckleWoynarovichFiniteSizeCorrectionsXXZ0Delta1HZero,
EckleTruongWoynarovichNonAnalyticCorrectionsForXXZInFiniteMagFieldANdCFTSpectrum,WoynaorwiczFiniteSizeCorrections}, 
the algebraic Bethe Ansatz based calculations of the matrix elements of the reduced density matrix \cite{KMTElementaryBlocksPeriodicXXZ} and, more generally, ground state correlation 
functions in this model \cite{IzerginKorepinQISMApproachToCorrNextDiscussion,KMNTOriginalSeries} or of the large-volume behaviour of
the matrix elements of local spin operators taken between two excited states close to the ground state \cite{KozDugaveGohmannSuzukiLargeVolumeBehaviourFFMassiveXXZ,IzerginKitMailTerSpontaneousMagnetizationMassiveXXZ,
KozKitMailSlaTerEffectiveFormFactorsForXXZ,KozKitMailSlaTerThermoLimPartHoleFormFactorsForXXZ} so as to name a few.  
Despite its importance due to the mentioned multiple applications, the existence and form of the limit  \eqref{ecriture convergence somme sur racines de Bethe limit thermo} was not proven in its full generality so far. 
The only works which did address this question were the ones of Gusev 
\cite{GusevCondensationGSNegativeDelta}, and more recently, 
the one of Dorlas and Samsonov \cite{DorlasSamsonovThermoLim6VertexAndConvergceToDensityInSomeCases6VrtX}.  
Gusev focused his analysis on the  $-1<\De \leq 0$ regime  with $0<D<\tf{1}{2}$ and developed convexity techniques, which can be seen as a  generalisation of the convexity arguments
invoked by Yang and Yang relatively to the existence and uniqueness of solutions to \eqref{ecriture Log BAE cas general tout delta}, 
so as to establish condensation \cite{GusevCondensationGSNegativeDelta} for the ground state
Bethe roots. However, his analysis relied on a boundedness property of the Bethe roots which he failed to establish. 
Thus although although putting the reasoning within a rigorous setting, Gusev's proof was still incomplete. 
Dorlas and Samsonov focused also on the case of the ground state Bethe roots and proved that indeed, for $0<D\leq \tf{1}{2}$, \eqref{ecriture convergence somme sur racines de Bethe limit thermo} does hold for
$-1<\De \leq 0$ and for $\De > \De_0>1$ where $\De_0$ was some explicit number. 
For $-1<\De \leq 0$ Dorlas and Samsonov's proof used   convex functional analysis 
-in a spirit slightly different from Gusev-. Their argument was however limited to this regime since convexity does not hold anymore for $\De >0$. The two authors also managed to prove the statement for $\De > \De_0 > 1$ by using the fixed point
theorem for an auxiliary operator which was contracting for this range of the anisotropy.
This contractivity  may be seen as a rigorous treatment of the perturbative regime in $\De$ large enough.

One  of the results of the present paper is the proof of \eqref{ecriture convergence somme sur racines de Bethe limit thermo} for all values of $ \De > -1$
and for the class of real valued, particle-hole, solutions to the logarithmic Bethe equations \eqref{ecriture Log BAE cas general tout delta} whose existence is 
established in Proposition \ref{Proposition existence solution Log BAE}. For simplicity, below, I only state the result in the case of the ground state Bethe roots. 
The general case can be found in Theorem \ref{Theorem densification racines de Bethe}. 
\begin{theorem*}
Let $N,L\tend +\infty$ in such a way that $ \tf{N}{L} \tend D $ with $D \in \intof{0}{ \tf{1}{2} }$. Let $q$ be the unique solution to \eqref{ecriture probleme integal pour q rho de q}
subordinate to $D$ and $\{\la_a\}_1^N$ correspond to the set of Bethe roots parametrising the ground state of $\op{H}^{(N)}_{\De}$.
Then, given any bounded Lipschitz function $f$ on $\R$, it holds
\beq
\f{1}{L} \sul{a=1}{L} f(\la_a) \; \tend \; \Int{-q}{q} \! f( \la ) \rho( \la \mid q ) \cdot \dd \la \;. 
\enq
\end{theorem*}
In fact, this proposition is a corollary of a much stronger result established in the core of the paper: the existence of an all-order asymptotic expansion for the counting functions associated 
with such Bethe roots, see Theorems \ref{Theorem Approximation effective de xi a densite un demi} and \ref{Proposition DA counting fct}. The counting function contains all the fine details of the distribution of the $\la_a$'s,
so that obtaining this asymptotic expansion goes much further that the simple limiting result \eqref{ecriture convergence somme sur racines de Bethe limit thermo}. 
The idea of introducing the counting function as a way to study the large-$L$ behaviour of a given solution to the Bethe equations goes back to the work of 
\cite{DeVegaWoynarowichFiniteSizeCorrections6VertexNLIEmethod,KlumperBatchelorNLIEApproachFiniteSizeCorSpin1XXZIntroMethod}.
The counting function formalism was further developed in the works \cite{DestriDeVegaAsymptoticAnalysisCountingFunctionAndFiniteSizeCorrectionsinTBAFirstpaper,DestriDeVegaAsymptoticAnalysisCountingFunctionAndFiniteSizeCorrectionsinTBAFiniteMagField,
KlumperBatchelorPearceCentralChargesfor6And19VertexModelsNLIE,KlumperWehnerZittartzConformalSpectrumofXXZCritExp6Vertex}
what allowed to derive the first few terms of the large-L asymptotic expansion of the counting function associated with various configurations of Bethe roots associated with the XXZ chain.
However, these derivations did build on various \textit{ad hoc} hypothesis which, technically speaking, boil down to a densification property of the type \eqref{ecriture convergence somme sur racines de Bethe limit thermo}
 or a close variant thereof. 
In this paper I circumvent the use of such \textit{ad hoc} hypothesis, hence bringing these formal asymptotic expansions to a rigorous level.   
It is important to stress that the method of proof introduced in the present paper neither relies on convexity arguments nor on fixed points theorem.  
The properties I use are rather general what makes the method applicable, \textit{in principle}, to many other instances of quantum integrable models.  

The paper is organised as follows. In Section \ref{Section Linear integral equations}, I review some properties of solutions to linear integral equations
that are relevant to the problem and establish certain auxiliary results that are of interest to the analysis. 
In Section \ref{Section solvability properties of BAE}, I  establish the solvability of the class of logarithmic Bethe equations of interest to the problem. 
Then, in Section \ref{Section Large L analysis of counting function}, I establish the main result of the paper, namely the existence of the asymptotic
expansion of the counting function. Section \ref{Section Applications to comutation of limit} is devoted to the applications of the results to various problems
that arose earlier in the literature. In particular, the densification proposition  is established there. 

Appendix \ref{Appendix Section Auxiliary results} contains an auxiliary result of interest to the analysis.

\section{Some properties of solutions to linear integral equations}
\label{Section Linear integral equations}

Let $J\subset \R$ be Lebesgue measurable. 
Let $\op{K}_{J}$ denote the integral operator on $L^2(J)$ acting as
\beq
\op{K}_{ J }[f] (\la) \; = \; \int_{ J } K(\la-\mu) f(\mu) \dd \mu \;. 
\enq
Here $K(\la)$ corresponds to one of the three integral kernels given in \eqref{ecriture noyau integral K tout delta}, depending on the value of $\De$. When $\De>1$, I will always assume that 
$\e{diam}(J)\leq \pi$, where $\e{diam}(J)$ is the diameter of $J$. Throughout the paper, the dependence on $\De$ of the operators and integral kernels will be kept implicit since this would not
bring more clarity to discussion while weighting down the intermediate formulae.

The purpose of this section is to recall some known facts about the operator $\e{id}+\op{K}_{J}$ and the solutions to specific integral equation driven by it. In particular, I will discuss its invertibility 
for any $J$ and review several properties of solutions of linear integral equation driven by this operator. Finally, I will prove an auxiliary result relative to the unique solvability of a non-linear problem
driven by the operator $\op{K}_J$. This unique solvability  will appear crucial in a later stage of the analysis.

Throughout the paper, given $\a>0$, $I_{\a}$ stands for the segment centred around $0$
\beq
I_{\a}=\intff{-\a}{\a}\;. 
\enq
There are certain choices of the interval $J$ which make the operator $\op{K}_{J}$ special, in that the linear integral equations driven by 
$\e{id}+\op{K}_{J}$ can be solved in a closed form by using Fourier transforms or Fourier series. These choices correspond to $J =\R$ when $-1<\De\leq 1$ and $J=\intff{ - \tf{\pi}{2} }{ \tf{\pi}{2} }$ when $\De >1$.
It is convenient to introduce  a parameter $\iota$ labelling the range of integration $I_{\iota}$ corresponding to those special cases:
\beq
\iota = \left\{ \ba{ccc}  \infty & \e{if} & -1<\De \leq 1 \\ 
		 \tf{\pi}{2} & \e{if} & \De >1 \ea \right.  \qquad \e{so}\;\e{that} \qquad I_{\iota} \; = \; 
\left\{ \ba{ccc}  I_{\infty}=\R  &  \e{if} & -1<\De \leq 1  \\
	    I_{\tf{\pi}{2}}=\intff{ -\tf{\pi}{2} }{ \tf{\pi}{2}  } & \e{if} & \De >1 \ea \right. \; . 
\label{definition parametre iota}
\enq

It is a well known result that will be recalled in Lemma \ref{lemme invertibilite I+K} that the operator $ \e{id} + \op{K}_{J} $ is invertible for all $J$. 
  The resolvent operator associated with $\op{K}_J$ will be denoted by $\op{R}_{J}$. It is defined as the operator on $L^2(J)$ such that $\big( \e{id} - \op{R}_{J} \big) \big(  \e{id} + \op{K}_{J} \big) =\e{id}$. 
The integral kernel of the resolvent will be denoted as $R_{J}(\la,\mu)$. It solves the linear integral equation 
\beq
R_{J}(\la,\mu) \, + \, \int_{ J }{} K(\la-\nu) R_{J}(\nu,\mu) \dd \nu \, = \, K(\la-\mu) \;. 
\enq

\subsection{Some explicit solutions and their properties}

\subsubsection{The density of Bethe roots}

The so-called density of Bethe roots is defined as the solution to the linear integral equation
\beq
\Big(\e{id} \, + \, \op{K}_{I_q}\Big)\big[\rho(*\mid q) \big] \; = \; \f{ 1}{ 2\pi } \mf{p}^{\prime}(\la) \;. 
\enq
The density can be computed in closed form when the support of integration is $I_{\iota}$.

\begin{lemme}
 
 The solution to the linear integral equation 
\beq
\ba{cccc c}  \Big( \e{id}+\op{K}_{\R} \Big)\big[\rho_{\infty}\big] (\la) & = & \f{ 1}{ 2\pi }  \mf{p}^{\prime}(\la)  \quad &  for & -1<\De \leq 1 \vspace{1mm}  \\ 
\Big( \e{id}+\op{K}_{I_{ \tf{\pi}{2} } } \Big)\big[\rho_{\tf{\pi}{2}}\big] (\la) & = & \f{ 1 }{ 2\pi } \mf{p}^{\prime}(\la)  \quad &  for & \De > 1  \ea
\enq
 takes the form 
\beq
\rho_{\infty}(\la) \; = \; \left\{  \ba{ccc }   \f{ 1 }{ 2 \zeta  \cosh\big( \tf{ \pi \la }{ \zeta } \big) }   & for & -1 <\De <1  \\ 
						\f{ 1 }{ 2  \cosh\big( \pi \la  \big) }  & for &   \De = 1 \ea \right. 
\quad and \qquad \rho_{ \tf{\pi}{2} }(\la) \; = \;  \f{1}{2\zeta} \sul{n\in \mathbb{Z} }{} \f{ 1 }{ \cosh\Big[ \f{\pi}{\zeta}(n\pi-\la)\Big]  } \;. 
\label{expression pour densite dans cas N est L sur 2}
\enq

\end{lemme}

The form of the solution, when $-1< \De \leq 1$ is readily obtained by taking the Fourier transform of the linear integral equation. When $\De>1$, one solves the 
linear integral equation by means of Fourier series expansions. This yields
\beq
\rho_{ \tf{\pi}{2} }(\la) \; = \; \f{1}{2\pi} \sul{n\in \mathbb{Z} }{} \f{ \ex{2\i n \la} }{ \cosh(n\zeta)  } \;. 
\label{ecriture serie fourier rho pi sur 2}
\enq
The expression \eqref{expression pour densite dans cas N est L sur 2} is obtained by applying the Poisson summation formula. 
These results appeared, for the first time, in \cite{WalkerFirstExplicitSolToLinIntEqnMassiceXXZ,Yang-YangXXZStructureofGS}. It is clear from \eqref{expression pour densite dans cas N est L sur 2} 
that both $\rho_{\infty}$ and $\rho_{\tf{\pi}{2}}$ are strictly positive functions.

\subsubsection{The resolvent kernel}

It is readily seen that $R_{I_{\iota}}(\la,\mu)$ only depends on the difference of its arguments. This integral kernel will be denoted below as $R(\la-\mu)$. 
\begin{lemme}
\label{Lemme noyau resolvent a densite un demi}
The solution $R(\la)$ to the equation $\Big( \e{id}+\op{K}_{I_{\iota}} \Big)\big[R\big] =K $ 
is an even function for any  $\De > -1$. 
\begin{itemize}
 \item[$\bullet$]  For $-1<\De<1$, $R$ admits the Fourier transform representation 
\beq
   R (\la) \; = \;  \Int{ \R }{} 
               \f{ \sinh \bigl[ (\pi/ 2 - \zeta) k \bigr] \ex{- \i k \la} }
		{ \cosh (\zeta k/ 2) \sinh \bigl[ (\pi/ 2 - \zeta / 2) k \bigr] }   \frac{\dd k}{4 \pi }\;, \quad \De=\cos(\zeta)\quad with \quad \zeta \in \intoo{0}{\pi} \;, 
\label{ecriture Fourier resolvent Delta moins 1 et 1}
\enq
has the large $\la$ estimates $R(\la) \; = \; \e{O}\Big( \ex{- \e{min}\big( \f{\pi}{\zeta}, \f{2\pi}{\pi-\zeta} \big) |\la| }  \Big)$ and is a strictly positive function when $0<\De<1$.
\item[$\bullet$] For $\De=1$, $R$ admits the Fourier transform representation 
\beq
  R (\la) \; = \;  \Int{ \R }{} \f{ \ex{-\f{|k|}{2} } \cdot  \ex{- \i k \la} }{ \cosh (k/ 2) }   \frac{\dd k}{4 \pi } \;,
\label{ecriture Fourier resolvent Delta egal 1}
\enq
has the large-$\la$ estimates $R(\la) \; = \; \e{O}\big( \la^{-2} \big)$ and is strictly positive on $\R$. 
\item[$\bullet$] For $\De>1$, $R$ has the Fourier series expansion 
\beq
 R (\la) \; = \; \f{1}{2\pi} \sul{n\in \mathbb{Z} }{} \f{ \ex{2\i n \la} \cdot \ex{ - |n| \zeta }   }{  \cosh(n\zeta)  } \;, \quad \De=\cosh(\zeta)\quad with \quad \zeta >0 \; , 
\enq
is $\pi$-periodic and strictly positive on $\R$. 

\end{itemize}

\end{lemme}

The only non-trivial statement is the one relative to the signs of $R$. When $\De>1$, using \eqref{ecriture serie fourier rho pi sur 2} and 
\beq
\f{\mf{p}(\la)}{2\pi} \; = \; \sul{n\in \mathbb{Z} }{} \f{\ex{-|n|\zeta} \ex{2\i n \la} }{ \pi } \quad \e{one}\;\e{gets} \qquad
R(\la) \; = \; \Int{ - \tf{\pi}{2} }{ \tf{\pi}{2} } \rho_{\tf{\pi}{2}}(\la-\mu) \mf{p}^{\prime}(\mu) \dd \mu
\label{ecriture resolvent comme convolution Delta plus grand que 1}
\enq
from where strict positivity is manifest in virtue of \eqref{expression pour densite dans cas N est L sur 2}. When $\De=1$, one can recast the Fourier transform representation of $R$ in the form 
$R(\la) = \frac{1}{2\pi}\int_{\R}{} \rho_{\infty}(\mu)\mf{p}^{\prime}(\la-\mu) \dd \mu $. The latter representation ensures strict positivity and 
readily yields the $\e{O}(\la^{-2})$ estimates for the decay of $R$ at infinity. Finally, after some calculations, when $0<\De<1$
one recasts $R$ as the convolution 
\beq
   R (\la) \; = \; \f{ \pi }{2\zeta (\pi-\zeta) } \Int{ \R }{} \f{ \sin\bigg( \f{\pi \zeta}{\pi-\zeta} \bigg)  }
   { \cosh\big[ \pi (\la- \mu)/\zeta\big] \sinh \bigg( \f{ \pi (\mu+\i\tf{\zeta}{2})  }{ \pi-\zeta } \bigg) \sinh \bigg( \f{ \pi (\mu-\i\tf{\zeta}{2})  }{ \pi-\zeta } \bigg) }   \cdot \dd \mu
\label{ecriture resolvent comme convolution cas delta entre zero et un}
\enq
which produces a manifestly strictly positive function. \qed 

The representation \eqref{ecriture resolvent comme convolution cas delta entre zero et un} was first found in \cite{Yang-YangXXZStructureofGS} while \eqref{ecriture resolvent comme convolution Delta plus grand que 1} 
is a straightforward application of the idea that leads to \eqref{ecriture resolvent comme convolution cas delta entre zero et un}. 

For the sake of further handlings, it is useful to introduce the integral operator $\op{L}_{J}$ on $L^{2}(J)$ defined as
\beq
\op{L}_{J}[f] (\la) \; = \; \int_{ J }{} R(\la-\mu) f(\mu) \dd \mu \;. 
\label{definition operateur L}
\enq
It will be established in Lemma \ref{lemme invertibilite I+K} that $ \e{id} - \op{L}_{J}$ is invertible. The resolvent of this operator will be denoted $\mf{L}_{J}$ and is defined by 
$\big(  \e{id} - \op{L}_{J} \big)^{-1}=\e{id} + \mf{L}_{J}$.

\subsection{General considerations}

\begin{lemme}
\label{lemme invertibilite I+K} 
 
Let $J\subset \R$ be such that $0<|J| < +\infty$ if $-1 < \De \leq 1$ and $\e{diam}(J)<\pi $ if $\De > 1 $. Then, the operators  $  \e{id} + \op{K}_{J} $ and 
$  \e{id} - \op{L}_{J}$  are both invertible. The integral kernel $R_{J}(\la,\mu)$ of the resolvent operator $\op{R}_{J}$ satisfies to the bounds
\beq
R(\la-\mu) \, < \, R_{J}(\la,\mu) < K(\la-\mu) < 0 \quad for \;\; -1<\De <0 
\label{ecriture inegalites sur les resolvent a interval fini De negatif}
\enq
and 
\beq
K(\la-\mu) > \, R_{J}(\la,\mu) >R(\la-\mu) \, >0  \quad for \;\;  \De > 0 \;. 
\label{ecriture inegalites sur les resolvent a interval fini De positif}
\enq
Furthermore, one has $\mf{L}_{ J^{\e{c}} }(\la,\mu) = \op{R}_{ J }(\la,\mu)$, where $J^{\e{c}} = I_{\iota}\setminus J$. 

\end{lemme}

\Proof 

For any $g \in L^{\infty}(J)$, one has the bounds
\beq
\norm{ K_{J}[g] }_{ L^{\infty}( \R ) } \; \leq \; \norm{ g }_{ L^{\infty}(J) } \cdot \sup_{\la \in \R} \Int{J+\la}{} | K(\mu)| \dd\mu
\;\leq \; \norm{ g }_{ L^{\infty}(J) } \cdot \sup_{I\in \mc{I}_{\De} }\norm{ K }_{ L^1(I) } \; \leq \; \norm{ g }_{ L^{\infty}(J) }
\cdot  \norm{ K }_{ L^1( I_{|J|/2}) }  
\label{ecriture borne sur norme L infty de KI}
\enq
where  $I_{|J|/2} = \intff{ - \tf{|J|}{2} }{ \tf{|J|}{2} } $ and 
\beq
\mc{I}_{\De}=\;\Big\{ I \subset \R \; : \; |I|=|J| \Big\} \quad \e{if} \; -1<\De \leq 1 \quad \e{and} \quad
\mc{I}_{\De}=\;\Big\{ I \subset \intff{-\tf{\pi}{2} }{ \tf{\pi}{2} } \; : \; |I|=|J| \Big\} \quad \e{when} \;\; \De >1 \;. 
\enq
The second bound is trivial for $-1<\De\leq 1$ and it holds for $\De>1$ since by $\pi$-periodicity of $K$ one can reduce the integration along an 
interval of diameter at most $\pi$ into one over an appropriate  subinterval of $\intff{-\tf{\pi}{2} }{ \tf{\pi}{2} }$. 
The bound holds since $|K|$ is even and decreasing on $\R^+$, resp. $\intfo{0}{\tf{\pi}{2}}$, when 
$-1<\De\leq 1$, resp. $\De>1$. 

Now due to 
\beq
\Int{ \R }{} K(\mu) \dd \mu \, = \,  \f{\pi-2\zeta}{\pi}   \quad \e{if} -1<\De \leq 1 \qquad \e{and} \quad \Int{ -\tf{\pi}{2} }{ \tf{\pi}{2} } K(\mu) \dd \mu \, = \, 1 \quad \e{if} \;\; \De > 1 
\enq
the hypothesis on $J$ ensure that $\norm{ K }_{ L^1( I_{|J|/2}) }  <1$ and thus that the Neumann series 
\beq
R_J(\la,\mu) \, = \,  K(\la-\mu) \, + \, \sul{n\geq 1}{} (-1)^{n-1}\int_{ J^n }{} K(\la-\nu_1) \pl{a=1}{n-1}K(\nu_a-\nu_{a+1}) \cdot K(\nu_n -\mu) \cdot \dd^n \nu
\enq
for the resolvent kernel converges uniformly on $\R^2$. This establishes the invertibility of $\big(  \e{id} + \op{K}_{I} \big)$. 
The one of $\big(  \e{id} - \op{L}_{I} \big)$ is proven analogously using the properties of the resolvent $R(\la)$ established in Lemma \ref{Lemme noyau resolvent a densite un demi}.

When $-1 < \De < 0$ the Neumann series for $R_{I}(\la,\mu)$ is a sum of strictly negative terms. Hence the upper bounds
given in \eqref{ecriture inegalites sur les resolvent a interval fini De negatif}. 
For $\De > 0$, following Yang and Yang \cite{Yang-YangXXZStructureofGS} it is enough to observe that the integral equation for $R_{J}(\la,\mu)$ can be recast as 
\beq
\big( \e{id}+\op{K}_{ I_{\iota} } \big)\big[ R_{J}(*,\mu) \big](\la) \, - \, \op{K}_{ J^{\e{c}} } \big[ R_{J}(*,\mu) \big](\la) \; = \; K(\la-\mu)  
\qquad i.e. \quad
\big( \e{id}-\op{L}_{ J^{\e{c}} } \big)\big[ R_{J}(*,\mu) \big](\la)  \; = \; R(\la-\mu)  
\label{ecriture eqn lin int pour Resolvent RI sur intervalle complementaire}
\enq
where I remind that $J^{\e{c}} = I_{\iota} \setminus J$. Since $R(\la)>0$ for $\De>0$, the Neumann series\symbolfootnote[3]{Its convergence follows from similar bounds and properties as those used 
for the one associated with $K(\la)$} for the resolvent kernel $\mf{L}_{ J^{\e{c}} }(\la,\mu)$ consists of strictly positive terms what entails 
$ R_{J}(\la,\mu) > R(\la-\mu)$. The upper bound follows from $K(\la-\mu)\geq 0$, $R_{J}(\la,\mu)\geq 0$ and $R_{J}(\la,\mu)=K(\la-\mu)-\op{K}_J\big[R_J(*,\mu)](\la)$. 
Finally, the equality $\mf{L}_{ J^{\e{c}} }(\la,\mu) = \op{R}_{ J }(\la,\mu)$ between the integral kernels follows from the fact that 
$\la \mapsto \mf{L}_{ J^{\e{c}} }(\la,\mu)$ is the unique solution to the linear integral equation appearing to the right of \eqref{ecriture eqn lin int pour Resolvent RI sur intervalle complementaire}. \qed

\begin{lemme}
\label{Lemme invertibilite RJ+ restreint a J-}

Let $\De>0$ and $J^{(\pm)}$ be two bounded and disjoint subsets of $I_{\iota}$. Let $\big[\op{R}_{ J^{(+)} }\big]_{ J^{(-)} } $ be the integral operator on $L^2\big( J^{(-)} \big) $ acting as
\beq
\big[\op{R}_{ J^{(+)} }\big]_{ J^{(-)} } [ f ](\la) \; = \; \Int{ J^{(-)} }{} R_{ J^{(+)} }(\la,\mu) f(\mu) \cdot \dd \mu \;. 
\enq
Then, the operator $\Big( \e{id} \;     -   \big[\op{R}_{ J^{(+)} }\big]_{ J^{(-)} }  \Big)$ is invertible. Its resolvent operator $ \big[ \msc{R}_{ J^{(+)} }  \big]_{ J^{(-)} } $ defined by 
\beq
\Big( \e{id} \;     -   \big[\op{R}_{ J^{(+)} }\big]_{ J^{(-)} }  \Big) \Big( \e{id} \;     +  \ \big[ \msc{R}_{ J^{(+)} }  \big]_{ J^{(-)} }  \Big) =\e{id} 
\enq
has a strictly positive resolvent kernel: $  \big[ \msc{R}_{ J^{(+)} }  \big]_{ J^{(-)} }(\la,\mu)>0$ for any $\la, \mu \in \R$.

\end{lemme}

\Proof 

For any $g \in L^{\infty}(J^{(-)})$ it holds 
\bem
\Norm{ \big[\op{R}_{ J^{(+)} }\big]_{ J^{(-)} }[g] }_{ L^{\infty}(\R) } \; \leq \; \norm{ g }_{ L^{\infty}(J^{(-)}) } \cdot \sup_{\la \in \R} \Int{ J^{(-)} }{ } |  R_{ J^{(+)} }(\la,\mu)| \dd\mu \\
 \; \leq \; \norm{ g }_{ L^{\infty}( J^{(-)} ) } \cdot \sup_{\la \in \R} \Int{ J^{(+)} +\la}{} | K(\mu)| \dd\mu \; \leq \; \norm{ g }_{ L^{\infty}( J^{(-)} ) }
\cdot  \norm{ K }_{ L^1\big( I_{| J^{(+)} |/2} \big) } 
\label{ecriture borne sur norme L infty de KI}
\end{multline}
where the second bound follows from the inequality \eqref{ecriture inegalites sur les resolvent a interval fini De positif}. 
The rest of the proof is carried out analogously to the one of Lemma \ref{lemme invertibilite I+K}.  \qed

\subsection{The magnetic Fermi boundaries}

When studying the observables of the XXZ chain in the thermodynamic limit $\tf{N}{L} \tend D$, one naturally ends up with operators $\e{id}+\op{K}_{I_{Q}}$ where the 
endpoint $Q$ is one of the unknowns of the problem. The endpoint $q(D)$ which will be pertinent for describing the thermodynamics at given $D$ and the so-called dressed momentum
$p\big(\la\mid q(D) \big) $ describing the momentum carried by an individual excitation over the model's ground state arise as the solution to the below problem for two unknowns $\big( Q , p(\la\mid Q) \big)$:
\beq
\Big( \e{id} \, + \,  \op{K}_{ I_Q } \Big)[p(*\mid Q)](\la) \; = \; \f{ \mf{p}(\la) }{2\pi} - \f{ D }{4\pi } \Big[ \th(\la-Q) + \th(\la+Q) \Big]\qquad \e{and} \qquad 
p(Q\mid Q)=\f{ D }{ 2 }  \;. 
\label{ecriture probleme pour la bord de Fermi magnetique}
\enq
It was shown by Yang and Yang \cite{Yang-YangXXZStructureofGS} that for any $D \in \intff{0}{ \tf{1}{2} }$ there exists a unique $q(D) \in \R^+$ such that 
$\big( q(D) , p\big(\la\mid q(D) \big) \, \big)$ solves the problem given above. Furthermore, for $\De>1$ one even has $q(D) \in \intff{0}{\tf{\pi}{2}}$. 
The proof is relatively straightforward and builds on the fact that $Q\mapsto \int_{-Q}^{Q}p^{\prime}(\la\mid Q) \dd \la$
is an increasing function of $Q$, \textit{c.f.} \cite{KozDugaveGohmannThermoFunctionsZeroTXXZMassless} for some details. 
The map $D\mapsto q(D)$ is smooth and strictly increasing diffeomorphism from $\intff{0}{\tf{1}{2}}$ onto $\intff{0}{+\infty}$. 

In the following, I will simply denote by $q$ the endpoint of integration solving \eqref{ecriture probleme pour la bord de Fermi magnetique}. 
This endpoint will be referred to as the magnetic Fermi boundary.

It turns out that in the intermediate steps of the analysis, it will become necessary to consider a slightly more general problem to \eqref{ecriture probleme pour la bord de Fermi magnetique},
namely one for three unknowns $ \big( Q_L, Q_R, f(\la \mid Q_L, Q_R ) \big) $
\beq
\Big( \e{id} \; + \;  \op{K}_{\intff{Q_L}{Q_R}}\Big)\big[ f(* \mid Q_L, Q_R ) \big](\la) \; = \; \f{ \mf{p}(\la) }{2\pi} - \f{ D }{4\pi } \cdot \Big[ \th(\la-Q_R) + \th(\la-Q_L) \Big]
\label{probleme pour f QR et QL eqn integrale}
\enq
and
\beq
f(Q_R \mid Q_L, Q_R )  \;  = \;     \f{ D }{ 2 }   \qquad  f(Q_L \mid Q_L, Q_R )   \;  =  \;    - \f{ D }{ 2 }    \;  
\label{probleme pour f QR et QL condition au bord}
\enq
 with $Q_{L}<Q_{R}$, and the additional constraint $|Q_R-Q_L|<\pi$ if $\De>1$.

\begin{prop}
\label{Proposition unique solvabilite de la borne de Fermi magnetique}

Given any $D \in \intff{0}{ \tf{1}{2} }$, the problem \eqref{probleme pour f QR et QL eqn integrale}-\eqref{probleme pour f QR et QL condition au bord}
admits the unique solution  $\big( q , -q, p(\la\mid q) \big)$, where $\big( q , p(\la\mid q) \big)$  corresponds to the unique solution to the problem \eqref{ecriture probleme pour la bord de Fermi magnetique}.

\end{prop}

\Proof 

It is evident that $\big( q , -q, p(\la\mid q) \big)$ provides one with a solution to the given problem. Hence, it remains to prove uniqueness. 
Let $ \big( Q_L, Q_R, f(\la \mid Q_L, Q_R ) \big) $ be a solution to \eqref{probleme pour f QR et QL eqn integrale}-\eqref{probleme pour f QR et QL condition au bord}. 
It is convenient to distinguish between the case where both $Q_{L}$ and $Q_{R}$ are infinite, one of them is or both are bounded. 
Note that the first two situations  can only arise when $-1 < \De \leq 1$.

\subsubsection*{ Both endpoints are infinite }

First assume that $Q_L=-\infty$ and $Q_R=+\infty$.  Then,  
\beq
f(\la \mid -\infty,  + \infty )= p(\la\mid +\infty) \equiv \Int{0}{\la} \rho_{\infty}(\la) \dd \la \quad \Rightarrow \quad p(\pm \infty \mid +\infty) = \pm \f{1}{4}
\enq
so that $D=\tf{1}{2}$ and either this cannot be or one simply recovers the solution to the problem \eqref{ecriture probleme pour la bord de Fermi magnetique}.

\subsubsection*{ One of the endpoints is infinite }

By symmetry, it is enough to deal with the case where $Q_L$ is bounded while $Q_R=+\infty$. For simplicity, denote simply the solution by $f(\la)$. 
An integration by parts ensures that $\big( \e{id} + \op{K}_{ \intfo{Q_L}{+\infty} }\big)[f^{\prime}] = \tf{ \mf{p}^{\prime} }{2\pi}$. Then, it follows from the bounds 
\eqref{ecriture inegalites sur les resolvent a interval fini De negatif} and \eqref{ecriture inegalites sur les resolvent a interval fini De positif}
that 
\beq
\f{\mf{p}^{\prime}(\la) }{ 2\pi } < f^{\prime}(\la) < \rho_{\infty}(\la) \quad \e{for} \quad -1 < \De < 0  \qquad \e{and} \qquad 
\f{\mf{p}^{\prime}(\la) }{ 2\pi } > f^{\prime}(\la) > \rho_{\infty}(\la) \quad \e{for} \quad 0 < \De \leq 1 \;. 
\label{ecriture borne sup sur densite}
\enq
These bounds ensure that $f$ is bounded on $\R$ and strictly increasing. As such it admits a limit at $\pm \infty$. Hence
\beq
\big| K(\la-\mu) f(\mu) \big| \; \leq \; C \ex{\la-\mu} \quad \e{for} \quad \la \, \leq \,  Q_L \qquad \e{so} \; \e{that} \qquad \lim_{\la \tend -\infty}\Big\{ \op{K}_{\intff{Q_L}{+\infty} }[ f ] (\la) \Big\} \; = \; 0
\enq
by dominated convergence. Thence
\beq
\lim_{\la\tend -\infty} f(\la) \; = \; -\f{D}{2} \, - \, \f{ \pi - \zeta }{ \pi } \Big( \f{1}{2} - D \Big)  \;. 
\label{limite f a moins infty sur intervalle semi infini}
\enq
Furthermore, due to the upper bound in \eqref{ecriture borne sup sur densite} and $  \lim_{\la \tend  + \infty} f(\la) \, = \,  \tf{D}{2} $ 
it holds, for $\la \geq Q_L$  
\beq
\Big| f(\la) -\tf{D}{2}\Big| \, \leq \, C \ex{ - \min\big( 2,\f{\pi}{\zeta} \big) \la } \, \leq \, C \ex{ -  \la }  
\enq
so that 
\beq
 \Big| \op{K}_{\intff{Q_L}{+\infty} }\big[ f -\tf{D}{2} \big] (\la) \Big| \; \leq \; 
C \Int{\R}{} \big| K(\mu) \big| \ex{ -(\la+\mu)  } \cdot \dd \mu \limit{\la }{+\infty} 0  \;. 
\enq
Thus recasting the integral equation for $f$ in the form
\beq
f(\la) \, + \, \Int{Q_L}{+\infty}K(\la-\mu) \Big[ f(\mu) - \f{D}{2} \Big] \cdot \dd \mu \; = \; \f{ \mf{p}(\la)  }{ 2\pi } \, - \, \f{ D }{ 2 \pi } \th\big( \la - Q_L \big) \;. 
\enq
Taking the $\la\tend +\infty$ limit of the above equation yields and equation for $D$ which implies that $D=\tf{1}{2}$. 
In its turn, due to \eqref{limite f a moins infty sur intervalle semi infini}, this implies that $\lim_{\la\tend -\infty} \big[ f(\la) \big]=-\tf{D}{2}$ hence contradicting $f(Q_L)=-D/2$
since $f$ is strictly increasing on $\R$ by \eqref{ecriture borne sup sur densite}.

\subsubsection*{ Both endpoints are bounded }

In this case, it is convenient to introduce 
\beq
Q \, = \, \f{Q_R-Q_L}{2} \;\;\; , \qquad a \, = \,  \f{Q_R+Q_L}{2}  \quad \qquad \e{and} \qquad  \wt{f}(\la ) \; = \; f(\la+a\mid Q_L, Q_R) \;. 
\enq
One gets that $\wt{f}(\la ) $ satisfies
\beq
\Big( \e{id} \; + \;  \op{K}_{I_Q}\Big)\big[ \wt{f} \big] \; = \; \f{ \mf{p}(\la+a) }{2\pi} - \f{ D }{4\pi } \cdot \Big[ \th(\la-Q) + \th(\la+Q) \Big]
\qquad \e{and} \qquad \wt{f}(Q)=-\wt{f}(-Q)=\f{D}{2} \;. 
\enq
The function $\wt{f}$ can be uniquely decomposed as $\wt{f}(\la) \; = \; \wt{f}_{i}(\la) \; + \; \wt{f}_{p}(\la) $ where $\wt{f}_{p}$ is even and $\wt{f}_{i}$ is odd. 
It is readily seen that the functions $\wt{f}_{p}$ and $\wt{f}_{i}$ satisfy the linear integral equations
\beq
\Big( \e{id} \; + \;  \op{K}_{I_Q}\Big)\big[ \left( \ba{c} \wt{f}_{i} \\ \wt{f}_{p} \ea \right)  \big] (\la) \; = \; \f{1}{4\pi } 
				    \left( \ba{cc}   \mf{p}(\la+a)  + \mf{p}(\la-a)   -  D \cdot \Big[ \th(\la-Q) + \th(\la+Q) \Big] \\
							  \mf{p}(\la+a)  - \mf{p}(\la-a) 	 \ea     \right) 
\label{ecriture probleme pour taup et taui}
\enq
and are subject to the constrains
\beq
\wt{f}_{p}(Q) \, = \, 0 \qquad \e{and} \qquad \wt{f}_{i}(Q) \, = \, \f{D}{2}\;. 
\enq
The integral operator appearing above acts component-wise on the entries of the vector. Consider the equation satisfied by 
the even part $\wt{f}_{p}$. It will be shown that the constraint $\wt{f}_{p}(Q) =0$ can only be satisfied if $a=0$. Once this is established, then \eqref{ecriture probleme pour taup et taui}
reduces to \eqref{ecriture probleme pour la bord de Fermi magnetique}, what ensures uniqueness. In the course of doing so, one should 
treat the two regimes $-1<\De\leq 0$ and $\De>0$  separately due to the change in the sign of the integral
kernel $K(\la-\mu)$. 

\subsubsection*{ $\bullet$  $-1<\De\leq 0$ }

By \eqref{ecriture inegalites sur les resolvent a interval fini De negatif}, the integral kernel of the resolvent operator $\op{R}_{ I_{Q} }$ to $\e{id}+\op{K}_{I_Q}$ is such that $R_{ I_{Q} }(\la,\mu) \leq 0 $. 
Thus, due to \eqref{ecriture probleme pour taup et taui},  $\wt{f}_{p}$ can be represented as
\beq
 \wt{f}_{p}  (\la) \; = \; \f{\mf{p}(\la+a)  - \mf{p}(\la-a)}{4\pi} \; - \; \Int{-Q}{Q}  R_{ I_{Q} }(\la,\mu) 
 \big[ \mf{p}(\mu+a)  - \mf{p}(\mu-a) \big] \cdot \f{ \dd \mu }{4\pi} \;. 
\enq
Since
\beq
\mf{p}(\la+a)  - \mf{p}(\la-a) \; = \; \Int{\la-a}{\la+a} \mf{p}^\prime(\mu) \cdot  \dd \mu  
\enq
has the same sign as $a$, it follows that, for any $\la \in \R^+$, 
\beq
\ba{ccccc} 
\e{if}    &   a>0    & \e{then} & \wt{f}_p(\la) \, \geq  \,   \f{1}{4\pi} \big[ \mf{p}(\la+a) - \mf{p}(\la-a) \big] \,> \, 0  \vspace{2mm} \\
\e{if}    &   a<0    & \e{then} & \wt{f}_p(\la) \, \leq  \,   \f{1}{4\pi} \big[ \mf{p}(\la+a) - \mf{p}(\la-a) \big] \, < \, 0 
\ea \; . 
\label{equation exprimant positivite de tau i}
\enq
Either of the two are inconsistent with the constraint $\wt{f}_{p}(Q) \,  = \, 0$. Therefore, necessarily, $a=0$ so that  the problem 
reduces to \eqref{ecriture probleme pour la bord de Fermi magnetique} and on that account is uniquely solvable.

\subsubsection*{ $\bullet$  $\De >0 $ }

We start by observing that the solution $\phi_{\iota;a}$, \textit{c.f.} \eqref{definition parametre iota} for the definition of $\iota$, to the linear integral equation 
\beq
\Big( \e{id} \, + \, K_{ I_{\iota} } \Big)[ \phi_{\iota;a} ](\la) \; =   \f{ \mf{p}(\la+a)-\mf{p}(\la-a) }{2\pi} \qquad \e{takes} \; \e{the} \; \e{form} \quad
\phi_{\iota;a}(\la) \; =  \Int{\la-a}{\la+a} \hspace{-1mm} \rho_{\iota}(\mu) \cdot \dd\mu \;. 
\label{ecriture rep int moment habille zone fermi infini}
\enq
This can be seen as follows. When $\iota=\infty$, the decay at infinity of the resolvent kernel $R(\la-\mu)$ and the boundedness of the driving term 
ensure that $\phi_{\infty;a}$ is bounded on $\R$. When $\iota=\tf{\pi}{2}$, it is readily inspected that $\phi_{\tf{\pi}{2};a}$  is $\pi$-periodic. 
Thus, for any $\iota$ either because the boundary terms vanish ($\iota=\infty$) or cancel out ($\iota=\tf{\pi}{2})$, taking the derivative of the linear integral equation in \eqref{ecriture rep int moment habille zone fermi infini} 
and then integrating by parts one gets a linear integral equation satisfied by $\phi_{\iota;a}^{\prime}$ whose solution reads  
$\phi_{\iota;a}^{\prime}(\la)=\rho_{\iota}(\la+a)-\rho_{\iota}(\la-a)$. The integral representation for $\phi_{\iota;a}$ then follows from the fact that 
it is an even function. 

\vspace{2mm}

$\wt{f}_p$ can be readily continued by means of the linear integral equation to the real axis. Then, building on the identity
\beq
\Big( \e{id} \; + \; \op{K}_{I_Q}\Big)[  \wt{f}_p ] (\la)  \; = \; \Big( \e{id} \; + \;  \op{K}_{ I_{\iota} }\Big)[ \wt{f}_p  ] (\la)   \; - \;   \op{K}_{I_Q^{\e{c}} }[  \wt{f}_p  ] (\la)
\qquad \e{with} \quad I_Q^{\e{c}} \, = \, I_{\iota} \setminus I_Q 
\enq
it is easily seen, in virtue of \eqref{ecriture rep int moment habille zone fermi infini}, that $\wt{f}_p$ satisfies the linear integral equation 
\beq
 \Big( \e{id} \; - \;  \op{L}_{ I_Q^{\e{c}} }\Big)[  \wt{f}_p  ] (\la)   \; = \; \f{1}{2}\phi_{\iota;a}(\la)  
\enq
where $\op{L}_{ I_Q^{\e{c}} }$ is as defined in \eqref{definition operateur L}. As a consequence, it holds,
\beq
 \wt{f}_p (\la) \; = \; \f{1}{2} \phi_{\iota;a}(\la) \; + \; \Int{ I_Q^{\e{c}} }{} \mf{L}_{ I_Q^{\e{c}} }(\la,\mu) \phi_{\iota;a}(\mu) \cdot \f{ \dd \mu }{2} \;. 
\enq
As a consequence, for any $\la \in \R^+$,
\beq
\ba{ccccc} 
\e{if}    &   a>0    & \e{then} & \wt{f}_p(\la) \, \geq  \,   \f{1}{2} \phi_{\iota;a}(\la) \, > \, 0 & \e{for} \; \; \la \,  > \, 0  \vspace{2mm} \\
\e{if}    &   a<0    & \e{then} & \wt{f}_p(\la) \, \leq  \,  \f{1}{2} \phi_{\iota;a}(\la) \, < \, 0 & \e{for} \; \; \la  \, > \, 0  
\ea \; . 
\enq
Again, either of the two are inconsistent with the constraint $\wt{f}_p(Q) \,  = \, 0$ so that, necessarily, $a=0$. \qed

\subsection{Auxiliary functions}

In this last subsection I introduce a few auxiliary special functions that will 
arise in a later stage of the analysis. These functions are defined as solutions to linear integral equations driven by 
$\e{id} \, + \, \op{K}_{I_Q}$ and thus depend on a free parameter $Q$.

The dressed phase $\vp(\la,\mu \mid Q )$ is defined as the solution to the linear integral equation 
\beq
\Big( \e{id} + \op{K}_{I_Q} \Big)[\vp(*,\mu \mid Q )](\la) \; = \; \f{\th(\la-\mu) }{2\pi } \;. 
\enq
Above, $*$ denotes the running argument of the function on which the integral operator acts. 
The dressed charge $Z(\la\mid Q )$ is defined as the solution to the linear integral equation 
\beq
\Big( \e{id} + \op{K}_{I} \Big)[Z(*\mid Q) ](\la) \; = \; 1  \;. 
\enq

The dressed energy $\veps(\la \mid Q)$ is defined as the solution to the linear integral equation 
\beq
\Big( \e{id} + \op{K}_{I} \Big)[\veps(*\mid Q) ](\la) \; = \; \mf{e}(\la) \qquad \e{with} \qquad \mf{e}(\la) \, = \,  h - 2 J \chi_{\De} \cdot \mf{p}^{\prime}(\la)
\label{definition dressed and bare energies}
\enq
where the constant $\chi_{\De}$ depends on the anisotropy as
\beq
\ba{lclc} 
\chi_{\De}=\sin \zeta   	 & \e{for}  &  \De=\cos(\zeta) &  0 <\zeta< \pi   \\
\chi_{\De}=1  			 & \e{for}  &  \De=1  \\
\chi_{\De}=\sinh(\zeta)  	& \e{for} &  \De= \cosh(\zeta) & \zeta >0 \ea  \, .
\enq

Finally, the thermodynamic counting function is defined in terms of the dressed momentum $p(\la\mid q)$ as 
\beq
\xi_0(\la\mid q) \; = \; p(\la\mid q) + \f{D}{2} \;.
\label{definition fction de cptge thermodynamique}
\enq

\begin{lemme}
\label{Lemme proprietes phase habillet et thermo counting fct}

The dressed phase is related to the dressed charge as
\beq
\vp(\la,Q\mid Q) \, - \,   \vp(\la,-Q\mid Q) \, + \, 1 \; = \; Z(\la\mid Q) \quad and \quad 1+\vp(Q,Q\mid Q) - \vp(-Q,Q\mid Q) \, = \, \f{1}{ Z(Q\mid Q) } 
\label{ecriture identites entre phase et charge habilles} 
\enq
and also satisfies
\beq
%
%
\Dp{\la}\vp(\la,\mu\mid Q) \; = \; R_{I_Q}(\la,\mu) \, + \,R_{I_Q}(\la,Q)\vp(Q,\mu\mid Q)\, - \, \,R_{I_Q}(\la,-Q)\vp(-Q,\mu\mid Q)\;. 
\label{ecriture derivee lambda phase habillee}
\enq

Also, the thermodynamic counting function satisfies $\xi_0^{\prime}(\la\mid q) = \rho(\la\mid q) >0$ and thus is a strictly increasing diffeormorphism from 
\beq
\R \qquad onto \qquad   \xi_0\big( \R \mid q\big) \; = \;   \Big[ -\f{ \pi - \zeta  }{ \pi } \big( \f{1}{2} - D\big) \, ; \, \f{ \pi - \zeta  }{ \pi } \big( \f{1}{2} - D\big)  \, + \, D \Big]   
\enq
for $-1 < \De \leq 1 $ and from $\R$ onto $\R$ when $\De>1$.

\end{lemme}

The proof of most statements is rather straightforward. In fact, the only non-trivial identity corresponds to the relationship with $Z^{-1}(Q\mid Q)$. The latter was established by 
Korepin and Slavnov \cite{KorepinSlavnovNonlinearIdentityScattPhase} and I refer to their paper for more details.

For magnetic fields below the critical field 
\beq
h_c \; = \; \left\{ \ba{ccc}  8J \cos^2(\tf{\zeta}{2}) & -1 < \De=\cos(\zeta) \leq 1 & \zeta \in \intfo{0}{\pi}   \vspace{2mm} \\
				8J \cosh^2(\tf{\zeta}{2}) &  \De=\cosh(\zeta) > 1 & \zeta>0 
\ea \right. 
\enq
the XXZ chain is an antiferromagnet. Furthermore, when $\De>1$, $0<h<h_{c}^{(L)}$ the model has a mass gap with $h_{c}^{(L)}=8J\sinh^2(\tf{\zeta}{2})$
whereas it is massless for $h_{c}^{(L)}\leq h < h_c$.

\begin{prop}
\label{Proposition propriete de dressed energy and fermi boundary}

Assume that the magnetic field $h$ satisfies to the bounds 
\beq
0<h<h_c \quad for \quad -1<\De <1 \qquad and \qquad h_{c}^{(L)}\leq h < h_c \quad for \quad \De >1\;. 
\enq
Then, there exists a unique solution $Q_F$, called the Fermi boundary, such that 
$\veps(Q_F\mid Q_F)=0$. Furthermore, it holds
\beq
\veps(Q\mid Q) \, < \,  0  \qquad for  \; \; Q \in \intoo{ - Q_F }{ Q_F } \quad and \quad 
\veps(Q\mid Q) \, > \,  0  \qquad for \; \; Q \in I_{Q_F}^{ \e{c} } \;. 
\enq
\end{prop}

This result has been established in \cite{KozDugaveGohmannThermoFunctionsZeroTXXZMassless} when $-1< \De < 1$ and the technique of proof can be readily extended to the regime 
$\De \geq 1$.  I refer the interested reader to that paper for more details.

\section{Solvability and boundedness}
\label{Section solvability properties of BAE}

\subsection{Existence of solutions to the logarithmic Bethe equations}

\begin{prop}
\label{Proposition existence solution Log BAE}

$  $

\vspace{2mm}
$\bullet$  $-1<\De \leq 1$

\noindent Let $\zeta  \in \intfo{ 0 }{ \pi }$ parametrise $\De=\cos(\zeta)$ and
\beq
h_1<\dots<h_n\; , \;\; h_a\in \intn{1}{N} \qquad , \quad  p_1<\dots<p_n \; , \;\; p_a \in \mathbb{Z}\setminus \intn{1}{N}\; , 
\label{ecriture domaine des entiers h et p}
\enq
be ordered integers such that 
\beq
\f{\pi - \zeta}{\pi} \Big( \f{1}{2}-\f{N-1}{L} \Big) > \f{ p_n-N }{ L } \quad , \quad  \f{ p_1-1 }{ L } >  - \f{\pi - \zeta}{\pi} \Big( \f{1}{2}-\f{N-1}{L} \Big) 
\quad and \quad  \f{\pi - \zeta}{\pi} \Big( \f{1}{2}-\f{N}{L} \Big) \geq  \f{ n }{ L }\;. 
\label{ecriture contraintes sur positions particules et trous}
\enq
Define the set of integers $\{\ell_a\}_1^N$ by
\beq
\ell_a = a \quad for \;\; a \in \intn{1}{N}\setminus\{h_1,\dots,h_n\} \qquad and \quad \ell_{h_a} = p_a \quad for \;\; a=1,\dots n \;. 
\label{definition entiers ella en termes ha et pa}
\enq
Then, the system of logarithmic Bethe equations \eqref{ecriture Log BAE cas general tout delta} with $\mf{p}$ and $\th$ as defined in \eqref{ecriture bare phase and momentum De entre 1 et -1} 
or \eqref{ecriture bare phase and momentum De egal 1} 
admits a solution such that all Bethe roots $\{ \la_a \}_1^N$ are real. Furthermore, for $\zeta \in \intfo{\tf{\pi}{2}}{\pi}$, this solution is unique. 

\vspace{2mm}

$\bullet$ $\De > 1$ 

\vspace{2mm}

\noindent  Let $\zeta  \in \intoo{ 0 }{ +\infty }$ parametrise $\De=\cosh(\zeta)$ and $h_a\in \intn{1}{N}$ and $p_a \in \mathbb{Z}\setminus \intn{1}{N}$ be any ordered integers, 
\textit{viz}. $h_1<\dots<h_n$ and $p_1<\dots<p_n$. 
Then, the system of logarithmic Bethe equations \eqref{ecriture Log BAE cas general tout delta} with $\mf{p}$ and $\th$ as defined in \eqref{ecriture bare phase and momentum De depasse 1}
admits a solution such that all Bethe roots $\{ \la_a \}_1^N$ are real.

\end{prop}

Note that the proposition only stipulates the existence of solutions in the case of general $\De > -1$. Uniqueness only holds, \textit{a priori}, for $0\geq \De >-1$. 
Furthermore, when $\De>1$, the proposition only states that the $\la_a$'s are real and says nothing about the domain to which they belong. 
The main peculiarity of the $\De>1$ regime is that two solutions of the logarithmic Bethe equations which differ by translations of $\pi$, namely $\la_a=\la^{\prime}_a + m_a\pi$
for $a=1,\dots,N$ and some $m_a \in \mathbb{Z}$ define \textit{equivalent} solution to the Bethe equations. 
Hence, distinct sets of integers $\ell_a$ and $\ell_a^{\prime}$ do not necessarily allow one, when $\De>1$, to distinguish between inequivalent solutions. 
So as to deal with only one representative one should, \textit{in fine} only focus on the solutions belonging to some fixed interval of length $\pi$, say $\intof{ - \tf{\pi}{2} }{ \tf{\pi}{2} }$. 
This is, however, something that should be done \textit{a posteriori}, after having built the solutions. 

\vspace{2mm}

The proof of this statement basically follows Yang-Yang's argument for the ground state Bethe roots when $ -1 <  \De  < 0$. 
The idea consist in introducing a function $\bs{\mu} \mapsto S_{\bs{\ell}}( \bs{\mu})$ on $\R^N$ such that the logarithmic Bethe Ansatz equations 
correspond to the conditions for a local minimum of $ S_{\bs{\ell}}$. The main new observation introduced here is that even though the function $ S_{\bs{\ell}}$
is \textit{not} convex for $\De >0 $, it still blows up at $\infty$ provided that \eqref{ecriture contraintes sur positions particules et trous} holds, and 
hence admits a minimum. 

\Proof

$\bullet$ $-1< \De \leq 1$

Following the reasoning of Yang and Yang, one introduces the function 
\beq
S_{\bs{\ell} }( \bs{\mu} )  \; = \;  \sul{a=1}{N} \f{P(\mu_a)}{2\pi} \; - \; \f{1}{4\pi L} \sul{a,b=1}{N} \Th(\mu_a-\mu_b)  \; - \; 
\sul{a=1}{N} \mu_a \f{ n_a }{  L  } \qquad \e{with} \quad n_a \, =  \,  \ell_a - \f{ N+1 }{ 2 }  
\label{definition action YY}
\enq
defined in terms of 
\beq
P(\la) = \Int{0}{\la} \mf{p}(\mu) \dd \mu   \; = \; (\pi-\zeta)|\la| + 
\left\{ \ba{cc c}   \e{O}( 1  )   \vspace{1mm} \\
		 \e{O}( \ln |\la|  )     \ea \right.  \quad \e{and} \quad  
\Th(\la) = \Int{0}{\la} \theta(\mu) \dd \mu   \; = \;   (\pi-2\zeta)|\la| +
\left\{ \ba{cc c}  \e{O}( 1  )    \vspace{1mm} \\
		  \e{O}( \ln |\la|  )    \ea \right.  \; . 
\label{definition fcts P et Theta pour fnelle Yang Yang}
\enq
The remainders appearing above are such that the first line holds pointwise in  $\zeta \in \intoo{0}{\pi}$ while the second line holds for $\zeta=0$. 
This convention will be carried on until the end of this proof.

It is easy to see that the logarithmic Bethe equations \eqref{ecriture Log BAE cas general tout delta} appear as conditions for the existence of a critical point $\bs{\la}$ of $S_{\bs{\ell}}( \bs{\mu} )$:
\beqa
{\f{ \Dp{} }{ \Dp{}\mu_a} \cdot S_{\bs{\ell}} ( \bs{\mu} ) }_{\mid \bs{\la}=\bs{\mu} }  \; = \;   \f{ \mf{p}(\la_a)  }{ 2\pi } - \f{1}{2\pi L} \sul{b=1}{N} \theta(\la_a-\la_b)  \;  - \; \f{n_a}{L }  \, = \, 0\;. 
\eeqa
Thus, it is enough to show that $S_{\bs{\ell}}( \bs{\mu} ) \tend + \infty$ as $\bs{\mu} \tend \infty$ so as to ensure the existence of a minimum of $S_{\bs{\ell}}( \bs{\mu} )$
and hence the existence of a solution to the logarithmic Bethe equations.

It appears convenient to study the behaviour of $S_{\bs{\ell}}( \bs{\mu} )$ when $\bs{\mu}$ goes to infinity along a ray. 
In such a situation, there exists $\sg \in \mf{S}_N$ such that $\bs{\mu}_{\sg} = (\mu_{ \sg(1) },\dots ,\mu_{ \sg(N) } )$ goes to infinity as
\beq
\bs{\mu}_{\sg}= t \bs{v} \qquad \e{with} \qquad \bs{v} = ( \underbrace{v_1,\dots,v_1}_{\a_1\, \e{terms} } ,\dots,  \underbrace{v_s,\dots,v_s}_{\a_s \, \e{terms} } )
\qquad v_1 < \dots < v_r \leq 0 < v_{r+1} < \dots < v_s \;. 
\enq
There $t \rightarrow + \infty $ and $\bs{v} \in \mathbb{S}^N$, the N-dimensional sphere. 
Given $\bs{\mu}$ as above, when $t\tend +\infty$, it holds 
\beq
S_{\bs{\ell}} ( \bs{\mu} ) \,  =  \,  - t \f{\pi-\zeta}{2\pi} \sul{\ell=1}{r} \a_{\ell}  v_{\ell}
+  t \f{\pi-\zeta}{2\pi}  \sul{\ell=r+1}{s} \a_{\ell}  v_{\ell}
-t\f{\pi-2\zeta}{2\pi L} \sul{j<\ell}{s} \a_{j} \a_{\ell} \big( v_{\ell}-v_j \big)
\;  -  \; \f{ t}{L} \sul{\ell=1}{s}v_{\ell} \ov{n}_{\ell}  
+ \left\{ \ba{c} \e{O}( 1 )  \\ 
	  \e{O}(\ln t) \ea \right.  \;.
\enq
 Here, we have set
\beq
\ov{n}_{\ell} = \sul{j=a_{\ell-1}+1}{a_{\ell}} n_{\sg(j)} \quad \e{and} \quad a_{\ell}=\sul{p=1}{\ell} \a_p \;, \quad  a_0 =0 \;. 
\enq
Therefore, using that 
\beq
\sul{j<\ell}{s} \a_{j} \a_{\ell} \big(v_{\ell}-v_j \big)= \sul{\ell=2}{s} \a_{\ell} v_{\ell} a_{\ell-1}
- \sul{j=1}{s-1}\a_j v_j \, \big(N-a_j \big) \;, 
\enq
one can recast the large $t$ asymptotics of $S_{\bs{\ell}} ( \bs{\mu} )$ as 
\bem
S_{\bs{\ell}} ( \bs{\mu} ) 
%
%
%
=t \; \bigg\{  \sul{\ell=1}{r-1}\pa{ v_{\ell}-v_{\ell+1} } \cdot \Big(  \sul{j=1}{\ell} \tau_{j}^{(-)} \Big)  \; + \; 
v_r \cdot \Big( \sul{j=1}{r} \tau_{j}^{(-)} \Big)   \\
\; + \; \sul{\ell=2+r}{s}\pa{ v_{\ell}-v_{\ell-1} } \cdot \Big( \sul{j=\ell}{s} \tau_{j}^{(+)} \Big)  \; + \; 
v_{r+1} \cdot \Big( \sul{j=1+r}{s} \tau_{j}^{(+)} \Big)  \bigg\} \; 
+ \left\{ \ba{c} \e{O}( 1 )  \\ 
	  \e{O}(\ln t) \ea \right. \; , 
\end{multline}
where 
\beq
\tau_{\ell}^{(\pm)} = \a_{\ell} \; \bigg\{ \pm \f{\pi-\zeta}{2\pi} \; + \; \f{N(\pi-2\zeta)}{2\pi L} \; - \; 
 \f{a_{\ell-1}+a_{\ell}}{2L} \f{\pi-2\zeta}{\pi} \bigg\} \; - \; \f{ \ov{n}_{\ell}  }{L}  \; . 
\enq
One will have $S_{\bs{\ell}} ( \bs{\mu} ) \tend +\infty$ for any $\bs{\mu} \tend \infty$ provided that 
\beq
\sul{j=1}{\ell} \tau_{j}^{(-)} \, < \, 0 \;\; \e{for} \; \e{all} \;\; \ell\in \intn{1}{r} \quad \e{and} \qquad 
\sul{j=\ell}{s} \tau_{j}^{(+)} \, > \, 0 \; \; \e{for} \; \e{all} \;\; \ell\in \intn{r+1}{s} \;  
\label{ecriture conditions sur les entiers tau j plus et moins}
\enq
this for all choices of point $\bs{v}\in \mathbb{S}^N$, \textit{viz}. for all admissible values of $r,s$ and of the  $\a_a$'s.  
These equation impose constraints on the integers $n_a$. Indeed, using that 
\beq
\sul{p=1}{\ell} \a_p \Big( a_{p-1}+a_p \Big) \; = \;  a_{\ell}^2  \;\;\;  , \qquad 
\sul{p=\ell}{s} \a_p \Big( a_{p-1}+a_p \Big) \; = \;  2N \, \wt{a}_{\ell} -\wt{a}_{\ell}^{\, 2}
\quad \e{with}  \quad  
\wt{a}_{\ell} \;  = \;  \sul{j=\ell}{s}\a_k=N-a_{\ell-1} 
\enq
one gets 
\beq
\sul{j=1}{\ell} \tau^{(-)}_j \; = \; 
a_{\ell} \Big( -\f{ \pi - \zeta }{2 \pi }  \; + \; \f{ N(\pi - 2 \zeta) }{ 2 \pi L }  \Big)
- a_{\ell}^2 \f{ (\pi - 2 \zeta) }{ 2 \pi L }   \; - \;  \sul{k=1}{a_{\ell}} \f{ n_{\sg(k)} }{ L } \;,
\enq
and also
\beq
\sul{j=\ell}{s} \tau^{(+)}_j=\wt{a}_{\ell}  \Big( \f{ \pi - \zeta }{2 \pi }  \; - \; \f{ N(\pi - 2 \zeta) }{ 2 \pi L }  \Big)
\;  +  \; \wt{a}_{\ell}^{\,2} \f{ N(\pi - 2 \zeta) }{ 2 \pi L }  \; - \;  \sul{k=1}{\wt{a}_{\ell} } \f{ n_{\sg(N-k+1)} }{ L } \;.
\enq
Now by running through all the possible types of inequalities \eqref{ecriture conditions sur les entiers tau j plus et moins}, one concludes 
that $S_{\bs{\ell}} ( \bs{\mu} ) \tend +\infty$ for any $\bs{\mu} \tend \infty$, provided that, for any $\mc{J} \subset \intn{1}{N}$,  the integers $n_a$ satisfy 
\beq
m \,  \Big( -\f{ \pi  - \zeta }{ 2 \pi }  \; + \; \f{ N( \pi - 2 \zeta) }{ 2 \pi L }  \Big)
- m^2  \f{ \pi-2\zeta}{2\pi L}  \; <  \;  \sul{a \in \mc{J} }{  } \f{ n_{a} }{ L } 
< m  \, \Big( \f{ \pi  - \zeta }{ 2 \pi }   \; - \; \f{ N( \pi - 2 \zeta) }{ 2 \pi L }  \Big)
+ m^2  \f{ \pi-2\zeta}{2\pi L}
\label{ecriture contrainte sur les entiers}
\enq
with $m=\# \mc{J}$. 

Since, when \eqref{ecriture contrainte sur les entiers} are satisfied, $S_{\bs{\ell}} ( \bs{\mu} ) \tend +\infty$ for any $\bs{\mu} \tend \infty$, it follows
that  $S_{\bs{\ell}}$ admits at least one minimum on $\R^N$ at some point $\bs{\la}$. The coordinates of the point realising this minimum  
satisfy to the logarithmic Bethe equations, hence ensuring the existence of solutions. 

\vspace{2mm}

It now remains to check that the constraints \eqref{ecriture contrainte sur les entiers} are always satisfied provided that the bounds 
\eqref{ecriture contraintes sur positions particules et trous} hold. Let $n_0$ be such that 
\beq
p_1< \cdots < p_{n_0} \leq 0 < N < p_{n_0+1} < \cdots < p_n \;. 
\enq
Then, for any  $\mc{J} \subset \intn{1}{N}$, $\# \mc{J}=m$, agreeing upon $m_0=\min(m,n_0)$, it holds
\beq
\sul{a \in \mc{J} }{  } n_{a}  \; \geq \; \sul{a=1}{m_0} \Big\{ p_a - \f{N+1}{2} \Big\} \;+\; \sul{a=1}{m-m_0} \Big\{ a - \f{N+1}{2} \Big\}
\; \geq \; m_0\big(p_1-m+m_0-1 \big) \; + \; \f{m(m+1)}{2} - \f{m(N+1)}{2} 
\enq
where the lowest bound has been obtained by using that $p_a \geq p_1 + a-1$. It thus follows that the lowest bound in \eqref{ecriture contrainte sur les entiers}
holds provided that 
\beq
-m \f{\pi-\zeta}{\pi} \Big( \f{1}{2}-\f{N}{L} + \f{m}{L} \Big) \; < \: \f{m_0}{L} \big(p_1-m+m_0-1 \big) \;. 
\label{condition suffisante sur entier p1}
\enq
Since $m_0\leq m$ for $m\in \intn{1}{n_0}$, these bounds are clearly satisfied as soon as \eqref{ecriture contraintes sur positions particules et trous} holds. 
Further, for $m> n_0$, the  bounds \eqref{condition suffisante sur entier p1}  will hold provided that they hold for the smallest possible choice of $p_1$ 
compatible with \eqref{condition suffisante sur entier p1}, namely by taking it equal to the \textit{lhs} of  \eqref{condition suffisante sur entier p1} when  at $m=m_0=1$. 
This translates itself on the condition for $n_0$:
\beq
-(m-n_0)\Big[ \f{\pi-\zeta}{\pi}\Big( \f{1}{2}-\f{N}{L}+ \f{m+n_0}{L}\Big) - \f{n_0}{L} \Big] - n_0(n_0-1) \f{\pi-\zeta}{\pi L } \; \leq \; 0 \; .  
\enq
Since $m\geq n_0$, for the above to hold, it is enough that the latter always holds for $0<\zeta\leq \tf{\pi}{2}$ and imposes the constraint 
\beq
 \f{\pi-\zeta}{2\zeta- \pi  } \Big( \f{1}{2}-\f{N}{L} \Big) \geq \f{n_0}{L}
\label{ecriture contrainte sur entier n0 cas moins contraignant}
\enq
which is clearly satisfied provided that $n$ is bounded according to \eqref{ecriture contraintes sur positions particules et trous}. 
One can repeat the same reasoning relatively to the upper bound in \eqref{ecriture contrainte sur les entiers} what leads to the sufficient constraints 
\beq
-m \f{\pi-\zeta}{\pi} \Big( \f{1}{2}-\f{N}{L} + \f{m}{L} \Big) \; > \: \f{m_{\ua} }{L} \big(p_n-N+m-m_{\ua} \big) \qquad \e{with} \quad m_{\ua} \, = \, \min (m, n-n_0) \;. 
\label{condition suffisante sur entier pn}
\enq
For the latter to hold, it is enough to have 
\beq
 \f{\pi-\zeta}{\pi} \Big( \f{1}{2}-\f{N}{L} + \f{1}{L} \Big) \; > \: \f{m_{\ua}}{L} \big(p_n-N + m_{\ua}-m \big)  \quad \e{and} \quad  \f{\pi-\zeta}{2\zeta- \pi  } \Big( \f{1}{2}-\f{N}{L} \Big) \geq \f{n}{L} \;.  
\enq
Clearly, both bounds hold if \eqref{ecriture contraintes sur positions particules et trous} holds.

Now, regarding to uniqueness,  observe that for $\zeta \in \intfo{\tf{\pi}{2}}{\pi}$ $S_{\bs{\ell} }(\bs{\mu})$ is strictly convex since it has a strictly positive defined Hessian matrix
\beq
\f{ \Dp{}^2 }{ \Dp{}\mu_k \Dp{}\mu_j} \cdot S_{\bs{\ell} }(\bs{\mu})   =  
\Big[ \f{ \mf{p}^{\prime}(\mu_k) }{ 2\pi } - \f{1}{ L} \sul{b=1}{N} K(\mu_k-\mu_b) \Big] \de_{jk}
\; + \; \f{K(\mu_j-\mu_k)}{ L}  \;. 
\label{ecriture Hessienne Action YY}
\enq
 Indeed, given $(h_1,\dots, h_N) \in \R^N$, one has that 
\beqa
\sul{j,k=1}{N}  h_j h_k \f{ \Dp{}^2 }{ \Dp{}\mu_k \Dp{}\mu_j} \cdot S_{\bs{\ell} }(\bs{\mu})  &=&  
\sul{k=1}{N}  h_k^2  \f{ \mf{p}^{\prime}(\mu_k) }{ 2\pi }  
\; - \; \sul{j,k=1}{N} \f{ (h_j-h_k)^2}{ 2 } \cdot  \f{K(\mu_j-\mu_k)}{ L}  >0\;. 
\label{Hessienne de YY action definie positive}
\eeqa
 Thus $S_{\bs{\ell} }(\bs{\mu}) $ admits at most a single minimum, in this range of $\zeta$'s.  
 
 \vspace{1mm}
 
 $\bullet$ $ \De > 1$

 \vspace{1mm}
 
 The analysis follows basically the same lines as for $-1<\De\leq 1$. Defining $S_{\bs{\ell} }$ as in \eqref{definition action YY} with, now, the functions $P$ and $\Th$ being given by 
\beq
P(\la) = \Int{0}{\la} \mf{p}(\mu) \dd \mu   \; = \; \f{|\la|^2}{2} +  \e{O}(   |\la|  ) 
  \quad \e{and} \quad  
\Th(\la) = \Int{0}{\la} \theta(\mu) \dd \mu   \; = \;   \f{|\la|^2}{2} +  \e{O}(   |\la|  ) 
\enq
one gets that the logarithmic Bethe equations do arise as necessary conditions for a local extremum of $S_{\bs{\ell} }$. Sending $\mu$ to $\infty$
exactly as it was done for $-1<\De \leq 1$ leads to the asymptotic behaviour 
\beq
S_{\bs{\ell}} ( \bs{\mu} ) \,  =  \,   \f{ t^2 }{8\pi L} \sul{a,b=1}{s} \a_{a}\a_b \Big\{  \Big(\f{L}{N}-2 \Big)\big(v_a^2+v_b^2 \big)+  \big(v_a+v_b\big)^2  \Big\}  \; + \; \e{O}( t ) \;.
\enq
Since $\tf{L}{N} \geq 2$, this ensures that  $S_{\bs{\ell} }$ blows up at infinity and thus admits at least one minimum.  \qed

\subsection{Boundedness of solutions to the logarithmic Bethe equations}

I now prove that for  $-1<\De\leq 1$  at  density $D\in \intfo{0}{\tf{1}{2}}$ and under certain restrictions on the allowed range for the $p_a$'s and $n$, the solutions to the logarithmic Bethe equations 
are bounded uniformly in $N,L$. Boundedness also holds for $\De>0$ and $D\in \intff{0}{\tf{1}{2}}$ under slightly different restrictions on the $p_a$'s. 
When $D=1/2$ and $-1<\De\leq 1$, the Bethe roots are not bounded any more. I establish bounds on the proportion of roots lying outside of a compact 
of large size.

\begin{prop}
\label{Proposition bornitude des racines de Bethe}
$ $

\vspace{1mm}

$\bullet$ Let $-1<\De  \leq 1$ 

\vspace{1mm}

Let $N,L$ be such that $\tf{N}{L} \tend D\in \intfo{0}{\tf{1}{2}}$. 
Parametrise  $\De=\cos(\zeta)$ with $\zeta \in \intfo{ 0 }{ \pi }$ and let $\la_1,\dots,\la_N$ correspond to a solution of the logarithmic Bethe Ansatz 
equations \eqref{ecriture Log BAE cas general tout delta}   where the integers $p_1<\dots <p_n$ are subject to the constraint 
\beq
\f{\pi - \zeta}{\pi} \Big( \f{1}{2}-\f{N-1}{L} \Big)  > \f{ p_n-N }{ L } \quad , \quad  \f{ p_1-1 }{ L } >  - \f{\pi - \zeta}{\pi} \Big( \f{1}{2}-\f{N-1}{L}  \Big) 
\label{ecriture borne sur entiers particule-trou}
\enq
and $n$ is fixed and $N$ independent. Then, there exists a $N,L$ independent constants $\La>0$  such that 
\beq
| \la_a |  \leq  \La \quad for \;  any \; a \in \intn{1}{N} \; such \; that \;  \ell_a \in \intn{1}{N} \qquad uniformly \; in \; (N,L) \;. 
\label{bornage uniforme des racines de Bethe} 
\enq

\vspace{1mm}

$\bullet$ Let $\De  > 1$. 

\vspace{1mm}

Let $N,L$ be such that $\tf{N}{L} \tend D\in \intff{0}{\tf{1}{2}}$  and et $\la_1,\dots,\la_N$ correspond to the solution of the logarithmic Bethe Ansatz 
equations \eqref{ecriture Log BAE cas general tout delta}   where the integers $p_1<\dots <p_n$ are such that $n$ is fixed, $L$ independent, and $|p_a/L|\leq C$
for some $L$-independent $C>0$.

\vspace{3mm}

 Then, there exists a $N,L$ independent constants $\La>0$  such that 
\beq
| \la_a |  \leq  \La \quad for \;  any \; a \in \{1,\dots, N\} \qquad uniformly \; in \; (N,L) \;. 
\label{bornage uniforme des racines de Bethe} 
\enq
\end{prop}

Note that given the constraints on the $p_a$'s, solutions to the logarithmic Bethe Ansatz equations do exist in virtue of Proposition \ref{Proposition existence solution Log BAE}.

\Proof

The proof splits in three steps since one has to distinguish between $ -1 <  \De \leq 0$,  $0<\De\leq 1$ and $\De>1$. 
The distinction between the first  two regimes is due to the change in the sign of the kernel $K(\la)$. 
The last regime has to be treated separately due to the change in the periodicity properties of the involved functions.

\vspace{2mm}

$\bullet$ \,  $\De>1$ 

\vspace{2mm}

Let $\{ \check{\la}_{a} \}_1^N$ be the reordered $\la_a$'s, namely\symbolfootnote[3]{There cannot be two $\la_a$'s that are equal since this would contradict 
that the $\ell_a$'s are pairwise distinct.} $\{ \check{\la}_{a} \}_1^N=\{ \la_a \}_1^N$ and $\check{\la}_{1}<\cdots < \check{\la}_{N}$. 
Also, let $\{ \check{\ell}_a \}$ denote the corresponding reordering of the $\ell_a$'s. 
Let 
\beq
\wh{\xi}(\la) \, = \, \f{1}{2\pi}\mf{p}(\la) \, - \, \f{1}{2\pi L} \sul{a=1}{N} \th(\la-\la_a) \, + \, \f{N+1}{2L}
\enq
be the counting function built up from this solution to the Bethe Ansatz equations. 

Since the $\la_a$'s are real, by $\pi$-periodicity of the functions, it follows that 
\beq
\big| \wh{\xi}^{\prime}(\la) \big| \,  \leq  \, \f{1}{2\pi}\mf{p}^{\prime}(0) \, + \, \f{1}{2} K(0) \;. 
\enq
Let $p$ be such that $(p+1)\pi > \check{\la}_N-\check{\la}_1\geq p \pi$. Then, since $\wh{\xi}(x+k\pi)-\wh{\xi}(x)= k \tf{(L-N)}{L}$, one has 
\beq
\f{\check{\ell}_N-\check{\ell}_1 }{ L } \; = \;  \wh{\xi}(\check{\la}_N) - \wh{\xi}(\check{\la}_1)\; = \; \wh{\xi}(\check{\la}_1+ p \pi) - \wh{\xi}(\check{\la}_1) \; + \hspace{-2mm} \Int{ \check{\la}_1 + p \pi }{ \check{\la}_N } \hspace{-2mm} \wh{\xi}^{\prime}(\la) \cdot \dd \la
 \; \geq  \; p \f{ L-N }{ L } \, - \, \f{1}{2}\mf{p}^{\prime}(0) \, - \,  \f{\pi}{2} K(0)  \;. 
\enq
This yields the upper bound on $p$ since the $\check{\ell}_a$'s are bounded in $L$. Thus there exists an $L$-independent constant $\wt{C}>0$ such that $\wt{C} > |\check{\la}_N - \check{\la}_1 |> 0$. 
Since $\th$ is increasing on $\R$, it holds 
\beq
\f{1}{2\pi} \mf{p}(\check{\la}_N) \, \leq  \,  \f{ 1 }{ L } \max \{ \ell_a \} \, - \, \f{N+1}{2L}\, + \, \f{N}{2\pi L} \th( \wt{C} ) \,  =  \, \f{ C^{\prime} }{2\pi} \quad \Rightarrow \quad 
\check{\la}_N \; \leq \; (\mf{p})^{-1}\big( C^{\prime} \big) =\La
\enq
The lower bound on $\check{\la}_1$ is obtained analogously.

\vspace{2mm}

$\bullet$ \,  $\De=\cos(\zeta)$ with $ \zeta \in \intff{0}{\tf{\pi}{2}} $ 

\vspace{2mm}

Reorganise the Bethe roots as 
\beq
\{\la_a\}_1^N \, = \, \{\nu_a\}_1^N \quad \e{with} \quad  \nu_{a} \, = \, \la_{q_a} \quad \e{and} \quad  \nu_{n+1}<\dots< \nu_{N}  \quad \e{with} \quad \ell_{q_a}\in \intn{1}{N} \;\; \; \e{for} \; a=n+1,\dots,N \; .  
\label{introduction parametres nu en termes de lambdas}
\enq
Since $\th$ is increasing in this region of $\zeta$'s and is bounded by $\pi-2\zeta$, one has
\beq
\f{  \mf{p}\big( \nu_{N}\big) }{ 2\pi } \; <  \;   \f{ 1}{ L }\big(\ell_{q_N}-N\big) \; + \; \f{N-1}{2L} \; + \;\f{\pi-2\zeta}{2\pi L} N \;. 
\enq
Since $\ell_{q_N} \leq N$
\beq
\f{  \mf{p}\big( \nu_{N}\big) }{ 2\pi } \; <  \;   \f{\pi-\zeta}{ \pi } \cdot \f{N}{L}  \, < \, \f{\pi-\zeta}{ \pi } (D+\eps)  \Leftrightarrow \nu_{N} \leq (\mf{p})^{-1}\Big( 2 (\pi-\zeta) (D+\eps) \Big) \;. 
\enq
where $\eps>0$ is such that $N/L < D+\eps <\tf{1}{2}$ and $L$ is taken large enough.  

The bound on $\nu_{n+1}$ is obtained analogously, leading to 
\beq
\f{  \mf{p}\big( \nu_{n+1}\big) }{ 2\pi } \; >   \;  - \f{\pi-\zeta}{ \pi } (D+\eps) \Leftrightarrow \nu_{n+1} \geq  - \mf{p}^{-1}\Big( 2 (\pi-\zeta) (D+\eps) \Big) \;. 
\enq
Thus, the claim holds with $\La=\mf{p}^{-1}\Big(2(\pi-\zeta) (D+\eps) \Big)$.

\vspace{2mm}

$\bullet$ \;  $\De=\cos(\zeta)$ with $\zeta \in \intfo{\tf{\pi}{2}}{\pi}$ 

\vspace{2mm}

It is a classical fact that if $\{\la_a\}_1^N$ solve the system of logarithmic Bethe equations associated with the choice of integers $\ell_a$, then the set of roots 
$\{\mu_a\}_1^N$ with $\mu_a =-\la_{N+1-a}$ solve  the system of logarithmic Bethe equations associated with the choice of integers $-\ell_{N+1-a}$. 
Therefore, the claim will follow as soon as one shows that $\limsup_{N,L \tend \infty} [ \nu_{ N } ] <+\infty$, where the $\nu_a$'s are as defined in 
\eqref{introduction parametres nu en termes de lambdas}.

The proof goes by contradiction. Hence, assume that $\limsup_{N,L \tend \infty} [ \nu_{N} ] = + \infty$. 
In other words, there exists a sequence $(L_m, N_m)$, $N_m, L_m \tend +\infty$, $\tf{N_m}{L_m}\tend D$ and the associated Bethe roots $\big\{  \la_a^{(m)} \big\}_1^{N_m}$
associated with the subsequences $\{ \ell_a^{(m)} \}_1^{N_m}$ and chains of $L_m$ sites such that $ \lim_{m\tend+\infty}\nu_{N}^{(m)} = + \infty$. 
 Fix $Q >0$ and set 
\beq
\mf{b}_{Q}^{(m)} = \f{1}{N_m-n} \# \Big\{   a \in \intn{n+1}{N_m} \; : \; \nu_a^{(m)} \geq Q \Big\} \;. 
\enq
Define further
\beq
\ov{\mf{b}}_{Q} = \limsup_{m\tend +\infty}  \mf{b}_{Q}^{(m)}  \qquad \e{and} \qquad 
\ov{\mf{b}} = \limsup_{Q\tend +\infty}  \ov{\mf{b}}_{Q} \;. 
\label{definition parametre b goth et b goth Q}
\enq
By construction, one has $1\geq \ov{\mf{b}} \geq 0$. Furthermore, by definition, there exists an increasing sequence $Q_{\ell}$, 
\newline $Q_{\ell} \limit{\ell}{+\infty} +\infty$, such that $\ov{\mf{b}}_{Q_{\ell}} \tend \ov{\mf{b}}$.

Assume that $1 \geq  \ov{\mf{b}}>0$ and pick any $\eps>0$ such that  $\ov{\mf{b}}-\eps>0$. Let $\ell_0$ be such that 
\beq
\forall \; \ell \geq \ell_0 \quad \ov{\mf{b}}-\f{\eps}{2} < \ov{\mf{b}}_{Q_{\ell}}  < \ov{\mf{b}} + \f{\eps}{2} \;. 
\enq
Also, given a fixed $\ell\geq \ell_0 $, there exists a sequence $r\mapsto m_r(\ell)$  and  $r_0(\ell) \in \mathbb{N}$ such that for any $r \geq r_0(\ell)$ one has
\beq
\ov{ \mf{b} } - \eps \; < \;  \ov{ \mf{b} }_{Q_{\ell}} -\f{\eps}{2}  \; < \;  \ov{ \mf{b} }_{Q_{\ell}}^{( m_r(\ell))} 
\;   <  \; \ov{ \mf{b} }_{Q_{\ell}}  + \f{\eps}{2}   \;< \;  \ov{\mf{b}} + \eps \;. 
\label{ecriture bornes sur bar b}
\enq
%
%
%
%
%
%
%
 %
%
%
%
%
%
One can slightly improve the \textit{rhs} of the bound. Indeed, since $\ov{\mf{b}}_{Q_{\ell_0}} \; < \;  \ov{\mf{b}} \, +  \, \tf{\eps}{2}$, 
there exists $m_0$ such that for any $m \geq m_0$, one has $\mf{b}_{Q_{\ell_0}}^{(m)} < \mf{b} \, +  \, \eps$. 
As $m_r(\ell)$ goes to infinity with $r$, there exists 
\beq
r^{\prime}_1(\ell) \quad \e{such} \; \e{that} \;\;  \forall \; r\,  \geq  \, r^{\prime}_1(\ell) \;  \quad \e{one} \; \e{has}
\quad m_r(\ell) \geq m_0 \;.
\enq
Thence, setting  $r_1(\ell) = \max (r_0(\ell) , r_1^{\prime}(\ell) )$, one has for any $r \geq r_1(\ell)$ that
\beq
 (\ov{\mf{b}} + \eps) (N_{ m_{r}(\ell) }-n) \; \geq  \;  \# \Big\{   a \in \intn{1+n}{N_{m_r(\ell)} } \; : \; \nu_a^{(m_r(\ell))} 
 \; \geq  \; Q_{\ell_0} \Big\}  \;. 
\label{estimations sur nombres de nu grands}
\enq
Since, the total number of the $\la$'s is $N_{m_r(\ell)}$, the above inequality implies that 
\beq
\# \Big\{   a \in \intn{1+n}{N_{m_r(\ell)} } \; : \; \nu_a^{(m_r(\ell) )} < Q_{\ell_0} \Big\} \; \geq \; N_{m_r(\ell)} \cdot  \e{max}\big\{ 0, 1 - \eps  - \ov{\mf{b}}  \big\} \; . 
\enq
Define
\beq
\ov{a} = \big[  N_{m_r(\ell) }(1 + \eps  - \ov{\mf{b}} ) \big] + 1 \qquad \e{and} \qquad 
\underline{a} =  \e{max}\big\{ 0,   \big[  N_{m_r(\ell) }(1 - \eps  - \ov{\mf{b}} ) \big]  \big\}  
\enq
where $[.]$ stands for the integer part. Then, 
\beq
\sul{ k=\ov{a} }{ N_{ m_r(\ell) } } \ell_{q_k} \; \leq \; \sul{k= \ov{a} + n }{ N_{ m_r(\ell) } } k 
\leq \f{1}{2}  ( N_{m_r(\ell)}  \, - \,  \ov{a}  \, + \,  1 ) (  N_{m_r(\ell)}  \, - \,  \ov{a} - n   \, + \,  1 ) \;. 
\enq
Therefore, after summing up the logarithmic Bethe equations involving $\nu_{\ov{a}}, \dots, \nu_{ N_{m_r(\ell)} }$, 
invoking the above bounds and using that $\th$ is odd, one gets  
\bem
 ( N_{m_r(\ell)}  \, - \,  \ov{a}  \, + \,  1 ) \f{ \ov{a} \, - \,  1 }{2 L_{m_r(\ell)} }   
\; \geq \; \f{1 }{ 2\pi } \sul{k=\ov{a}}{N_{m_r(\ell)}} \mf{p} \Big( \nu_k^{ (m_r(\ell)) } \Big)   
\; - \; \f{1}{2\pi L_{m_r(\ell)}} \sul{k=\ov{a}}{N_{m_r(\ell)}} \sul{p=\underline{a}+1}{\ov{a}-1} \theta \Big( \nu_k^{ (m_r(\ell)) } - \nu_p^{ (m_r(\ell)) } \Big) \\ 
\; - \; \f{1}{2\pi L_{m_r(\ell)} } \sul{k=\ov{a}}{N_{m_r(\ell)}} \sul{p=1}{ \underline{a} } \theta \Big( \nu_k^{ (m_r(\ell)) } - \nu_p^{ (m_r(\ell)) } \Big) 
\; - \; \f{1}{2\pi L_{m_r(\ell)} } \sul{k=\ov{a}}{N_{m_r(\ell)}} \sul{p=1}{ n } \theta \Big( \nu_k^{ (m_r(\ell)) } - \nu_p^{ (m_r(\ell)) } \Big)  \;. 
\end{multline}
Using that $- \vartheta (\la)$ is strictly increasing so that for $k > p > n$ one has  $- \theta \Big( \nu_k^{ (m_r(\ell)) } - \nu_p^{ (m_r(\ell)) } \Big) \,  > \, - \theta ( 0 )= 0 $
and that by \eqref {ecriture bornes sur bar b}
\beq
\check{\la}_k > Q_{\ell} \quad \e{for} \; \; k \in \intn{ \ov{a} }{ N_{m_r(\ell)}  }  \qquad \e{and} \qquad 
\check{\la}_k < Q_{\ell_0} \quad \e{for} \; \; k \in \intn{1}{ \underline{a} } 
\enq
by \eqref{estimations sur nombres de nu grands}, one arrives to the bound
\beq
\f{ \ov{a} \, - \,  1 }{2 L_{m_r(\ell)} }    \; \geq \; 
\f{ \mf{p}( Q_{\ell} ) }{2\pi }  \; - \; \f{ \underline{a}  }{2\pi L_{m_r(\ell)} }     \theta (Q_{\ell} - Q_{\ell_0})  \, - \, \f{ 2\zeta-\pi }{2\pi L_{m_r(\ell)} } n  \;. 
\enq
Taking the $r \tend +\infty$ limit, it follows from  

\beq
\lim_{r \tend +\infty} \big( \tf{ \ov{a} }{ L_{m_r(\ell)} } \big) \,  = \,  D \cdot (1+ \eps - \ov{\mf{b}})
\quad \e{and}\quad \lim_{r \tend +\infty} \big( \tf{ \underline{a} }{ L_{m_r(\ell)} }  \big) \, = \,  D \cdot \max\big\{ 0,  (1- \eps - \ov{\mf{b}}) \big\} \; ,
\enq
that 
\beq
 \f{ D}{2} \big( 1+ \eps - \ov{\mf{b}}  \big)  \; \geq  \; 
 \f{ \mf{p}( Q_{\ell} ) }{2\pi }  
\; - \; \f{D}{2\pi  } \max\big\{ 0,  (1- \eps - \ov{\mf{b}}) \big\} \;  \vartheta( Q_{\ell} - Q_{\ell_0} ) \;. 
\enq
Sending first $\ell \tend +\infty$ and then $\eps\tend 0^+$ leads to 
\beq
D \f{ 1  - \ov{\mf{b}}  }{2  } \; \geq  \; \f{\pi -\zeta }{2 \pi }   \; - \; D \f{\pi - 2 \zeta }{2 \pi  } (1 - \ov{\mf{b}} )   \qquad \e{viz}. \qquad 
D \cdot (1-\ov{\mf{b}})    \geq \f{ 1 }{ 2 } \;. 
\label{ecriture inequation sur D}
\enq
However, the last inequality cannot hold since, by hypothesis, $0< D <\tf{1}{2}$. 

Thus, one necessarily has $\ov{b}=0$. This however does not yet guarantee that the roots are bounded from above since a small portion thereof can escape to $+\infty$. 
Thus assume that one has $\check{\la}_{ N_{m_r(\ell)} }^{(m_r(\ell))} \tend +\infty$. Then, staring from the logarithmic Bethe equations for $\check{\la}_{ N_{m_r(\ell)} }^{(m_r(\ell))} $, and using similar bounds 
one gets 
\beq
\f{1}{L_{m_r(\ell)} } \Big( \f{ N_{m_r(\ell)}  \, - \,    1  }{ 2 }    \Big)    \; \geq \; 
 \f{ \mf{p} \big( \nu_{N_{m_r(\ell)}} \big) }{2\pi }    \; - \; \f{ \underline{a} }{2\pi L_{m_r(\ell)} } \theta \big( \nu_{N_{m_r(\ell)}} - Q_{\ell_0} \big)
 \, - \, \f{n(2\zeta-\pi)}{2\pi L_{m_r(\ell)}  }   
\enq
with $\underline{a}$ and other quantities as defined above with the difference that, now, $\ov{\mf{b}}=0$. One can send $r\tend + \infty$ on the level of these bounds
and then relax $\eps \tend 0$ leading to \eqref{ecriture inequation sur D} with  $\ov{\mf{b}}=0$. This yields the sought contradiction.  \qed

In the rest of this subsection, I focus on the case $-1 <\De \leq 1$ when $D=\tf{1}{2}$ and obtain bounds on the proportion of Bethe roots for the ground state which lie away from
some segment $I_{\La}$. 
More precisely, one has the
\begin{prop}
\label{Proposition estimation fraction la GS qui sechappent a infini}
Let $-1 <\De \leq 1$ and $N/L=D=\tf{1}{2}$. Let $\{\la_a\}_1^N$ be a solution to the logarithmic Bethe Ansatz equations  associated with the ground state choice of integers $\ell_a=a$, with $a\in \intn{1}{N}$. Then, for any 
$\eps>0$ there exists $\La_{\eps}, L_0>0$, such that for any $L\geq L_0$,
\beq
\wh{c}_{\La_{\eps}} \leq \eps \qquad where \qquad \wh{c}_{\La} \, = \, \f{1}{L} \# \Big\{ a \in \intn{1}{N} \; : \; \la_a\in \R \setminus I_{\La} \Big\} 
\label{borne et definition de la fraction c Lambda}
\enq
corresponds to the fraction of Bethe roots lying away from the segment $I_{\La}$. 
\end{prop}

\Proof

\subsubsection*{$\bullet$ $0\leq \De \leq  1$}

Straightforward bounds analogous to those developed in the proof of Proposition \ref{Proposition bornitude des racines de Bethe} lead to the upper bounds
\beq
\f{ N-a }{ L }\,+\, \f{ 1 }{ 2L } \; < \; \Int{ \la_a}{+\infty}\mf{p}^{\prime}(\mu) \cdot \f{\dd \mu}{2\pi} \qquad \e{and} \qquad
\f{ a }{  L } \; < \;  \f{1}{2L}  \Int{-\infty}{ \la_a} \mf{p}^{\prime}(\mu) \cdot \f{\dd \mu}{2\pi} \; . 
\enq
Assume that $\la_a>\La$. Then, one gets the upper bound 
\beq
\f{ N- a }{ L } \; < \; \Int{ \La }{+\infty}\mf{p}^{\prime}(\mu) \cdot \f{\dd \mu}{2\pi}  \qquad \e{and} \; \e{setting} \qquad
\wh{k}_{\La}^+\; =\; \bigg[ L \Int{ \La }{+\infty}\mf{p}^{\prime}(\mu) \cdot \f{\dd \mu}{2\pi}  \bigg]+1
\enq
where $[*]$ denotes the integer part, it is clear that there can be at most $\wh{k}_{\La}^+$ distinct integers $\ell_a \in \intn{1}{N}$ satisfying to this constraint.
Likewise, if $\la_a < -\La$, one has 
\beq
\f{ a }{ L } \; < \; \Int{-\infty}{ -\La }\mf{p}^{\prime}(\mu) \cdot \f{\dd \mu}{2\pi}  + \f{1}{2L} \;. \qquad \e{Thus} \; \e{setting} \qquad
\wh{k}_{\La}^-\; =\; \bigg[ L \Int{-\infty}{ -\La }\mf{p}^{\prime}(\mu) \cdot \f{\dd \mu}{2\pi}  \bigg]+2 
\enq
it becomes clear that there can be at most $\wh{k}_{\La}^-$ distinct integers $\ell_a \in \intn{1}{N}$ satisfying to the above constraint.
Hence, all-in-all, one gets the bound
\beq
\wh{c}_{\La} \leq \f{3}{L} \Int{ \R \setminus I_{\La} }{} \mf{p}^{\prime}(\mu) \cdot \f{\dd \mu}{2\pi} \;. 
\enq
The conclusion then follows from the fact that the \textit{rhs} of the inequality approaches $0$ when $L, \La \tend +\infty$.

\subsubsection*{$\bullet$ $-1 <  \De <0 $}

If the sequence $\{ \la_a\}_1^N$ is bounded uniformly in $N$, then the statement holds simply by taking $\La_{\eps}>M$, where $M$ is a bound on the magnitude of the Bethe roots. 
Else, one reasons as in the proof of Proposition \ref{Proposition bornitude des racines de Bethe} so as to establish that $\ov{\mf{b}}=0$ with $\ov{\mf{b}}$ 
as defined in \eqref{definition parametre b goth et b goth Q}. One gets similar bounds relatively to the proportion of roots lower that some fixed parameter $Q$. 
The rest is straightforward. \qed

\section{Asymptotic expansion of the counting function}
\label{Section Large L analysis of counting function}

\subsection{Leading asymptotics of the counting function}

Given a solution $\{\la_a\}_1^N$ to the logarithmic Bethe equations associated with the choice of integers $\{\ell_a\}_{1}^{N}$, it is convenient to introduce the associated counting function
\beq
\wh{\xi}(\la) \; = \; \f{1}{2\pi}\mf{p}(\la) \; - \; \f{1}{2\pi L} \sul{a=1}{N} \th(\la-\la_a) \; + \; \f{N+1}{2L} \;. 
\label{definition fct comptage}
\enq
By construction, this function is such that $\wh{\xi}(\la_a)=\ell_a$ for $a=1,\dots,N$. The purpose of this section is to provide an alternative to \eqref{definition fct comptage} characterisation of the counting function
valid in the large-$L$ regime. This characterisation is obtained by means of a non-linear integral equation. The very idea goes back to the works of De Vega, Woynarowich \cite{DeVegaWoynarowichFiniteSizeCorrections6VertexNLIEmethod}
and Batchelor, Kl\"{u}mper \cite{KlumperBatchelorNLIEApproachFiniteSizeCorSpin1XXZIntroMethod} and was further developed in the works 
\cite{DestriDeVegaAsymptoticAnalysisCountingFunctionAndFiniteSizeCorrectionsinTBAFirstpaper,KlumperBatchelorPearceCentralChargesfor6And19VertexModelsNLIE}. The non-linear integral equations obtained
in the earlier literature were obtained by \textit{assuming} that the counting function satisfies certain properties such as being strictly increasing on the real axis and, on compact subsets of $\R$, having a bounded 
from below derivative, $\wh{\xi}^{\prime}>\varkappa>0$, this unformly in $L$. 
The main point of the method is that once a non-linear integral equation is taken for granted just as certain amount of properties of its solution, then it is relatively easy to 
compute, order-by-order, the coefficients of its large-$L$ asymptotic expansion. 
The assumptions which allow one to derive the non-linear integral equation for the counting function and which also allow one to derive its large-$L$ asymptotic expansion 
could, in the best case scenario, be verified \textit{a posteriori}, namely on the level of the obtained form for the large-$L$ asymptotic expansion. This only allowed
for a consistency test of the calculations.

\noindent The main input of the analysis that I develop below is to set techniques allowing one
\begin{itemize}
\item[i)]  to prove that, for $L$-large enough, the counting function can indeed be characterised as the unique solution to a non-linear integral equation and that it does indeed satisfy to the expected properties, in particular, 
that it is strictly increasing on $\R$;
\item[ii)] to demonstrate that the counting function admits a large-$L$ asymptotic expansion up to  $\e{o}(L^{-1})$ corrections.
\end{itemize}
In other words, the framework developed below allows one to step out of the formal handling of asymptotic expansions. 
Prior to stating the result and going into the details of the proof, I need to introduce several building blocks of the non-linear integral equations satisfied by 
$\wh{\xi}$. 

Given $\wh{D}=\tf{N}{L}\in \intff{ 0 }{ \tf{1}{2} }$, here and in the following $\wh{q}$ 
will denote the unique solution to the magnetic Fermi boundary problem \eqref{ecriture probleme pour la bord de Fermi magnetique} associated with $\wh{D}$. In its turn, $q$ will denote its thermodynamic limit, 
\textit{viz}. the unique solution to the Fermi boundary problem \eqref{ecriture probleme pour la bord de Fermi magnetique} associated with $D=\lim \wh{D}$.

\begin{defin}
\label{Definition operateur NLIE} 

Let $f$ be real analytic function such that it  is a biholomorphism on some open neighbourhood 
of a segment of $\R$ and such that its range contains, for some $\a_0>0$ small enough, the closed set $\wh{G}(\a_0)$ depicted on Fig.~\ref{Figure domaine G alpha hat}.   
Then, one can construct the three below non-linear integral operators
\beq
\mf{R}_{N;1}\big[ \, f  \, \big](\la) \; = \;  - \,  \sul{\eps = \pm }{}\Int{ \wh{\Ga}_{\eps} }{} 
\f{  R_{I_{\wh{q}} }\big( \la , f^{-1}(z)  \big) }{ f^{\prime}\big( f^{-1}(z) \big) } \cdot \ln\Big[1- \ex{2\i \pi \eps L z} \Big] \cdot \f{ \dd z}{2\i \pi L } 
\enq
and agreeing upon 
\beq
\wh{q}_R[f] =  f^{-1}\Big(  \f{ N + \tf{1}{2} }{ L }  \Big) \qquad , \qquad \wh{q}_L[f] =  f^{-1}\Big(  \f{1}{ 2 L }  \Big) 
\enq
one has
\beq
\mf{R}_{N;2}\big[ \, f  \, \big](\la) \; =   - \hspace{-2mm} \Int{ \wh{q} }{ \wh{q}_R[f] } \hspace{-2mm} R_{ I_{\wh{q}} }(\la,\mu) \Big[f(\mu) - f\big( \wh{q}_R[f] \big) \Big] \cdot \dd \mu \qquad and \qquad 
\mf{R}_{N;3}\big[ \, f  \, \big](\la) \; =   - \hspace{-2mm} \Int{ \wh{q}_L[f] }{ -\wh{q} }\hspace{-2mm}  R_{ I_{\wh{q}} }(\la,\mu) \Big[f(\mu) - f\big( \wh{q}_L[f] \big) \Big] \cdot \dd \mu   
 \;. 
\enq
These operators, build up a "master" non-linear integral operator as
\beq
\mf{R}_{N}\big[ \, f \, \big](\la) \; = \;    \sul{a=1}{3}\mf{R}_{N;a}\big[ f \big](\la)\;. 
%
%
\enq

\end{defin}

\begin{figure}[ht]
\begin{center}

\begin{pspicture}(7,7)

\psline[linestyle=dashed, dash=3pt 2pt]{->}(1,4)(6.7,4)
\psdots(1.5,4)(6,4) 
\rput(1.1,3.5){$\f{1}{2L}$}
\rput(6.8,3.4){$\f{N+\tf{1}{2}}{L}$}

\rput(0.7,5.9){$\f{1}{2L}+\i\a$}
\rput(6.8,6){$\f{N+\tf{1}{2}}{L}+\i\a$}

\rput(0.5,1.8){$\f{1}{2L}-\i\a$}
\rput(6.8,1.6){$\f{N+\tf{1}{2}}{L}-\i\a$}

\psline{-}(1.5,2.5)(1.5,5.5)
\pscurve{-}(1.5,5.5)(1.65,5.9)(2,6)

\psline{-}(2,6)(5.5,6)
\pscurve{-}(5.5,6)(5.85,5.9)(6,5.5)

\psline{-}(6,5.5)(6,2.5)
\pscurve{-}(6,2.5)(5.85,2.1)(5.5,2)

\psline{-}(5.5,2)(2,2)
\pscurve{-}(2,2)(1.65,2.1)(1.5,2.5)




\rput(3.5,6.5){ $ \wh{\Ga}_{+} $ }
\rput(4,1.5){ $ \wh{\Ga}_{-} $ }

\psline[linewidth=2pt]{->}(5,2)(5.1,2)
\psline[linewidth=2pt]{->}(4,6)(3.9,6)

\end{pspicture}
\caption{ Closed set $\wh{G}(\a)$ delimited by its boundary  $\wh{\Ga}_{ +} \cup \wh{\Ga}_{ - }$. \label{Figure domaine G alpha hat} }
\end{center}
\end{figure}

\noindent Just as in Section \ref{Section solvability properties of BAE}, I will consider
\beq
h_1<\dots<h_n\, ,  \,  h_a \in \intn{1}{N} \qquad \e{and} \quad  p_1<\dots<p_n \, , \,  p_a \in \mathbb{Z}\setminus \intn{1}{N} 
\enq
where $n \in \mathbb{N}$ is $N$-independent. However, the integers $h_a$ and $p_a$ can depend on $N$ provided that they satisfy to certain bounds:

\begin{itemize}
\item[$\bullet$] for $-1<\De\leq 1$, $\De=\cos(\zeta)$, in addition, the integers $\{p_a\}_1^n$ are assumed to satisfy to the additional constraint 
\beq
\f{\pi - \zeta}{\pi} \Big( \f{1}{2}-\f{N-1}{L} \Big)   > \f{ p_n-N }{ L } \quad , \quad  \f{ p_1-1 }{ L } >  - \f{\pi - \zeta}{\pi} \Big( \f{1}{2}-\f{N-1}{L}  \Big) \;. 
\label{ecriture condition pour particules-trous delta moins que 1}
\enq
\item[$\bullet$] For $\De>1$, the integers $\{p_a\}_1^n$ are solely assumed to be bounded as 
\beq
 \big| \tf{p_a}{L} \big| \leq C \,  \qquad \e{for} \; \e{some}\; L-\e{independent} \;\;  C>0. 
\label{ecriture condition pour particules-trous delta plus que 1}
\enq
\end{itemize}

In the course of the proof, it will be useful to introduce the sets 
\beq
%
%
%
\mc{S}_{\eta }(J) \; = \; \Big\{ z \in \Cx \; : \; | \Im(z) | \, < \,  \eta \; , \; \Re(z) \in J \Big\} 
\enq
and
\beq
\mc{S}_{\eta, \eps} (J) \; = \; \Big\{ z \in \Cx \; : \; | \Im(z) |  \, < \,  \eta  \; , \;  d( \Re(z) , J) \,  < \,  \eps \Big\}
\label{definition voisinages boites autour de intervalle}
\enq
where $J \subset \R$ is a subset of $\R$ and $d$ is the distance between subsets of $\R$ induced by the Euclidian distance.

Finally, it will also appear convenient to introduce the parameter $\varkappa_{\De}$ defined by 
\beq
\ba{c l c c } -1 < \De < 1  \qquad & \varkappa_{\De} \, = \, \tf{\zeta}{4} \qquad &   \De=\cos(\zeta)  &   0<\zeta<\pi  \vspace{2mm} \\
 \De=1  \qquad &   \varkappa_{\De} \, = \, \tf{1}{4}   &     \vspace{2mm}\\ 
 \De > 1  \qquad & \varkappa_{\De}=\tf{\zeta}{4} & \De=\cosh(\zeta)   & \zeta>0    \ea   \;.
\label{definition varkappa Delta}
\enq

\begin{theorem}
\label{Theorem convergence et eqn NL pour ctg fct}

 Given $\De>-1$, let $n\in \mathbb{N}$ be fixed  and the integers  $\{\ell_a\}_1^N$ be defined  
in terms of integers $\{p_a\}_1^n$ and $\{h_a\}_1^n$ according to \eqref{definition entiers ella en termes ha et pa}.
Assume that these integers satisfy, depending on the value of $\De$, the conditions \eqref{ecriture condition pour particules-trous delta moins que 1}-\eqref{ecriture condition pour particules-trous delta plus que 1}. 
Let $N,L$ be such that  $0 \leq \wh{D}=\tf{N}{L} \leq \tf{1}{2}$ and  go to infinity so that $\wh{D} \tend D $ with 
\beq
D \in \intfo{0}{\tf{1}{2}}  \quad for \; \; -1 < \De \leq 1 \quad and \quad D \in \intff{0}{\tf{1}{2}}  \quad for \; \De > 1 \;. 
\enq
Finally, let $\{\la_a\}_1^N$ denote a solution to the logarithmic Bethe equations subordinate to the associated integers $\{\ell_a\}_1^N$. 

Then 
\begin{itemize}
\item[i)] there exists $L_{0}$ large enough such that, for any $L\geq L_0$ the associated counting function \eqref{definition fct comptage} is a strictly increasing diffeomorphism from 
\beq
\R \qquad  onto  \qquad \bigg[ - \f{ \pi-\zeta }{ \pi }\Big( \f{1}{2}-\f{N}{L}\Big) + \f{1}{2L}  \, ; \,  \f{N}{L} +\f{1}{2L} +\f{ \pi-\zeta }{ \pi }\Big( \f{1}{2}- \f{N}{L} \Big)    \bigg]   
\enq
when $-1<\De \leq 1$ while, for $\De >1$, $\wh{\xi}$ is a diffeomorphism from $\R$ onto $\R$. 

\vspace{2mm}

\item[ii)] There exists an open neighbourhood $\wh{U}$ of $\intff{-q}{q}$ and an $L$-independent open neighbourhood $V_{D}$ of $\intff{0}{D}$ 
containing $\wh{G}(\a)$ delimited by the curves $\wh{\Ga}_{\pm}$, \textit{c.f.} Fig. \ref{Figure domaine G alpha hat}, for some $\a_0>0$ such that $\wh{\xi} : \wh{U} \mapsto  V_D$ is a biholomorphism.

\vspace{2mm}

\item [iii)] There exists $C>0$ such that  
\beq
\norm{ \wh{\xi}(*) - \xi_0(*\mid \wh{q}\, ) }_{ L^{\infty}\big( \mc{S}_{ \varkappa_{\De} }(\R) \big)  } \; \leq \; \f{C}{L} \qquad with \qquad \xi_0( \la \mid \wh{q} \, ) \, = \, p( \la \mid \wh{q} \, ) \, + \, \f{ \wh{D} }{ 2 }\;. 
\label{ecriture estimation distance xi hat a xi zero}
\enq
Above, $*$ denotes the running variable of the functions.

\vspace{2mm}

\item[iv)] The counting function solves the non-linear integral equation 
\beq
\wh{\xi}(\la) \; = \; \xi_0(\la\mid \wh{q} \, ) \; + \;\f{1}{ L} \cdot \Phi_{\wh{q}}^{(0)}\big( \la \mid \{\wh{x}_{p_a} \}_1^n \, ; \, \{\wh{x}_{h_a} \}_1^n \big)
\; + \; \mf{R}_N\big[ \, \wh{\xi}  \, \big](\la)\quad , \;\;
\label{ecriture NLIE pour la ctg fct}
\enq
with
\beq
\wh{x}_a \, = \, \wh{\xi}^{\, -1}\Big( \f{a}{L} \Big)\qquad a \in \Big\{ h_1,\dots, h_n, p_1,\dots, p_n\Big\} \; . 
\enq
The function $\Phi_{\wh{q}}^{(s)}$ is expressed in terms of the dressed phase and charge as
\beq
 \Phi_{\wh{q}}^{(s)}\big( \la \mid \{ y_a \}_1^m \, ; \, \{ z_{a} \}_1^n \big) \; = \;  \f{1}{2 } \big[ 1 + s Z(\la\mid \wh{q}\,) \big]
 \, - \; \sul{a=1}{m} \vp\big(\la, y_{a}\mid \wh{q} \big) \, + \, \sul{a=1}{n}  \vp\big(\la, z_{a}\mid \wh{q} \big)  
\label{definition fct Phi spin et q hat dependent}
\enq
and the operator $\mf{R}_N$ is as given in Definition \ref{Definition operateur NLIE}.

\item[v)] The non-linear integral equation \eqref{ecriture NLIE pour la ctg fct} admits a unique solution in the class of functions satisfying to ii) and iii) 
when $\De>1$ or $-1<\De\leq 1$ and, on top of the previous conditions, the $p_a/L$ all belong to an $L$-independent compact subset of 
\beq
\bigg] - \f{ \pi-\zeta }{ \pi }\Big( \f{1}{2}-D\Big)  \, ; \,  D  + \f{ \pi-\zeta }{ \pi }\Big( \f{1}{2} - D \Big)    \bigg[ \;. 
\enq

\end{itemize}

\vspace{2mm}

The constant $C$ in \eqref{ecriture estimation distance xi hat a xi zero} and $L_0$ appearing above only depend on $n$ and are uniform in $\wh{D}$ belonging to compact subsets of $\intfo{0}{\tf{1}{2}}$
for $-1 < \De \leq 1 $ and throughout the segment $\intff{0}{\tf{1}{2}}$ for $\De >1$.

\end{theorem}

In virtue of Proposition \ref{Proposition existence solution Log BAE}, given integers $\{h_a\}_1^n$ and $\{p_a\}_1^n$ satisfying to
the constraints \eqref{ecriture condition pour particules-trous delta moins que 1}-\eqref{ecriture condition pour particules-trous delta plus que 1}, depending on the value of $\De$,  
 there are always solutions to the logarithmic Bethe Ansatz equations. When $ -1< \De \leq 0 $, the solution was shown to be unique but, 
\textit{a priori}, when $ \De>0$, there could exist more than one solution to the logarithmic Bethe Ansatz equations
associated with this given choice of integers. The statement of the theorem does hold for any such solution.

\Proof

For each value of $N,L$, one is provided with integers $\{h_a\}_1^n$ and $\{p_a\}_1^n$, which, as mentioned, in most cases do depend on $N$ and $L$. In virtue of Proposition \ref{Proposition existence solution Log BAE},
these give rise to a sequence $\{\la_a\}_1^N$, indexed by $L$, of solutions to the logarithmic Bethe Ansatz equations. 
By Proposition \ref{Proposition bornitude des racines de Bethe}, equation \eqref{bornage uniforme des racines de Bethe}, 
there exists an $L$-independent $\La>0$ such that:
\begin{itemize}
\item[$\bullet$] $|\la_a| \leq \La$  for $a\in \intn{1}{N}$ if $\De >1$, \vspace{2mm}
\item[$\bullet$] $|\la_a| \leq \La$  for $a\in \intn{1}{N}$ such that $L\, \wh{\xi}(\la_a) \in \intn{1}{N}$ in the regime  $-1 < \De  \leq 1$,
\end{itemize}
\vspace{1mm}
this uniformly in $N,L$. This sequence of Bethe roots $\{\la_a\}_1^N$ defines the associated counting function through \eqref{definition fct comptage}, 
therefore giving rise to a sequence (in respect to $L$) of counting functions $\wh{\xi}$. These counting functions are readily seen to be holomorphic functions on the strip $\mc{S}_{ 2 \varkappa_{\De} }(\R)$. 
Furthermore, straightforward bounds show there exists an $L$-independent constant $B>0$ such that
\beq
\norm{ \wh{\xi} }_{ L^{\infty}\big( \mc{S}_{\varkappa_{\De}  }(I_{2\La} ) \big) } \; \leq \;  B \;. 
\label{ecriture bornage global de xi hat}
\enq
Since $\wh{\xi}$ is a sequence of holomorphic functions on $ \mc{S}_{ \varkappa_{\De} }(I_{2\La} )$ that are uniformly bounded in $L$,
by Montel's theorem, it admits a converging subsequence  $\wh{\xi}_{\e{e}}$  which converges, in the sup norm topology on compacts subsets, to a holomorphic function $ \xi_{\e{e}}$ on   $  \mc{S}_{ \varkappa_{\De} }(I_{2\La} ) $.

The strategy of the proof consists in characterising the limit $\xi_{\e{e}}$ of such a convergent subsequence. It will be shown that 
$ \xi_{\e{e}} $ necessarily coincides with the so-called thermodynamic counting function $\xi_0(\cdot \mid q)$, \textit{c.f.} \eqref{ecriture estimation distance xi hat a xi zero}. Worded differently, 
any converging  subsequence of $\wh{\xi}$ has the \textit{same} limit. 
Since, by Montel's theorem, any subsequence of $\wh{\xi}$ admits a converging subsequence, $\wh{\xi}$ necessarily converges to $\xi_0(\cdot \mid q)$. 
Once that the convergence is established, the form of the non-linear integral equation \eqref{ecriture NLIE pour la ctg fct} satisfied by $\wh{\xi}$ follows
rather easily from the properties of the limit $\xi_0$. A straightforward investigation of the non-linear integral equation ensures the uniqueness of solutions for $L$ large-enough
and the bounds \eqref{ecriture estimation distance xi hat a xi zero}.

In order to lighten the notations, I will subsequently drop the subscript $\e{e}$ in all the considerations namely the subsequence and its limit will still be denoted by $\wh{\xi}$ and $\xi$. 
Since $\xi^{\prime}$ is holomorphic on $I_{2\La}$, it admits, a finite number of zeroes and thus may only change signs a finite number of times.

In the first part of the proof, I will assume that $\xi^{\prime}>0$ and show that this property is enough so as to characterise the limit. 
In the second part of the proof, I will establish the uniqueness of solutions to the non-linear 
integral equation. Finally, in the third part of the proof, I will rule out the possibility that $\xi^{\prime}$ has zeroes or changes sign on $\intff{-\La}{\La}$.

\subsubsection*{ $\bullet$  $\xi^{\prime}>0$ on $\intff{-\La}{\La}$ }

In virtue of Proposition \ref{Proposition invertibilite de fL en fct de f}, there exists $\eta, \eps>0$ such that 
\beq
\xi \; : \; \mc{S}_{2\eta,  2\eps}( I_{\La} ) \tend \xi\big(  \mc{S}_{2\eta,  2\eps}( I_{\La} )  \big) \; \; \e{is} \; \e{a} \; \e{biholomorphism} \;\e{satisfying} \; \; \; 
\xi \big(  \mc{S}_{2\eta,  2\eps}( I_{\La} )  \cap \mathbb{H}^{\pm} \big)  \; = \;  \xi\big(  \mc{S}_{2\eta, 2\eps}( I_{\La} ) \big) \cap \mathbb{H}^{\pm} \;. 
\nonumber
\enq
Furthermore, for $L$ large enough, it holds that $\wh{\xi} \big(   \mc{S}_{2\eta, 2\eps}( I_{\La} )  \big)  \supset  \xi\big( \mc{S}_{\eta, \eps} ( I_{\La} ) \big) \supset \wh{\xi}(I_{\La})$ and 
\beq
\wh{\xi} \, : \, \mc{S}_{2\eta, 2\eps}( I_{\La} ) \cap \wh{\xi}^{-1}\Big(   \xi\big( \mc{S}_{\eta, \eps}( I_{\La} ) \big)   \Big)  \; \tend  \; \xi\big( \mc{S}_{\eta, \eps} ( I_{\La} ) \big)
\label{domaine de depart et image de xi hat}
\enq
is a biholomorphism. Besides $\wh{\xi}$ is strictly increasing on $I_{\La+\eps}$.

Since all the Bethe roots $\la_a$ such that $L\wh{\xi}(\la_a) \in \intn{1}{N}$ are contained in $I_{\La} $ where $\wh{\xi}$ is increasing 
and owing to  $\e{min}\big\{ \ell_a \big\} \leq n$ and  $\e{max}\big\{ \ell_a \big\} \geq N-n$, it follows that 
\beq
\Big[ \f{n}{L} ; \f{N-n}{L} \Big] \subset \wh{\xi}\big( I_{\La} \big) \;. 
\enq
Moreover,  for $L$ large enough 
\beq
 \wh{\xi} \big(  I_{\La} \big) \subset \Big\{ x \in \R \; : \;  d\big( x, \xi \big(  I_{\La} \big)  \big)  <   \norm{ \wh{\xi}-\xi }_{ L^{\infty}\big( \mc{S}_{  \varkappa_{\De}  }(I_{2\La}) \big) }   \Big\}
\subset  \xi \big( I_{\La+\eps/2 } \big) \;. 
\enq
Then, passing to the limit in \eqref{domaine de depart et image de xi hat}, it holds that 
\beq
\intff{0}{D} \subset  \xi \big( I_{\La+\eps/2 } \big) \quad \Longrightarrow  \quad \intff{-v}{D+v} \subset   \xi \big( I_{\La+\eps } \big) 
\enq
for some $v>0$ small enough. This inclusion ensures that 
\beq
\Big[ \f{1}{2L} \, ; \,  \f{ N+\tf{1}{2} }{ L }  \Big] \subset \xi\big( \mc{S}_{\eta, \eps}(I_{\La}) \big) \;. 
\enq
Thus, due to \eqref{domaine de depart et image de xi hat}, the points 
\beq
\wh{q}_L \; = \; \wh{\xi}^{-1}\Big( \f{1}{2L} \Big) \qquad \e{and} \qquad \wh{q}_R \; = \; \wh{\xi}^{-1}\Big( \f{N+1/2}{L} \Big)
\enq
are well defined and belong to $\mc{S}_{ \eta , \eps}( I_{\La} ) $. Moreover,  when $\De>1$, one has the bound 
\beq
\wh{q}_R-\wh{q}_L  \leq \pi
\label{ecriture borne sup sur hat q R et L}
\enq
as ensured by the 
fact that $\wh{\xi}$ is strictly increasing on $I_{\La+\eps}$ and satisfies $\wh{\xi}(x+\pi)- \wh{\xi}(x)=(L-N)/L \leq \tf{N}{L}$ owing to $2N\leq L$.

Furthermore, since $\xi\big( \mc{S}_{\eta, \eps}(I_{\La}) \big)$ is an open $L$-independent neighbourhood 
of the compact interval 
\beq
\intff{ - \tf{v}{2} }{ D + \tf{v}{2} }\supset \Big[  \f{1}{2L} ; \f{ N+\tf{1}{2} }{ L }  \Big] \;, 
\enq
by compactness, there exists $\a>0$ such that the domain $\wh{G}(\a)$ as depicted in Fig.~\ref{Figure domaine G alpha hat} 
satisfies  $\wh{G}(\a) \subset \xi\big( \mc{S}_{\eta, \eps}( I_{\La} ) \big) $. 

Finally, for any $\tf{a}{L} \in \xi\big( \mc{S}_{ \eta, \eps }( I_{\La} )  \big)$,  it is possible to define $\wh{x}_a=\wh{\xi}^{-1}\big( \tf{a}{L} \big)$. 
This holds, in particular, for $a=h_p$, $p\in \intn{1}{n}$.
Furthermore, owing to  $|\la_a| \leq \La$, if $\ell_a\in \intn{1}{N}$ the strict increase of $\wh{\xi}$ on $I_{\La}$ ensures that 
\beq
\Big\{  \la_a \; : \; \ell_{a} \in \intn{1}{N}  \Big\} \, = \,    \{\wh{x}_a \}_1^N\setminus \{ \wh{x}_{ h_a } \}_1^n  \; .
\enq
Finally, define $\{ \wh{x}_{p_a} \}_1^n = \{\la_a\}_1^N \setminus \Big\{ \{\wh{x}_a \}_1^N\setminus \{ \wh{x}_{ h_a } \}_1^n \Big\} $. One needs to recourse to such a definition since
it could be that $\tf{p_a}{L} \not \in \xi\big( \mc{S}_{\eta,\eps}(I_{\La}) \big) $. However, for those $ \tf{p_a}{L} \in \xi\big( \mc{S}_{\eta,\eps}(I_{\La}) \big) $ one does have
$\wh{x}_{p_a} = \wh{\xi}^{-1}(\tf{p_a}{L})$.

These properties being established, it follows from a straightforward computation of residues and the strict increase of $\wh{\xi}$ on $I_{\La+\eps}$ that
\beq
\wh{\xi}(\la) \; = \; \f{ \mf{p}(\la) }{ 2\pi } \; - \; \f{1}{2\pi L} \sul{a=1}{n} \Big[ \th\big(\la-\wh{x}_{p_a}\big) \, - \, \th\big(\la-\wh{x}_{h_a}\big)   \Big]
\; + \; \f{N+1}{2L} \; - \; \Oint{ \wh{\mc{C}} }{}  \f{ \th(\la-\mu) \wh{\xi}^{\prime}(\mu)  }{ \ex{2\i\pi L \wh{\xi}(\mu) } -1  } \cdot \f{ \dd \mu  }{2\pi } 
\enq
where the contour $\wh{\mc{C}}$ is defined as $\wh{ \mc{C} } \,  =  \, \wh{\xi}^{-1}\Big(\wh{\Ga}_+\cup\wh{\Ga}_- \Big)$ and $\wh{\Ga}_{\pm}$ have been depicted in Fig.~\ref{Figure domaine G alpha hat}. The expression can be further rearranged. Namely, 
setting $\wh{ \mc{C} }_{\eps} \,  =  \, \wh{\xi}^{-1}\Big(\wh{\Ga}_{\eps} \Big)$, one has
\bem
\; - \; \Oint{ \wh{\mc{C}} }{}  \f{ \th(\la-\mu) \wh{\xi}^{\prime}(\mu)  }{ \ex{2\i\pi L \wh{\xi}(\mu) } -1  } \cdot \f{ \dd \mu  }{2\pi } \; = \; 
-\Int{ \wh{q}_L }{ \wh{q}_R } \wh{\xi}^{\prime}(\mu)  \th(\la-\mu) \cdot \f{ \dd \mu  }{2\pi } 
\; + \; \sul{\eps = \pm }{}  \eps \Int{ \wh{\mc{C}}_{\eps} }{}  \f{ \th(\la-\mu) \wh{\xi}^{\prime}(\mu)  }{ \ex{ - 2\i\pi \eps L \wh{\xi}(\mu) } -1  } \cdot \f{ \dd \mu  }{2\pi } \\
\; = \; -\f{ N  }{ 4\pi L} \Big\{  \th\big( \la - \wh{q}_R \big) \, + \, \th\big( \la - \wh{q}_L \big) \Big\} \; - \; 
\Int{ \wh{q}_L }{ \wh{q}_R }  K(\la-\mu) \wh{\xi}_{\e{sym}}(\mu)  \cdot \dd \mu \; + \; \mf{r}_1\big[ \wh{\xi} \big](\la) \; . 
\end{multline}
In the second line, appears the function
\beq
\wh{\xi}_{\e{sym}}(\la) \; = \;  \wh{\xi}(\la) \, - \,  \f{N+1}{2L}
\label{definition version sym de fct de cptge}
\enq
and $\mf{r}_1$ is the remainder 
\beq
\mf{r}_1\big[ \wh{\xi} \, \big](\la) \; = \;  - \; \sul{\eps = \pm }{}\Int{ \wh{\Ga}_{\eps} }{} 
\f{  K\big( \la - \wh{\xi}^{-1}(z)  \big) }{ \wh{\xi}^{\prime}\big( \wh{\xi}^{-1}(z) \big) } \cdot \ln\Big[1- \ex{2\i \pi \eps L z} \Big] \cdot \f{ \dd z}{2\i \pi L } \;. 
\label{definition reste operateur integral r1}
\enq
Note that the second line is obtained by carrying out an integration by parts followed by a change of variables in what concerns the form of the remainder $\mf{r}_1\big[ \wh{\xi} \big](\la)$
given in \eqref{definition reste operateur integral r1}.

Thus, all-in-all, one gets the representation:
\bem
\wh{\xi}_{\e{sym}}(\la) \; + \; \Int{ \wh{q}_L }{ \wh{q}_R }  K(\la-\mu) \, \wh{\xi}_{\e{sym}}(\mu)  \cdot \dd \mu 
\; = \; \f{ \mf{p}(\la) }{ 2\pi } -\f{ N  }{ 4\pi L} \Big\{  \th\big( \la - \wh{q}_R \big) \, + \, \th\big( \la - \wh{q}_L \big) \Big\} \\
\; - \; \f{1}{2\pi L} \sul{a=1}{n} \Big[ \th\big(\la-\wh{x}_{p_a}\big) \, - \, \th\big(\la-\wh{x}_{h_a}\big)   \Big]
 \; + \; \mf{r}_1\big[ \wh{\xi} \big](\la) \; .
\label{ecriture NLIE for hat xi form pour premiere limite}
\end{multline}
All is now in position to write down the form of the non-linear integral equation \eqref{ecriture NLIE pour la ctg fct} satisfied by $\wh{\xi}$. 
The latter, along with the characterisation of the limits of $\xi$ and $\wh{q}_R$ and $\wh{q}_L$ will then allows one to 
infer the bounds \eqref{ecriture estimation distance xi hat a xi zero} given in point $ii)$. 

Starting from \eqref{ecriture NLIE for hat xi form pour premiere limite},  one splits the integration versus $K$ as
\beq
 \Int{ \wh{q}_L }{ \wh{q}_R }  K(\la-\mu) \, \wh{\xi}_{\e{sym}}(\mu)  \cdot \dd \mu  \; = \; 
 \Bigg\{ \Int{ \wh{q}_L }{ -\wh{q} } \; + \; \Int{ -\wh{q} }{ \wh{q} } \; + \;   \Int{ \wh{q} }{ \wh{q}_R } \, \Bigg\} K(\la-\mu) \, \wh{\xi}_{\e{sym}}(\mu)  \cdot \dd \mu     \; . 
\enq
and observes that it is possible to recast the combination of $\th$ functions as 
\bem
\f{ N  }{ 4\pi L} \Big\{  \th\big( \la - \wh{q}_R \big) \, + \, \th\big( \la - \wh{q}_L \big) \Big\} \;=\;
\f{ N  }{ 4\pi L} \Big\{  \th\big( \la - \wh{q}  \, \big) \, + \, \th\big( \la + \wh{q}   \, \big) \Big\} \\ 
-\wh{\xi}_{\e{sym}}(\, \wh{q}_R)  \Int{ \wh{q} }{ \wh{q}_R }  K(\la-\mu)   \cdot \dd \mu 
-\wh{\xi}_{\e{sym}}( \, \wh{q}_L)  \Int{ \wh{q}_L }{ -\wh{q} }  K(\la-\mu)   \cdot \dd \mu \;. 
\end{multline}
These rewritings allow one to recast  \eqref{ecriture NLIE for hat xi form pour premiere limite} in the form 
\bem
\Big( \e{id}+\op{K}_{ I_{\wh{q}} }  \Big)\big[\, \wh{\xi}_{\e{sym}}](\la) \; = \;  \f{ \mf{p}(\la) }{ 2\pi } -\f{ N  }{ 4\pi L} \Big\{  \th\big( \la - \wh{q} \,  \big) \, + \, \th\big( \la + \wh{q} \, \big) \Big\} \\
\; - \; \f{1}{2\pi L} \sul{a=1}{n} \Big[ \th\big(\la-\wh{x}_{p_a}\big) \, - \, \th\big(\la-\wh{x}_{h_a}\big)   \Big]
 \; + \; \sul{a=1}{3} \mf{r}_a\big[ \, \wh{\xi} \,  \big](\la) 
\label{ecriture NLIE forme quasi finale}
\end{multline}
where $\mf{r}_1\big[ \, \wh{\xi} \,  \big] $ is as given in \eqref{definition reste operateur integral r1} while 
\beq
\mf{r}_2\big[ \, \wh{\xi} \,  \big](\la) \; = \;  -\Int{ \wh{q} }{ \wh{q}_R }   K(\la-\mu) \, \big[ \wh{\xi} (\mu) \, - \, \wh{\xi} ( \, \wh{q}_R )  \big]  \cdot \dd \mu  
\enq
and 
\beq
\mf{r}_3\big[ \, \wh{\xi} \,  \big](\la) \; = \; - \Int{ \wh{q}_L }{ -\wh{q} }   K(\la-\mu) \, \big[ \wh{\xi} (\mu) \, - \, \wh{\xi} ( \, \wh{q}_L )  \big]  \cdot \dd \mu   \;. 
\enq
The form of the non-linear integral equation given in the body of the theorem then follows upon acting with the inverse operator $ \e{id}-\op{R}_{ I_{\wh{q}} } $
on both sides of \eqref{ecriture NLIE forme quasi finale} and observing that the first line in the \textit{rhs} gives rise, upon this action, to $p( \la\mid\,  \wh{q} \, \big)$.

\subsubsection*{ $\bullet$ Characterisation of the limit when $\xi^{\prime}>0$ on $\intff{-\La}{\La}$}

Equation \eqref{ecriture NLIE for hat xi form pour premiere limite} is enough so as to characterise the limit $\xi$. Indeed,  define $q_L=\xi^{-1}(0)$ and $q_R=\xi^{-1}(D)$. Then, one has  
\bem
 |\wh{q}_{L} - q_L|\; \leq \; \big|  \wh{\xi}^{-1}\big( \f{1}{2L} \big) - \xi^{-1}\big( \f{1}{2L} \big)   \big| \, + \, 
 \big|  \xi^{-1}\big( \f{1}{2L} \big) - \xi^{-1}\big( 0 \big)   \big|  \\ 
 \; \leq \; C_1 \norm{ \wh{\xi}-\xi }_{ L^{\infty}\big( \mc{S}_{ \varkappa_{\De} }(I_{2\La})\big) } \; + \;  \f{C_2}{ 2 L}
 \norm{ \big(\xi^{-1}\big)^{\prime} }_{ L^{\infty}\big(  \xi\big( \mc{S}_{\eta,\eps}( I_{\La} ) \big)  \big) } =\e{o}(1) \;. 
\label{bornes sur qhat L vs qL}
\end{multline}
Note that, in the second chain of bounds, I used that 
\beq
\norm{\wh{\xi}^{-1}-\xi^{-1}  }_{ L^{\infty}\big( \xi\big( \mc{S}_{\eta, \eps}( I_{\La} ) \big) \big) } \; \leq \; C \cdot 
\norm{\wh{\xi}-\xi  }_{ L^{\infty}\big( \mc{S}_{ \varkappa_{\De} }(I_{2\La})  \big) } ,
\enq
as ensured by Proposition \ref{Proposition invertibilite de fL en fct de f}. 
Similarly to \eqref{bornes sur qhat L vs qL}, one concludes that $ |\wh{q}_{R} - q_R|=\e{o}(1)$. These estimates brought together with \eqref{ecriture borne sup sur hat q R et L} 
ensure that, for $\De>1$, $q_R-q_L \leq \pi$.

It remains to bound $\mf{r}_1\big[ \wh{\xi} \big](\la) $. For any $\la \in \R$ one has 
\beq
\big| \mf{r}_1\big[ \wh{\xi} \big](\la) \big| \; \leq \;  \norm{ K }_{L^{\infty}\big(\mc{S}_{2\eta}(\R) \big) } \cdot \Big\{ \inf_{ \mc{S}_{2\eta,2\eps}( I_{\La} ) } |\wh{\xi}^{\prime} |  \Big\}^{-1}
\cdot \sul{\eps=\pm }{}  \Int{ \wh{\Ga}_{\eps} }{} \Big|\ln \big[ 1 - \ex{2\i\pi \eps L z} \big]  \Big|   \cdot   \f{ | \dd z |  }{ 2\pi L } \; \leq \; \f{C^{\prime} }{ L^2 } \;. 
\label{equation bornage  reste 1}
\enq
The last bound follows from  $\inf_{ \mc{S}_{2\eta,2\eps}( I_{\La} ) } |\wh{\xi}^{\prime} | >  \frac{1}{2} \inf_{ \mc{S}_{2\eta,2\eps}( I_{\La} ) } |\xi^{\prime} | >0$  for $L$ large enough. 
Also the $1/L^{2}$ decay in \eqref{equation bornage  reste 1} follows after an asymptotic estimation of the integral by a variant of Watson's lemma: the boundaries  of $\wh{\Ga}_{\eps}$ 
generate an algebraic decay in $L$ starting with $\e{O}(L^{-1})$ while all other contributions to the integral are exponentially small. The uniform convergence of $\wh{\xi}_{\e{sym}}$ to $\xi_{\e{sym}}=\xi-\tf{D}{2}$ and the above bounds
are enough so as to take the $L\tend +\infty$ on the level of \eqref{ecriture NLIE for hat xi form pour premiere limite}. This yields the system of equation for three unknowns:
the function $\xi_{\e{sym}}$
\beq
\Big( \e{id} \, + \, \op{K}_{\intff{q_L}{q_R} }\Big)\big[ \xi_{\e{sym}} \big](\la) \; = \; \f{ p_0(\la) }{ 2\pi } -\f{ D  }{ 4\pi } \Big\{  \th\big( \la - q_R \big) \, + \, \th\big( \la - q_L \big) \Big\} 
\label{ecriture problem pour xi s eqn integral}
\enq
and the endpoints of integration $q_L, q_R$  
\beq
\xi_{\e{sym}}(q_R)=-\xi_{\e{sym}}(q_L)=\f{D}{2} \;.
\label{ecriture problem pour xi s conditions bords}
\enq

Also, the additional condition $q_R-q_L \leq \pi$ is imposed for $\De>1$. In virtue of Proposition \ref{Proposition unique solvabilite de la borne de Fermi magnetique}, there exists a unique 
solution to \eqref{ecriture problem pour xi s eqn integral}-\eqref{ecriture problem pour xi s conditions bords} given by $q_R=-q_L=q$ and $\xi_{\e{sym}}(\la)=p(\la\mid q)$, with $q$ the magnetic Fermi boundary associated with $D$.  
Thus 
\beq
\wh{q}_R \, =\,  q  \, + \,  \e{o}(1) \qquad  \e{and} \qquad  \wh{q}_L \, = \,  - q  \, + \,  \e{o}(1) \; . 
\label{ecriture cvges ordre zero hat q R et L}
\enq

\subsubsection*{ $\bullet$ Estimates in $L$}

One is now almost in position to establish the bounds \eqref{ecriture estimation distance xi hat a xi zero} given in point $iii)$.
What still remains is to improve the control on the rate at which $\wh{q}_{R}$, resp. $\wh{q}_{L}$, approaches $q$, resp. $-q$. 
First observe that $\wh{q}=q(\wh{D})$ where the smooth, strictly increasing, diffeomorphism $s\mapsto q(s)$ has been introduced  in the paragraph that followed
\eqref{ecriture probleme pour la bord de Fermi magnetique}. Since $\wh{D}\tend D$, this representation for $\wh{q}$ ensures that $\wh{q}\tend q$. 
Therefore, $\wh{q}_R-\wh{q}=\e{o}(1)$ and $\wh{q}_L+\wh{q}=\e{o}(1)$  due to \eqref{ecriture cvges ordre zero hat q R et L}.

Direct bounds ensure that 
\beq
\norm{ \mf{R}_{N;2}\big[ \, \wh{\xi} \,  \big] }_{ L^{\infty}(\mc{S}_{\varkappa_{\De}}(\R) )} \; = \; 
\norm{ R }_{ L^{\infty}\big( \mc{S}_{\varkappa_{\De}}(\R) \times  I_{2\La}   \big) } \cdot \norm{ \wh{\xi}^{\prime} }_{ L^{\infty}( I_{2\La} ) } \cdot 
\Int{ \wh{q} }{ \wh{q}_R } |\mu-\wh{q}_R|\, \dd \mu = \e{O}\Big( (\,\wh{q}_R-\wh{q}\,)^2  \Big) \;,  
\enq
 where one uses that, analogously to \eqref{ecriture bornage global de xi hat},  $\wh{\xi}^{\prime}$ is bounded by  some $N,L$  independent constant. 
In a similar way, one shows that 
\beq
\norm{ \mf{R}_{N;3}\big[ \, \wh{\xi} \,  \big] }_{ L^{\infty}(\mc{S}_{\varkappa_{\De}}(\R)) } \; = \;  \e{O}\Big( (\,\wh{q}_L+\wh{q}\,)^2  \Big) \;. 
\enq
Finally, analogously to \eqref{equation bornage  reste 1}, one infers that $ \norm{ \mf{R}_{N;1}\big[ \, \wh{\xi} \,  \big] }_{ L^{\infty}(\mc{S}_{\varkappa_{\De}}(\R) )} = \e{O}\big( L^{-2} \big)$. 
These estimates on the remainders and the form of the Taylor expansions, 
\beqa
\xi_0(\, \wh{q}_R\mid \wh{q}) & = & \f{N}{L} \, + \, \xi_0^{\prime}( \wh{q} \mid \wh{q} )\cdot ( \, \wh{q}_R-\wh{q} \, ) \, + \, \e{O}\Big(  (\, \wh{q}_R-\wh{q}\,)^2 \Big) \\
\xi_0(\, \wh{q}_L\mid \wh{q}) & = &  \xi_0^{\prime}( -\wh{q} \mid \wh{q} )\cdot ( \, \wh{q}_L+\wh{q} \, ) \, + \, \e{O}\Big(  (\, \wh{q}_L+\wh{q}\,)^2 \Big) \;, 
\eeqa
allow one to use the non-linear integral equation  \eqref{ecriture NLIE pour la ctg fct} as a means to obtain an equation for $\wh{q}_R+q$ and $\wh{q}_L-q$. Using it to recast 
$\wh{\xi}(\,\wh{q}_R\,)$ one gets the estimates:
\bem
\wh{\xi}(\,\wh{q}_R\,)\; = \; \f{ N +\tf{1}{2}}{L}\,= \,  \f{N}{L} \, + \, \xi_0^{\prime}( \wh{q} \mid \wh{q} )\cdot ( \, \wh{q}_R-\wh{q} \, ) \, + \, \e{O}\Big(  (\, \wh{q}_R-\wh{q}\,)^2 \Big) \\
\, +\,  \f{1}{ L} \cdot \Phi_{\wh{q}}^{(0)}\big( \wh{q}_R \mid \{\wh{x}_{p_a} \}_1^n \, ; \, \{\wh{x}_{h_a} \}_1^n \big)  \; + \; \e{O}\Big( \f{1}{L^2}+ (\, \wh{q}_R-\wh{q}\,)^2  +  (\, \wh{q}_L+\wh{q}\,)^2 \Big)
\end{multline}
which can be recast as
\beq
\xi_0^{\prime}( \wh{q} \mid \wh{q} )\cdot ( \, \wh{q}_R-\wh{q} \, ) \cdot \big(1 \, + \, \e{o}(1) \big) \; = \; \e{O}\Big( \f{1}{L } \Big) + \e{O}\Big(  (\, \wh{q}_L+\wh{q}\,)^2 \Big) \;. 
\label{ecriture equation definissant q har R a ordre dominant}
\enq
Analogous reasoning relatively to an evaluation at $\wh{q}_L$ lead to 
\beq
\xi_0^{\prime}( -\wh{q} \mid \wh{q} )\cdot ( \, \wh{q}_L+\wh{q} \, ) \cdot \big(1 \, + \, \e{o}(1) \big) \; = \; \e{O}\Big( \f{1}{L } \Big) + \e{O}\Big(  (\, \wh{q}_R-\wh{q}\,)^2 \Big) \;. 
\enq
This last equation provides one with an estimate of the remainder  $\e{O}\Big(  (\, \wh{q}_L+\wh{q}\,)^2 \Big)$ in \eqref{ecriture equation definissant q har R a ordre dominant}:
\beq
\e{O}\Big(  (\, \wh{q}_L+\wh{q}\,)^2 \Big) \, = \,   \e{O}\Big( \f{1}{L^2 } \Big) + \e{O}\Big(  (\, \wh{q}_R-\wh{q}\,)^4 \Big)
\enq
so that one can recast \eqref{ecriture equation definissant q har R a ordre dominant} in the form
\beq
\xi_0^{\prime}( \wh{q} \mid \wh{q} )\cdot ( \, \wh{q}_R-\wh{q} \, ) \cdot \bigg(1 \, + \, \e{o}(1)\, + \, \e{O}\Big(  (\, \wh{q}_R-\wh{q}\,)^3 \Big)  \bigg) \; = \; \e{O}\Big( \f{1}{L } \Big) \; \;  , 
\qquad i.e. \quad \wh{q}_R-\wh{q} \, = \, \e{O}\big( L^{-1} \big)\;. 
\enq
Likewise, one infers that  $\wh{q}_L+\wh{q}  \, = \, \e{O}\big( \tf{1}{L } \big)$. 

It remains to focus on $\Phi^{(0)}_{\wh{q}}$. When $-1< \De \leq 1$, it is given by a  finite sum of bounded functions on $\ov{\mc{S}}_{\varkappa_{\De}}(\R)$, so that 
putting together all the obtained estimates leads directly to the bounds \eqref{ecriture estimation distance xi hat a xi zero}.
The situation demands slightly more care when $\De > 1$ owing to the quasi-periodicity property of $\vartheta(\la,\zeta)$: $\vartheta(\la+\pi,\zeta)=\vartheta(\la,\zeta)+2\pi$ which holds 
for $|\Im(\la)|<\zeta$
A direct inspection of the linear integral equation satisfied by the dressed phase leads to the same quasi-periodicity properties: 
$\vp( \la + \pi,\mu\mid \wh{q}\, ) \, = \, \vp( \la  ,\mu\mid \wh{q}\, ) +2 \pi$ for $|\Im(\la)|<\zeta$. Since $\Phi^{(0)}_{\wh{q}}$ is expressed as the difference of an equal number of $\vp$
functions, it is  $\pi$-periodic and, as such, bounded on $\ov{\mc{S}}_{\varkappa_{\De}}(\R)$. This ensures the estimates in $iii)$ for this range of $\De$. 

%
%
%
%
%
%
%
%

\subsubsection*{ $\bullet$  Diffeomorphism property}

Lemma \ref{Lemme proprietes phase habillet et thermo counting fct} ensures that $\xi_0^{-1}\big(\intff{0}{\wh{D}}\big)=\intff{ - \wh{q} }{ \wh{q} }$. 
Furthermore, since $\wh{D}\tend D$ with either $D<\tf{1}{2}$ for  $-1< \De \leq 1 $ or $D\in \intff{0}{\tf{1}{2}}$ for $\De>1$, one has for $L$ large-enough that $\wh{q}<2 q < +\infty$.
Then, by Proposition \ref{Proposition invertibilite de fL en fct de f}, one readily deduces the biholomorphism property from \eqref{ecriture estimation distance xi hat a xi zero}. 

 Then straightforward bounds in the non-linear term 
$\wh{\mf{R}}^{\prime}_N[\wh{\xi}]$ based on the fact that 
\beq
\sup_{ \mu \in \mc{S}_{2\eta, 2\eps}(I_{\wh{q}} ) } \Big\{  \big| \Dp{\la}R_{I_{\wh{q}} }(\la,\mu) \big| + \big| R_{I_{\wh{q}}}(\la,\mu) \big| \Big\} \, \leq \, C \cdot g_{\De}(\la) \qquad \e{with} \quad g_{\De}(\la) \; = \; 
\left\{ \ba{ccc}  1   & \e{for} & \De >1 \hspace{1mm} \\
		\tf{ 1 }{ ( \la^2+1 ) }& \e{for}   & \De =1  \hspace{1mm}\\ 
		  \tf{1}{ \cosh(2\la) }  & \e{for} & -1 <\De < 1  
		  \ea \right. 
\enq
ensure that 
\beq
\wh{\xi}^{\prime}(\la) \; = \; \xi_0^{\prime}(\la\mid \wh{q}) \; + \; \e{O}\Big( \f{1}{L} g_{\De}(\la)  \Big)
\enq
with a remainder uniform in $L$ and $\la \in \R$. Since $\tf{g_{\De}}{ \xi_0^{\prime} }$ is bounded on $\R$, this ensures that, for $L$ large enough 
$\wh{\xi}^{\prime}>0$. Therefore, $\wh{\xi}$ is a strictly increasing diffeomorphism from $\R$ onto $\wh{\xi}(\R)$. The explicit form  
of the range for $-1<\De \leq 1$ follows from computing the limits of $\wh{\xi}(\la)$ at $\la \tend \pm \infty$ starting from the definition \eqref{definition fct comptage} of the counting function. 
When $\De>1$, the finite  difference growth $\wh{\xi}(x+\pi)- \wh{\xi}(x)=(L-N)/L$ ensures that $\wh{\xi}$ is a strictly increasing diffeomorphism form $\R$ onto $\R$.

\subsubsection*{ $\bullet$  Uniqueness of solutions to the non-linear integral equation}

By hypothesis, for $-1<\De\leq 1$, there exists $0<v$ small enough such that $\tf{1}{2}-D-v>0$ and
\beq
\f{ p_a }{ L } \in \mc{K}_v \, = \,  \Big[ -\f{\pi-\zeta}{\pi}\Big(\f{1}{2}-D - v)\Big) \, ; \, \f{\pi-\zeta}{\pi}\Big(\f{1}{2}-D-v \Big) + D  \Big] \qquad \e{for} \qquad a=1,\dots, n\, . 
\enq
Then, by Lemma \ref{Lemme proprietes phase habillet et thermo counting fct}, for $-1 < \De \leq 1 $ one has $\xi_0^{-1}( \mc{K}_v \mid \wh{q} \, ) \subset I_{\ga}  $
for some $\ga>0$ and uniformly in $L$. Likewise, when $\De>1$, given $C>0$ such that $|\tf{p_a}{L}|<C$, one has $\xi_0^{-1}\big( I_{2C} \mid \wh{q}\,\big)\subset I_{\ga} $ for some $\ga>0$ and 
uniformly in $L$.

Suppose that one is given two solutions $\wh{\xi}_{1}, \wh{\xi}_{2}$ to the non-linear integral equation \eqref{ecriture NLIE pour la ctg fct} and satisfying to all the 
other requirements, \eqref{ecriture estimation distance xi hat a xi zero} in particular. In virtue of Proposition \ref{Proposition invertibilite de fL en fct de f}, 
there exists $\eta, \eps>0$ such that  for any $z \in \xi_0\big( \mc{S}_{\eta,\eps}(I_{\ga})  \mid \wh{q}\,\big)$, one has 
\beq
\wh{\xi}_{a}^{-1}(z) \; = \; \Oint{ \Dp{} \mc{S}_{2\eta,2\eps}(I_{\ga})   }{} \hspace{-3mm}  \f{  \la \cdot \wh{\xi}^{\prime}_a(\la) }{ \wh{\xi}_a(\la) - z } \cdot \f{ \dd \la }{2\i \pi } 
\qquad \e{and} \quad \inf_{ \la \in  \Dp{} \mc{S}_{2\eta,2\eps}(I_{\ga})  } \inf_{z \in \mc{S}_{\eta,\eps}(I_{\ga}) } | \wh{\xi}_a(\la) - z | > c_0 >0
\label{ecriture rep. int. pour xihat a inverse}
\enq
for $a=1,2$ and some constant $c_0$ only depending on $\xi_{0}(*\mid q)$. Likewise, one has
\beq
\norm{\wh{\xi}_{a}^{-1} \, - \,  \xi_{0}^{-1}(*\mid \wh{q} )  }_{ L^{\infty}\big(  \xi_0\big( \mc{S}_{\eta,\eps}(I_{\ga}) \mid \, q \big) \, \big) } \; \leq  \; C \cdot 
\norm{\wh{\xi}_{a} \, - \,  \xi_{0}(*\mid \wh{q} ) }_{ L^{\infty}\big(   \mc{S}_{ \varkappa_{\De} }(I_{2\ga}) \big) }  \qquad a=1,2 \;, 
\label{ecriture bornage distance xi hat -1 a xi0 -1}
\enq
and, after  straightforward bounds in \eqref{ecriture rep. int. pour xihat a inverse}, one gets that, for some constant $C>0$,
\beq
\norm{\wh{\xi}_{1}^{-1} \, - \,  \wh{\xi}_{2}^{-1} }_{ L^{\infty}\big(  \xi_0\big( \mc{S}_{\eta,\eps}(I_{\ga}) \mid \, q \, \big)  \, \big) } \; \leq  \; C \cdot 
\norm{\wh{\xi}_{1} \, - \,  \wh{\xi}_{2} }_{ L^{\infty}\big(   \mc{S}_{  \varkappa_{\De} }(I_{2\ga}) \big) }  \;. 
\label{ecriture bornage difference des inverses de xi hat 1 et 2}
\enq

I stress that, in virtue of the condition satisfied by the $p_a$'s, provided that $L$ is large enough, for all 
\beq
\ell \in \Big\{ h_1,\dots, h_n, p_1,\dots,p_n  \Big\}\cup \intff{0}{N+1} \qquad \e{it} \; \e{holds} \qquad \f{\ell}{L} \in \xi_0\big( \mc{S}_{\eta,\eps}(I_{\ga}) \mid \wh{q} \,  \big) \;. 
\enq

All is now in place to estimate the norm $\norm{\wh{\xi}_{1} \, - \,  \wh{\xi}_{2} }_{ L^{\infty}\big(   \mc{S}_{  \varkappa_{\De} }(I_{\ga}) \big) }$ by using the non-linear integral equation satisfied by $\wh{\xi}_{a}$. 
Agreeing upon $\wh{x}_{\ell;a}=\wh{\xi}^{-1}_a\big( \tf{\ell}{L} \big)$, in virtue of  \eqref{ecriture bornage difference des inverses de xi hat 1 et 2},
it holds 
\bem
\Norm{  \Phi_{\wh{q}}^{(0)}\big( * \mid \{\wh{x}_{p_a;1} \}_1^n \, ; \, \{\wh{x}_{h_a;1} \}_1^n \big) \, - \, \Phi_{\wh{q}}^{(0)}\big( * \mid \{\wh{x}_{p_a;2} \}_1^n \, ; \, \{\wh{x}_{h_a;2} \}_1^n \big)  }_{ L^{\infty}\big(   \mc{S}_{  \varkappa_{\De} }(I_{2\ga}) \big) } \\
\, \leq   \; \f{C n }{L} \norm{ \Dp{2}\vp  }_{L^{\infty}\big(   \mc{S}_{  \varkappa_{\De} }(I_{2\ga}) \times \R \big)  } \cdot \norm{\wh{\xi}_{1} \, - \,  \wh{\xi}_{2} }_{ L^{\infty}\big(   \mc{S}_{  \varkappa_{\De} }(I_{2\ga}) \big) }  \;. 
\end{multline}
Further, define
\beq
\wh{q}_{R;a} \, = \, \wh{\xi}_{a}^{-1}\Big( \f{N+1}{2L} \Big) \qquad \e{and} \qquad \wh{q}_{L;a} \, = \, \wh{\xi}_{a}^{-1}\Big( \f{1}{2L} \Big) \;. 
\enq
In order to bound $\mf{R}_{N;2}\big[ \, \wh{\xi}  \, \big](\la)$, it is enough to observe that, in virtue of \eqref{ecriture bornage distance xi hat -1 a xi0 -1}, 
\beq
\big| \wh{q}_{R;a}- \wh{q}   \big| \; \leq \;  \f{C}{L} \qquad \e{and} \qquad \big| \wh{q}_{R;1}-  \wh{q}_{R;2}   \big|  \; \leq \; C \cdot \norm{\wh{\xi}_{1} \, - \,  \wh{\xi}_{2} }_{ L^{\infty}\big(   \mc{S}_{  \varkappa_{\De} }(I_{2\ga}) \big) } 
\; \leq \; \f{  1  }{ L } \cdot 2 C^{\prime}\;. 
\enq
Then, it holds, for any $\la \in \mc{S}_{ \varkappa_{\De} }(I_{2\ga})$ 
\bem
\Big| \mf{R}_{N;2}\big[ \, \wh{\xi}_1  \, \big](\la) \, - \,  \mf{R}_{N;2}\big[ \, \wh{\xi}_2  \, \big](\la) \Big| \; \leq \; 
\bigg| \Int{ \wh{q} }{ \wh{q}_{R;1} } \hspace{-1mm} R_{ I_{\wh{q}}  }(\la,\mu) \big[\wh{\xi}_1(\mu) - \wh{\xi}_2(\mu) \big] \dd \mu  \bigg| \; + \; 
\bigg| \Int{ \wh{q}_{R;1} }{ \wh{q}_{R;2} } \hspace{-1mm} R_{ I_{\wh{q}} }(\la,\mu) \big[\wh{\xi}_2(\mu) - \wh{\xi}_2(\wh{q}_{R;2}) \big] \dd \mu  \bigg| \\
\; \leq \; \norm{ R_{ I_{\wh{q}} } }_{ L^{\infty}\big( \mc{S}_{ \varkappa_{\De} }^2(I_{2\ga})\big) } \cdot
\bigg\{   \big| \wh{q} - \wh{q}_{R;1}\big|  \cdot  \norm{ \wh{\xi}_{1} \, - \,  \wh{\xi}_{2} }_{ L^{\infty}\big(   \mc{S}_{ \varkappa_{\De} }(I_{2\ga}) \big) }  
\; + \;  \big| \wh{q}_{R;2}  - \wh{q}_{R;1}\big|^2  \norm{ \wh{\xi}_{2}^{\prime} }_{  L^{\infty}\big( \mc{S}_{\eps,\eta}(I_{\ga}) \big) }  \bigg\}\; \leq \; 
\f{ \wt{C} }{ L } \cdot  \norm{ \wh{\xi}_{1} \, - \,  \wh{\xi}_{2} }_{ L^{\infty}\big(   \mc{S}_{ \varkappa_{\De} }(I_{2\ga}) \big) }   \;. 
\nonumber
\end{multline}
An identical type of bound can be obtained for $\mf{R}_{N;3}$. Finally,  using that, for $L$ large enough
and independent of $a$
\beq
\norm{ \wh{\xi}_{a}^{\prime} }_{  L^{\infty}\big( \mc{S}_{2\eta,2\eps}(I_{\ga}) \big)  } \, \geq \, \f{1}{2} \norm{ \xi^{\prime}_0(*\mid \wh{q}\,) }_{  L^{\infty}\big( \mc{S}_{2\eta,2\eps}(I_{\ga}) \big)  } >0
\enq
one bounds $\mf{R}_{N;1}$ as 
\beq
 \norm{  \mf{R}_{N;1}\big[ \, \wh{\xi}_1  \, \big] \, - \,  \mf{R}_{N;1}\big[ \, \wh{\xi}_2  \, \big] }_{ L^{\infty}\big(   \mc{S}_{ \varkappa_{\De} }(I_{2\ga}) \big) }   \; \leq \; 
\f{ C^{\prime} }{ L } \cdot  \norm{ \wh{\xi}_{1} \, - \,  \wh{\xi}_{2} }_{ L^{\infty}\big(   \mc{S}_{ \varkappa_{\De} }(I_{2\ga}) \big) }  \;. 
\enq
By taking the difference of the two non-linear integral equations satisfied by $\wh{\xi}_{1}$ and $\wh{\xi}_{2}$, the various bounds obtained earlier lead to 
\beq
 \norm{ \wh{\xi}_{1} \, - \,  \wh{\xi}_{2} }_{ L^{\infty}\big(   \mc{S}_{ \varkappa_{\De} }(I_{2\ga}) \big) } \; \leq \;  
 \f{C^{\prime\prime}}{L} \cdot  \norm{ \wh{\xi}_{1} \, - \,  \wh{\xi}_{2} }_{ L^{\infty}\big(   \mc{S}_{ \varkappa_{\De} }(I_{2\ga}) \big) } \;. 
\enq
The latter can only hold for $L$ large enough provided that $ \wh{\xi}_{1} \, = \,  \wh{\xi}_{2}$, thus entailing uniqueness of solutions.

\subsubsection*{ $\bullet$ $\xi^{\prime}$ is not necessarily positive on $\intff{-\La}{\La}$.}

In this last part of the proof, I study the case where, \textit{a priori} the limit $\xi^{\prime}$  of the extracted subsequence is not positive and 
show that such a situation cannot arise. 

Prior to going into the details of the analysis, one should observe that this situation cannot arise for $-1 < \De \leq 0$ since then one has the trivial bound
$2\pi \wh{\xi}^{\prime}(\la) \, > \,   \mf{p}^{\prime}(\la) >0$ on $I_{\La}$. 

 When $\De>1$, it is readily seen that the functions $\wh{\xi}^{\prime}$ and its limit $\xi^{\prime}$ are $\pi$ periodic.
In principle, one could have that $\La>\tf{\pi}{2}$. Yet, then it is enough to observe that the 
Bethe roots can be decomposed as $\la_a=\wt{\la}_a+n_a \pi$, where $n_a\in \mathbb{Z}$ is such that $ \wt{\la}_a \in \intof{ - \tf{\pi}{2} }{ \tf{\pi}{2} } $. 
Making explicit the dependence of the counting functions on the Bethe roots, it holds
\beq
\wh{\xi}^{\, \prime}\big( \la \mid \{\la_a\}_1^N \big) \, = \, \wh{\xi}^{\, \prime}\big( \la \mid \{ \wt{\la}_a\}_1^N \big) \;. 
\enq
Therefore, since both functions will admit the same limit of the extracted sequence, it is enough to reason on the level of the $\wt{\la}_{a}$ which all belong to the interval $\intff{ - \tf{\pi}{2} }{ \tf{\pi}{2} }$. 
Once that it will be established that $\xi^{\prime}>0$ on $\intff{ - \tf{\pi}{2} }{ \tf{\pi}{2} }$, its strict positivity on intervals of large diameter will follow. 
Hence, below, when $\De>1$, we shall assume that $0 \leq \La \leq \tf{\pi}{2}$.

\vspace{3mm}

The function $\xi^{\prime}$ is holomorphic on $  \mc{S}_{ \varkappa_{\De} }(I_{2\La} ) $, and thus admits a finite number of zeroes on $\intff{-\La}{\La}$. Let 
\beq
-\La = \mf{z}^{(0)} < \mf{z}^{(1)}< \dots < \mf{z}^{(r)} < \mf{z}^{(r+1)}=\La 
\enq
be such that 
\beq
\Big\{ z\in \intff{-\La }{ \La } \; : \;  \xi^{\prime}(z)=0 \; \; \e{or} \; \;  z=\pm \La \Big\} \; = \; \big\{  \mf{z}^{(a)}  \big\}_1^{(r+1)} \;. 
\enq
Also, given $\de>0$ and small enough, let
\beq
\kappa^{(k)} \, =  \, \e{sgn} \Big(  \xi^{\prime}_{\mid \intoo{ \mf{z}^{(k)} }{ \mf{z}^{(k+1)} }} \Big) 
\quad \e{and} \quad 
I^{(k)}_{\de}=\intff{ \mf{z}^{(k)} +\de }{ \mf{z}^{(k+1)} - \de } \qquad \e{for} \quad k=0,\dots,r\;.  
\enq
 By Proposition \ref{Proposition invertibilite de fL en fct de f} there exist $\eps, \eta>0$ such that 
\beq
\xi \; : \; \mc{S}_{2\eta,  2\eps}( I^{(k)}_{\de}  ) \tend \xi\big(  \mc{S}_{2\eta,  2\eps}( I^{(k)}_{\de} )  \big) 
\enq
is a biholomorphism that satisfies 
\beq
\xi \Big(  \mc{S}_{2\eta,  2\eps}( I^{(k)}_{\de} )  \cap \mathbb{H}^{\pm } \Big)  \; = \;  \xi\Big(  \mc{S}_{2\eta, 2\eps}( I^{(k)}_{\de} ) \Big) \cap \mathbb{H}^{\pm \kappa^{(k)} } \;,  
\qquad \wh{\xi} ( I^{(k)}_{\de} ) \subset \xi \Big(  \mc{S}_{\eta,  \eps}( I^{(k)}_{\de} )  \Big) \subset \wh{\xi} \Big(  \mc{S}_{2\eta,  2\eps}( I^{(k)}_{\de} ) \Big)
\label{ecriture propriete emboitement ensembles images par biholomorphisme}
\enq
and 
\beq
\wh{\xi} \, : \, \mc{S}_{2\eta, 2\eps}( I^{(k)}_{\de} ) \cap \wh{\xi}^{-1}\Big(   \xi\big( \mc{S}_{\eta, \eps}( I^{(k)}_{\de} ) \big)   \Big)  \; \tend  \; \xi\big( \mc{S}_{\eta, \eps} ( I^{(k)}_{\de} ) \big)
\enq
is a biholomorphism. Furthermore, one has $\e{sgn}\big( \, \wh{\xi}^{\prime} \, \big)=\kappa^{(k)}$  on $I^{(k)}_{\de}$.

Let $\mf{m}^{(k)}$ be the multiplicity of the zero $\mf{z}^{\,(k)}$. By Rouch\'{e}'s theorem, since $\norm{ \wh{\xi}^{\prime} - \xi^{\prime}  }_{ L^{\infty}\big(\mc{S}_{ \varkappa_{\De} }(I_{2\La})  \big)  }\tend 0$, 
given any $\de>0$, there exists $L_{\de}$ such that for any $L>L_{\de}$ the function $\wh{\xi}^{\prime}$ has $\mf{m}^{(k)}$ zeroes, counted with multiplicities, inside of the disk 
$ \mc{D}_{ \mf{z}^{(k)},\tf{\de}{4} } $. The distinct zeroes will be denoted by $\wh{\mf{z}}^{\, (k)}_{a}$, with $a=1,\dots, \wh{n}^{(k)}$.
\textit{A priori} some of these zeroes can have non-zero imaginary parts while other will be real. Thus, I assume that the zeroes are ordered in such a way that 
\beq
\wh{\mf{z}}^{  \, (k)}_{1} < \dots < \wh{\mf{z}}^{\, (k)}_{ \wh{n}^{(k)}_{\e{Re}} } \qquad  \e{and} \qquad  \Im\big( \, \wh{\mf{z}}^{\, (k)}_{a} \, \big) \not=0 \quad \e{for} \quad 
a=\wh{n}^{(k)}_{\e{Re}}+1,\dots, \wh{n}^{(k)}\;. 
\label{ecriture ordonnement racines reelle de xi hat sur voisinage z k}
\enq
Further, take $L$ large enough so that 
\beq
\big|\wh{\xi}(\mf{z}^{(k)}+\de) \, -  \, \wh{\xi}(\mf{z}^{(k+1)}-\de) \big| \, >\, \f{1}{2} \big| \xi(\mf{z}^{(k)}+\de) \, - \,  \xi(\mf{z}^{(k+1)}-\de) \big|  > 0 \;. 
\label{equation borne inf sur la variation de xi hat sur Ik}
\enq
%
%
%
Then,  define $ \wh{q}_{R/L}^{\, (k)} \in   I^{(k)}_{\de} $ as the two solutions to $\ex{ 2\i \pi L \wh{\xi}(\la) } = - 1$ such that 
\beq
\wh{q}_{R}^{ \, (k)}-\mf{z}^{(k+1)} \;\; \e{is} \;\e{maximal} \qquad \e{and} \qquad 
-\mf{z}^{(k)}  + \wh{q}_{L}^{ \, (k)}  \;\; \e{is} \;\e{minimal}  \;. 
\enq
In virtue of \eqref{equation borne inf sur la variation de xi hat sur Ik}, for $L$ large enough, both $ \wh{q}_{R/L}^{\, (k)} $ are well-defined, exists, and are distinct. 
Furthermore, for any given $\de>0$, there exists $L$ large enough such that 
\beq
\sul{k=0}{r}\Big\{ \,  \big|  \,  \wh{q}_{R}^{ \, (k)}-\mf{z}^{(k+1)}  \big| \; + \;   \big| \,   \wh{q}_{L}^{\, (k)}-\mf{z}^{(k)} \big| \, \Big\} \; \leq \; 4 (r+1) \de \;. 
\label{ecriture borne sur les devration des q hat R et L des zeroes de xi prime}
\enq
Finally, let 
\beq
X\; = \; \Big\{ x \in \intff{ -\La }{ \La } \; : \; \ex{ 2\i \pi \wh{\xi}(x) } \;= \; 1 \Big\}\quad , \quad X^{(\e{in})} \; = \; X \cap \Big\{  \bigcup_{k=0}^{r} I_{\de}^{(k)} \Big\}
\quad , \quad X^{(\e{out})} \; = \; X \setminus X^{(\e{in})} \;. 
\label{definition X in et out}
\enq
Similarly, setting $Y=\{\la_a\}_1^N$,
\beq
Y^{(\e{in}) } \; = \; Y\cap X^{(\e{in}) } \quad \e{and}\; \e{define} \quad  Y^{(\e{out}) } \; = \; Y\setminus Y^{(\e{in})} \; . 
\label{definition Y in et out}
\enq

The purpose of the below paragraph is to estimate $\# X^{(\e{out})}$. The first step consists in estimating the number of roots in the interval $\intff{  \wh{q}_{R}^{\, (k)} }{  \wh{q}_{L}^{\, (k+1)} }$ with $k=-1,\dots r$ 
under the convention that 
\beq
 \wh{q}_{R}^{\, (-1)} = -\La \quad \e{and} \quad  \wh{q}_{L}^{\, (r+1)}=\La \;.
\enq
$\wh{\xi}^{\prime}$ has constant sign on each of the intervals 
\beq
\intoo{\, \wh{q}^{\, (k)}_R }{ \wh{\mf{z}}^{\, (k)}_{1} } \; ,\;\;   \intoo{ \, \wh{\mf{z}}^{\,(k)}_{1}  }{ \wh{\mf{z}}^{\,(k)}_{2} }\; , \;\;  \dots \; ,  \; \; \intoo{  \, \wh{\mf{z}}^{\,(k)}_{ \wh{n}^{(k)}_{\e{Re}} } }{\wh{q}^{\, (k+1)}_L } \;. 
\enq
Due to the ordering \eqref{ecriture ordonnement racines reelle de xi hat sur voisinage z k}  of the real roots one gets the upper bound
\bem
\# \Big\{ X^{(\e{out})} \cap  \intff{  \wh{q}_{R}^{\, (k)} }{  \wh{q}_{L}^{\, (k+1)} } \Big\}\, \leq \, 
\Big[ L \big| \wh{\xi}\big( \, \wh{q}^{\, (k)}_R  \big) -  \wh{\xi}\big( \,  \wh{\mf{z}}^{\, (k)}_{1} \big)  \big| \Big] \; + \; \dots \; +\;
\Big[ L \big|  \wh{\xi}\big(  \, \wh{\mf{z}}^{\, (k)}_{ \wh{n}^{(k)}_{\e{Re}} } \big) - \wh{\xi}\big(  \, \wh{q}^{\, (k)}_L  \big)   \big| \Big]   \\
\; \leq \; L \cdot \norm{ \wh{\xi}^{\prime} }_{ L^{\infty}\big(  \intff{  \wh{q}_{R}^{\, (k)} }{  \wh{q}_{L}^{\, (k+1)} }  \big) } \cdot \big| \wh{q}_{R}^{\, (k)} - \wh{q}_{L}^{\, (k+1)}  \big| 
\; \leq \; 2 L \cdot \norm{ \xi^{\prime} }_{ L^{\infty}\big(  I_{ \La }  \big) } \cdot  \big| \wh{q}_{R}^{\, (k)} - \wh{q}_{L}^{\, (k+1)}  \big| \;. 
\end{multline}
In the first line, $[*]$ denotes the integer part. Furthermore, so as to get the last bound, $L$ is taken large enough so that 
$ \norm{ \wh{\xi}^{\prime} }_{ L^{\infty}\big(  I_{ \La }  \big) } \;\leq \; 2  \norm{ \xi^{\prime} }_{ L^{\infty}\big(  I_{ \La }  \big) }$. 
Then, summing over $k$, and using \eqref{ecriture borne sur les devration des q hat R et L des zeroes de xi prime}, one arrives to
\beq
\#  X^{(\e{out})} \; \leq \; 8 (r+1) \de  L \cdot \norm{ \xi^{\prime} }_{ L^{\infty}\big( I_{ \La }  \big) }  \;. 
\label{bornage du cardinal de X out}
\enq

\begin{figure}[ht]
\begin{center}

\begin{pspicture}(7,7)

\psline[linestyle=dashed, dash=3pt 2pt]{->}(1,4)(6.7,4)
\psdots(1.5,4)(6,4) 
\rput(0.9,3.5){$ \wh{\xi}\big(\, \wh{p}_L^{\,(k)} \big) $}
\rput(6.8,3.4){$\wh{\xi}\big(\, \wh{p}_R^{\,(k)} \big)$}

\rput(0.7,6.1){$\wh{\xi}\big(\, \wh{p}_L^{\,(k)} \big) + \i\a$}
\rput(6.8,6.1){$\wh{\xi}\big(\, \wh{p}_R^{\,(k)} \big) + \i\a$}

\rput(0.5,1.8){$\wh{\xi}\big(\, \wh{p}_L^{\,(k)} \big) - \i\a$}
\rput(6.8,1.6){$\wh{\xi}\big(\, \wh{p}_R^{\,(k)} \big) - \i\a$}

\psline{-}(1.5,2.5)(1.5,5.5)
\pscurve{-}(1.5,5.5)(1.65,5.9)(2,6)

\psline{-}(2,6)(5.5,6)
\pscurve{-}(5.5,6)(5.85,5.9)(6,5.5)

\psline{-}(6,5.5)(6,2.5)
\pscurve{-}(6,2.5)(5.85,2.1)(5.5,2)

\psline{-}(5.5,2)(2,2)
\pscurve{-}(2,2)(1.65,2.1)(1.5,2.5)




\rput(3.5,6.5){ $ \wh{\Ga}_{+}^{(k)} $ }
\rput(4,1.5){ $ \wh{\Ga}_{-}^{(k)} $ }

\psline[linewidth=2pt]{->}(5,2)(5.1,2)
\psline[linewidth=2pt]{->}(4,6)(3.9,6)

\end{pspicture}
\caption{ Domain $\wh{G}^{\,(k)}(\a)$ delimited by its boundary  $\wh{\Ga}_{ +}^{(k)} \cup \wh{\Ga}_{ - }^{(k)}$. 
When $\kappa^{(k)}=1$ one has $\wh{p}_L^{\,(k)}=\wh{q}_L^{\,(k)}$ and $\wh{p}_R^{\,(k)}=\wh{q}_R^{\,(k)}$, while, when
$\kappa^{(k)}=-1$ one has $\wh{p}_L^{\,(k)}=\wh{q}_R^{\,(k)}$ and $\wh{p}_R^{\,(k)}=\wh{q}_L^{\,(k)}$. 
\label{Figure domaine G alpha hat dependant de k} }
\end{center}
\end{figure}
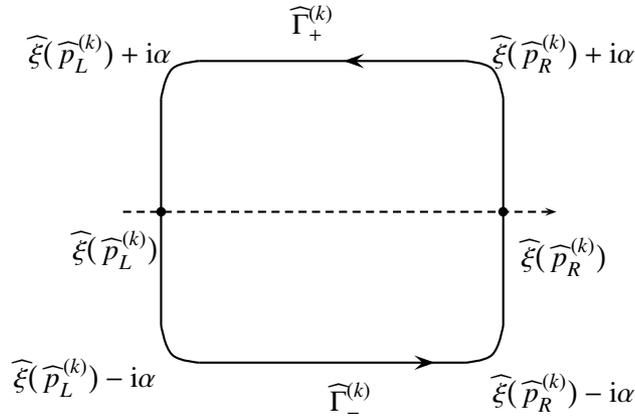

In order to write down a non-linear integral equation satisfied by $\wh{\xi}^{\prime}$, I still need to define auxiliary contours. 
Let $ \wh{G}^{\,(k)}(\a)$ and its boundary $ \wh{\Ga}_{ +}^{(k)} \cup \wh{\Ga}_{ - }^{(k)} $ be defined as in Fig.~\ref{Figure domaine G alpha hat dependant de k}. 
By compactness, it follows that there exists $\a>0$
such that $ \wh{G}^{\,(k)}(\a) \subset \xi\big(  \mc{S}_{\eps,\eta}( I_{\de}^{(k)} ) \big)  $ for any $k=0,\dots, r$.

It now remains to characterise the sum over Bethe roots occurring in $\wh{\xi}^{\prime}$. The latter can be recast as 

\bem
- \f{1}{ L} \sul{a=1 }{N} K(\la-\la_a)  \; = \; -\f{1}{ L} \sul{ y \in Y^{(\e{out})} }{} K(\la-y) \,- \, \f{1}{ L} \sul{ y \in Y^{(\e{in})} }{} K(\la-y)  \\
\; = \; - \, \f{1}{ L} \sul{ y \in Y^{(\e{out})} }{} K(\la-y) \, + \,  \f{1}{ L} \sul{ x \in X^{(\e{in})}\setminus Y^{(\e{in})} }{} K(\la-x) 
\; - \; \sul{k=0}{r} \Oint{ \wh{\mc{C}}^{\,(k)} }{}  \f{ \wh{\xi}^{\prime}(\mu) K(\la-\mu) }{ \ex{2\i\pi L \wh{\xi}(\mu) } -1 }  \cdot  \dd \mu 
\end{multline}
where  $\wh{C}^{(k)}=\wh{\xi}^{-1}\big( \wh{\Ga}_{ +}^{(k)} \cup \wh{\Ga}_{ - }^{(k)} \big)$.
 Now, agreeing upon $\wh{\mc{C}}^{\,(k)}_{\pm} = \wh{\mc{C}}^{\,(k)} \cap \mathbb{H}^{\pm} $, one can decompose the integral over  $\wh{\mc{C}}^{\,(k)}_{\pm}$ as 
\bem
- \; \Oint{ \wh{\mc{C}}^{\,(k)} }{}  \f{ \wh{\xi}^{\prime}(\mu) K(\la-\mu) }{ \ex{2\i\pi L \wh{\xi}(\mu) } -1 }  \cdot   \dd \mu   \; = \; 
- \kappa^{(k)} \Int{ \wh{q}_L^{\, (k)} }{ \wh{q}_R^{ \, (k)} } K(\la-\mu)  \wh{\xi}^{\prime}(\mu) \cdot \dd \mu \; + \; \kappa^{(k)} \sul{ \eps = \pm }{} \eps
\Int{ \wh{\mc{C}}^{\,(k)}_{\eps} }{} \f{ \wh{\xi}^{\prime}(\mu) K(\la-\mu) }{ \ex{- 2\i\pi \eps \kappa_k L \wh{\xi}(\mu) } -1 }  \cdot  \dd \mu  \\
\; = \; - \kappa^{(k)} \Int{ \mf{z}^{(k)} }{ \mf{z}^{(k+1)} } K(\la-\mu)  \wh{\xi}^{\prime}(\mu) \cdot \dd \mu \; + \; \mf{r}^{(k)}[\wh{\xi}](\la) \;. 
\end{multline}
There, I have introduced
\beq
\mf{r}^{(k)}[\wh{\xi}](\la) \; = \; - \kappa^{(k)} \bigg\{ \Int{ \wh{q}_L^{\, (k)} }{ \mf{z}^{(k)}  } \, + \,  \Int{ \mf{z}^{(k+1)}  }{ \wh{q}_R^{\, (k)} } \bigg\} K(\la-\mu)  \wh{\xi}^{\prime}(\mu) \cdot \dd \mu \; + \; \sul{ \eps = \pm }{} \eps
\Int{ \wh{\Ga}^{\,(k)}_{\eps} }{} \f{ K\big(\la-\wh{\xi}^{-1}(z) ) }{ \ex{ -2\i\pi \eps L z } -1 }  \cdot  \dd z  \; . 
\enq
Finally, introduce the intervals
\beq
J^{(\pm)}\; = \; \bigcup_{ k \; : \;  \kappa^{(k)} = \pm 1 }{} \intoo{  \mf{z}^{\, (k)} }{ \mf{z}^{\, (k+1)} }  \;. 
\enq
All the above leads to the non-linear integral equation satisfied by $\wh{\xi}$
\beq
\Big( \e{id} \; + \; \op{K}_{ J^{(+)} }   -   \op{K}_{ J^{(-)} }  \Big)\big[ \, \wh{\xi}^{\prime} \big](\la)
\; = \; \f{ \mf{p}^{\prime}(\la) }{ 2\pi } \; - \, \wh{\phi}_{\e{out} }(\la) \, + \,  \wh{\phi}_{\e{in} }(\la)
\; + \; \sul{k=0}{r}\mf{r}^{(k)}[\wh{\xi}](\la)   \;. 
\enq
There, I have set
\beq
\wh{\phi}_{\e{out} }(\la) \; = \; \f{1}{ L} \sul{ y \in Y^{(\e{out})} }{} K(\la-y)\qquad \e{and} \qquad 
\wh{\phi}_{\e{in} }(\la) \; = \; \f{1}{ L} \sul{ x \in X^{(\e{in})}\setminus Y^{(\e{in})} }{} K(\la-x) \; . 
\enq
Let $\op{R}_{J^{(+)} }$ be the resolvent operator\symbolfootnote[3]{Recall that when $\De>1$ it holds $\e{diam}( J^{(+)}\cup  J^{(-)} ) < \pi $, so that indeed the involved operators are invertible in virtue of 
Lemma \ref{lemme invertibilite I+K}} to $ \e{id}  + \op{K}_{ J^{(+)} }$. By Lemma \ref{lemme invertibilite I+K}, the associated resolvent kernel is strictly positive: 
$R_{ J^{(+)} }(\la,\mu) >0$. Furthermore, by \eqref{ecriture inegalites sur les resolvent a interval fini De positif}
\beq
 P_{ \! J^{(+)} }^{\prime}(\la) \; = \;  \Big( \e{id} \; - \; \op{R}_{ J^{(+)} }\Big) \Big[ \f{ \mf{p}^{\prime} }{ 2\pi }  \Big](\la) \, \geq  \,  \Big( \e{id} \; - \; \op{L}_{ \R }\Big) \Big[ \f{ \mf{p}^{\prime} }{ 2\pi }  \Big](\la) 
 \, = \,\rho_{\infty}(\la) >0 \quad \e{on} \; \R\;. 
\label{ecriture borne inf sur moment habille intervalle J+}
\enq
Thus, adopting the notations of Lemma \ref{Lemme invertibilite RJ+ restreint a J-}, one gets
\beq
\Big( \e{id} \;     -   \big[\op{R}_{ J^{(+)} }\big]_{ J^{(-)} }  \Big)\big[ \, \wh{\xi}^{\prime} \big](\la)
\; = \;  P_{ \! J^{(+)} }^{\prime}(\la)  \; - \, \wh{\psi}_{\e{out} }(\la) \, + \,  \wh{\psi}_{\e{in} }(\la)
\; + \; \sul{k=0}{r}\wt{\mf{r}}^{\, (k)}[\wh{\xi}](\la)   
\enq
where 
\beq
\wh{\psi}_{\e{out} }(\la) \; = \; \f{1}{  L} \sul{ y \in Y^{(\e{out})} }{} R_{ J^{(+)} }(\la,y)\qquad \e{and} \qquad 
\wh{\psi}_{\e{in} }(\la) \; = \; \f{1}{  L} \sul{ x \in X^{(\e{in})}\setminus Y^{(\e{in})} }{} R_{ J^{(+)} }(\la,x) \; . 
\label{definition hat psi in et out}
\enq
Finally,
\beq
\wt{\mf{r}}^{\, (k)}[\wh{\xi}](\la) \; = \; - \kappa^{(k)} \bigg\{ \Int{ \wh{q}_L^{\, (k)} }{ \mf{z}^{(k)}  } \, + \,  \Int{ \mf{z}^{(k+1)}  }{ \wh{q}_R^{\, (k)} } \bigg\} R_{ J^{(+)} }(\la,\mu)  \wh{\xi}^{\prime}(\mu) \cdot \dd \mu 
\; + \; \sul{ \eps = \pm }{} \eps \Int{ \wh{\Ga}^{\,(k)}_{\eps} }{} \f{   R_{ J^{(+)} }(\la,\wh{\xi}^{-1}(z) ) }{ \ex{ 2\i\pi \eps L z } -1 }  \cdot  \dd z  \; . 
\enq
It follows from Lemma  \ref{Lemme invertibilite RJ+ restreint a J-} that the operator $ \e{id}     -   \big[\op{R}_{ J^{(+)} }\big]_{ J^{(-)} } $ is invertible and that its resolvent operator $ \big[\msc{R}_{ J^{(+)} }\big]_{ J^{(-)} } $
has a strictly positive integral kernel $ \big[\msc{R}_{ J^{(+)} }\big]_{ J^{(-)} } (\la,\mu)>0$. Therefore, one has
\beq
 \wh{\xi}^{\prime} (\la)
\; = \; \Big( \e{id} \;     +  \big[\msc{R}_{ J^{(+)} }\big]_{ J^{(-)} }  \Big) \Big[ P_{ \! J^{(+)} }^{\prime}  \; - \, \wh{\psi}_{\e{out} } \, + \,  \wh{\psi}_{\e{in} }
\; + \; \sul{k=0}{r}\wt{\mf{r}}^{\, (k)}[\wh{\xi}] \Big](\la)   \;. 
\label{ecriture hat xi prime dans l'hypothese de plusieurs intervalles de positivite pour la limite xi}
\enq
By the above, the operators $ \big[\msc{R}_{ J^{(+)} }\big]_{ J^{(-)} }$ and \textit{a fortioti} $ \e{id}     +    \big[\msc{R}_{ J^{(+)} }\big]_{ J^{(-)} } $ are strictly positive. Since, $P_{ \! J^{(+)} }^{\prime}(\la)>0$, $\wh{\psi}_{\e{in} }(\la)\geq 0$, 
this leads to the lower bound
\beq
 \wh{\xi}^{\prime} (\la) \; \geq  \; P_{ \! J^{(+)} }^{\prime}(\la) \; + \;  \Big( \e{id} \;     +   \big[\msc{R}_{ J^{(+)} }\big]_{ J^{(-)} }  \Big) \Big[   \; - \, \wh{\psi}_{\e{out} }
\; + \; \sul{k=0}{r}\wt{\mf{r}}^{\, (k)}[\wh{\xi}] \Big](\la)   \;. 
\label{ecriture borne inf sur xi hat}
\enq
Since  $\# \big\{ Y^{(\e{out})} \setminus  X^{(\e{out})} \big\} \leq n$ and $\# \big\{ Y^{(\e{out})} \cap  X^{(\e{out})} \big\}$ can be bounded by $\# X^{(\e{out})} $, the bound \eqref{bornage du cardinal de X out} ensures that 
\beq
 \Big| \wh{\psi}_{\e{out} }(\la) \Big| \; \leq \; \norm{ R_{ J^{(+)} } }_{ L^{\infty}\big( \R^2   \big) } \f{ \#  Y^{(\e{out})} }{ 2\pi L} \; \leq \; 
 C \Big( \de +\f{n}{L}\Big) 
\qquad \e{leading} \, \e{to} \quad \limsup_{L\tend + \infty} \big|   \Big( \e{id} \;     +   \big[\msc{R}_{ J^{(+)} }\big]_{ J^{(-)} }  \Big) \big[   \; - \, \wh{\psi}_{\e{out} }\big] (\la) \big| \; \leq C^{\prime} \de
\enq
by continuity of  $\Big( \e{id} \;     +   \big[\msc{R}_{ J^{(+)} }\big]_{ J^{(-)} }  \Big)$ and with a bound that is uniform on $\R$. 

Owing to $\norm{ \wh{\xi} }_{ L^{\infty}( I_{ \La }) }  \leq  2 \norm{ \xi }_{ L^{\infty}( I_{ \La }) } $
and $\wh{q}_{R/L}^{\, (k)} \in \intff{-\La}{\La}$, it follows readily that the contours $\wh{\Ga}^{\,(k)}_{\eps}$ are all bounded in $L$ and thus 
for any $\la \in I_{\La}$  
\beq
\bigg|  \sul{k=0}{r}\sul{ \eps = \pm }{} \eps \Int{ \wh{\Ga}^{\,(k)}_{\eps} }{} \f{ R_{ J^{(+)} }\big(\la,\wh{\xi}^{-1}(z) \big) }{ \ex{ 2\i\pi \eps L z } -1 }  \cdot  \dd z  \bigg| \; \leq \; 
\norm{ R_{ J^{(+)} } }_{ L^{\infty}\big( I_{\La}  \times \mc{S}_{ \varkappa_{\De} }(\R) \big) } \cdot \f{C}{L}
\enq
where the $1/L$ decay follows from a straightforward estimation of the integral while the bound on the integral kernel $R_{ J^{(+)} }$
follows from the inclusion $ \wh{\Ga}^{(k)}_{\pm} \subset  \xi\big( \mc{S}_{\eta,\eps}(I_{\de}^{(k)}) \big) $ what leads to  that
$ \wh{\xi}^{-1}\big( \wh{\Ga}^{(k)}_{\pm} \big) \subset   \mc{S}_{2\eta,2\eps}(I_{\de}^{(k)} ) \subset \mc{S}_{ \varkappa_{\De} }( I_{\La + 2\eps} )$ as ensured by \eqref{ecriture propriete emboitement ensembles images par biholomorphisme}. 
Thus, in virtue of \eqref{ecriture borne sur les devration des q hat R et L des zeroes de xi prime}, one gets 
\beq
\limsup_{L\tend + \infty} \Big( \e{id} \;     +   \big[\msc{R}_{ J^{(+)} }\big]_{ J^{(-)} }  \Big) \Big[  \sul{k=0}{r}\wt{\mf{r}}^{\, (k)}[\wh{\xi}] \Big](\la) \; = \; \e{O}(\de) \;. 
\enq
This being settled, it remains to take the $L\tend + \infty$ limit superior  of the inequality \eqref{ecriture borne inf sur xi hat} followed by sending $\de \tend 0^+$. 
One gets, for any $\la \in I_{\La}$
\beq
 \xi^{\prime} (\la) \; \geq  \; P_{ \! J^{(+)} }^{\prime}(\la) \; \geq \; \rho_{\infty}(\la) \, >\, \inf_{ I_{\La} }  \rho_{\infty} \; > \; 0   \;. 
\enq
This contradicts that, either $\xi^{\prime} (\la) < 0 $ on $I_{\La} $ or that $ \xi^{\prime}$ has zeroes on $I_{\La}$.  \qed

\subsection{Various corollaries of interest}

Theorem \ref{Theorem convergence et eqn NL pour ctg fct} has several important corollaries. First of all, it guarantees the uniqueness of solutions to the logarithmic Bethe  
Ansatz equations \eqref{ecriture Log BAE cas general tout delta} throughout the regime $ \De > -1$ and, in particular, for $\De>0$ when convexity arguments cannot be used. 
To the best of my knowledge, uniqueness of such solutions has never been proven earlier.

\begin{cor}
 
Under the hypothesis of Theorem  \ref{Theorem convergence et eqn NL pour ctg fct} and point v) in particular, there exists $L_0$ such that the system of logarithmic 
Bethe Ansatz equations associated with the given choice of integers $\{\ell_a\}_1^N$ admits a unique real-valued solution for any $L\geq L_0$. 
 
\end{cor}

\Proof 

If there would be two solutions to the logarithmic Bethe equations associated with the given choice of integers $\{\ell_a\}_1^N$, than one would then be able to build two distinct counting functions $\wh{\xi}^{(a)}$, $a=1,2$. 
These, however, will both satisfy the non-linear integral equation when $L$ will be large enough. As follows from Theorem  \ref{Theorem convergence et eqn NL pour ctg fct} point v), the non-linear integral equation 
admits a unique solution, contradicting that $\wh{\xi}^{(1)}\not= \wh{\xi}^{(2)}$ for $L$ large enough. \qed 

\vspace{2mm}

A second important consequence of Theorem \ref{Theorem convergence et eqn NL pour ctg fct}  is the existence of an all order large-$L$ asymptotic expansion of the 
counting function.

\begin{prop}
 \label{Proposition DA counting fct}

 Let $h_1<\dots < h_n$, $h_a \in \intn{ 1 }{ N^{\prime} }$ and $p_1<\dots < p_n$ , $p_a \in \mathbb{Z}\setminus\intn{ 1 }{ N^{\prime} }$ be and increasing sequence of integers 
such that 
 
\begin{itemize}
\item[$\bullet$] for $-1<\De\leq 1$, $\De=\cos(\zeta)$, 
\beq
\f{\pi - \zeta}{\pi} \Big( \f{1}{2}-\f{N^{\prime}-1}{L} \Big)   > \f{ p_n-N^{\prime} }{ L } \quad , \quad  \f{ p_1-1 }{ L } >  - \f{\pi - \zeta}{\pi} \Big( \f{1}{2}-\f{N^{\prime}-1}{L}  \Big) \;; 
\enq
\item[$\bullet$] for $\De>1$, the integers $\{p_a\}_1^n$ are uniformly bounded in $L$: $\big| \tf{p_a}{L} \big| \leq C$ for some $L$-independent   C>0. 
\end{itemize}

 Let $\{\la_a\}_1^{N^{\prime}}$ be the solution to the logarithmic Bethe equations subordinate to this choice of integers and let $N^{\prime}=N+s$ with $s$ bounded uniformly in $L$.  
Then, the associated counting function $\wh{\xi}$ admits the large-$L$ asymptotic expansion 
\beq
\wh{\xi}(\la) \; = \; \xi_0(\la\mid \wh{q} \, ) \; + \; \sul{p=1}{r} \f{1}{L^p}\xi_p^{(s)}\big( \la \mid \{ \wh{x}_{p_a} \}_1^{n} ;  \{ \wh{x}_{h_a} \}_1^n \big)
\; + \; \e{O}\big( L^{-(r+1)} \big)\;. 
\label{ecriture DA xi a tout ordre}
\enq
There $\wh{x}_{a}$ are defined through $\wh{\xi}(\wh{x}_a)=\tf{a}{L}$.

 This asymptotic expansion is such that the remainder just as all functions, $\xi_p^{(s)}$, $p=1,\dots,r$ are holomorphic on $\mc{S}_{2\varkappa_{\De} }(\R)$ where $\varkappa_{\De}$
 has been introduced in \eqref{definition varkappa Delta}. 
Furthermore, the remainder is uniform on $\ov{\mc{S}}_{\eta}(\R)$ for any $0<\eta<2\varkappa_{\De}$. The function $\xi_0$ is as defined in \eqref{definition fction de cptge thermodynamique} 
with $\wh{q}$ the solution to the magnetic Fermi boundary problem associated with $\wh{D}=\tf{N}{L}$. Also, 
\beq
 \xi_1^{(s)}\big( \la \mid  \{ \wh{x}_{p_a} \}_1^{n} ;  \{ \wh{x}_{h_a} \}_1^n \big) \; = \;  \Phi^{(s)}_{\wh{q}}\big( \la \mid  \{ \wh{x}_{p_a^{\prime}} \}_1^{ n_p^{(s)} }  ;  \{ \wh{x}_{ h_a^{\prime} } \}_1^{ n_h^{(s)} } \big)
\label{ecriture xi0 et xi1 dans DA xi hat}
\enq
where 
\beq
\big\{ p_a^{\prime} \big\}_1^{ n_p^{(s)} } \, = \, \left\{ \ba{cc}   \big\{ p_a \big\}_1^{n} \cup \big\{ N+a \big\}_1^{s}  &  if \; s \geq 0 \vspace{1mm} \\ 
							     \big\{ p_a \big\}_1^{n}   &  if \; s < 0   \ea \right. 
\quad and \quad 
\big\{ h_a^{\prime} \big\}_1^{ n_h^{(s)} } \, = \, \left\{ \ba{cc}   \big\{ h_a \big\}_1^{n}  &  if \; s \geq 0 \vspace{1mm}  \\ 
							     \big\{ h_a \big\}_1^{n} \cup \big\{ N-a +1\big\}_1^{|s|}   &  if \; s < 0   \ea \right. \;. 
\label{definition des racines p prime et h prime}
\enq
Also, the function $\Phi_{ \wh{q} }^{ (s) }$ is as defined in \eqref{definition fct Phi spin et q hat dependent} and 
\beq
 \xi_2^{(s)}\big( \la \mid \{ \wh{x}_{p_a} \}_1^{n}  ;  \{ \wh{x}_{h_a} \}_1^n \big) \; = \; \f{1}{2 \xi_0^{\prime}(\wh{q} \mid \wh{q} \, ) } \sul{\eps=\pm}{} 
 \eps R_{I_{\wh{q}} }(\la,\eps \wh{q} \, ) \cdot \bigg[  \Big( \xi_1^{(s)}\big( \eps \wh{q} \mid \{ \wh{x}_{p_a} \}_1^{n} ;  \{ \wh{x}_{h_a} \}_1^n \big)  - \f{1}{2}    \Big)^2  - \f{1}{12}   \bigg] \;. 
\label{ecriture xi2 dans DA xi hat}
\enq
Finally, the unique solutions $\wh{q}_{L/R}$ to $\wh{\xi}(\wh{q}_{L})=\tf{1}{(2L)}$ and $\wh{\xi}(\wh{q}_{R})=\tf{(N+\tf{1}{2})}{L}$ admit the large-$L$ asymptotic expansion
\beq
\wh{q}_R - \wh{q} \; = \; \sul{k=1}{r}\f{ q_+^{(k)} }{ L^k }   \; + \; \e{O}\Big( \f{1}{ L^{r+1} } \Big) \hspace{1.2cm} and \hspace{1.2cm}
\wh{q}_L + \wh{q} \; = \; \sul{k=1}{r}\f{ q_-^{(k)} }{ L^k }   \; + \; \e{O}\Big( \f{1}{ L^{r+1} } \Big)
\label{ecriture DA order 2 des endpoints}
\enq
where the first terms of the expansion take the form  
\beq
q_{\pm}^{(1)} \; = \; \f{ \tf{1}{2}- \xi_1^{(s)}\big( \pm \wh{q} \, \big) }{  \xi_{0}^{\prime}(\wh{q} \mid \wh{q} \, ) }  \;. 
\label{ecriture coeff DA q L and R ordre 1}
\enq
Also, the second order deviations in respect to $\wh{q}$ read
\beq
q_{\pm}^{(2)} \; = \;   - \f{ 1 }{  \big( \xi_{0}^{\prime}(\wh{q}\mid \wh{q})\,  \big)^2 } \bigg\{  \big[ \tf{1}{2}- \xi_1^{(s)}\big( \pm \wh{q} \, \big)  \big] \, \big(\xi_{1}^{(s)}\big)^{\prime}( \pm \wh{q} \, )  
				    \, + \, \xi_{2}^{(s)}(\pm \wh{q} \, )\, \xi_{0}^{\prime}( \wh{q} \mid \wh{q}) 
				    \, + \,  \f{ \xi_{0}^{\prime\prime }(\pm \wh{q} \mid \wh{q} \, )  }{ 2 \xi_0^{\prime}(\pm \wh{q}\mid \wh{q} \, ) } \,  \big[ \tf{1}{2}- \xi_1^{(s)}\big( \pm \wh{q} \, \big)  \big]^2 \bigg\} \;. 
\label{ecriture coeff DA q L and R ordre 2}
\enq
The dependence of $\xi_{1}^{(s)}$ and $\xi_2^{(s)}$ on the auxiliary rapidities has been omitted both in \eqref{ecriture coeff DA q L and R ordre 1} and \eqref{ecriture coeff DA q L and R ordre 2}. 

\end{prop}

The decomposition $N^{\prime}=N+s$ might appear slightly artificial in that one can absorb all the dependence on $s$ by simply doing the substitution 
$(\wh{D},\wh{q} , s) \,  \hookrightarrow \, (\wh{D}^{\prime},\wh{q}^{\, \prime},0)$ where  $\wh{D}^{\prime}=\tf{N^{\prime}}{L}$ and $\wh{q}^{\, \prime}$ is the magnetic Fermi boundary 
associated with $\wh{D}^{\prime}$. However presenting the expansion in the above form allows one to immediately take into account various terms of the asymptotic expansion when 
$\wh{D}\tend D$ sufficiently fast in $L$ for instance
\beq
\wh{D}-D = \e{O}\bigg(\f{1}{L^{r+1} } \bigg) \qquad \e{while} \qquad \wh{D}^{\prime} - D =  \e{O}\bigg(\f{1}{L } \bigg) \;. 
\enq

Further, note that all the terms of the asymptotic expansion depend implicitly on $\wh{q}$. It is readily seen that the latter admits an all order asymptotic expansion in $ \wh{D}-D$. 
Presenting the asymptotic expansion of $\wh{\xi}$ in the form \eqref{ecriture DA xi a tout ordre} allows one to absorb all the orders in $ \wh{D} - D$ and thus have a slightly simpler
to obtain, from the computational point of view, asymptotic expansion. If one would make the additional hypothesis $\wh{D}-D=\e{O}\big( L^{-(r+1)} \big)$ then one would also have 
$\wh{q}-q=\e{O}\big( L^{-(r+1)} \big)$ and it would be possible to simplify \eqref{ecriture DA xi a tout ordre} further by replacing $\wh{q}$ by $q$. 
The presence of terms $\wh{D}-D$ in the asymptotic expansion of the counting function was first noticed in \cite{EckleTruongWoynarovichNonAnalyticCorrectionsForXXZInFiniteMagFieldANdCFTSpectrum}.

\Proof 
The first two terms of the asymptotic expansion, just as its form, are readily obtained by recasting the non-linear integral equation \eqref{ecriture NLIE pour la ctg fct} in the form
\beq
\wh{\xi}(\la) \; = \; \wh{\xi}_0(\la\mid \wh{q}) \,  + \, \f{1}{L}\Phi^{(s)}_{\wh{q}}\big( \la \mid  \{ \wh{x}_{ p_a^{\prime} } \}_1^{ n_p^{(s)} }  ;  \{ \wh{x}_{ h_a^{\prime} } \}_1^{ n_h^{(s)} } \big) \, + \, 
\mf{R}_N[\wh{\xi}](\la)  
\enq
and then using straightforward bounds on the "remainder" operator $\mf{R}_N$. Note that the easiest way to obtain this form of the non-linear integral equation
is to follow the reasoning leading to \eqref{ecriture NLIE pour la ctg fct} but use the $N$-dependent contours $\Ga^{(\eps)}$ -as given in Figure \ref{Figure domaine G alpha hat}- instead of
the $N^{\prime}$ dependent ones which would be used if one followed that part of the proof word for word. The additional (if $s<0$) roots  that are picked up  or the missing (if $s>0$)  roots 
that are omitted  by the integration contour $\Ga_+ \cup \Ga_-$ are absorbed by re-defining the particle-hole integers $(p_a,h_a)\hookrightarrow (p_a^{\prime},h_a^{\prime})$
with $p_a^{\prime}$ and 
$h_a^{\prime}$ as given in \eqref{definition des racines p prime et h prime}. Their contribution is then gathered in the function 
$\Phi^{(s)}_{\wh{q}}\big( \la \mid  \{ \wh{x}_{ p_a^{\prime} } \}_1^{ n_p^{(s)} }  ;  \{ \wh{x}_{ h_a^{\prime} } \}_1^{ n_h^{(s)} } \big)$.

\subsubsection*{$\bullet$ The asymptotic expansion up to $r=2$ }

The starting point for pushing the asymptotic expansion one order further, 
\textit{viz}. up to $r=2$, is to observe that the properties of the solution to the non-linear integral equation lead to 
an overall estimate
\beq
\big| \wh{q} - \wh{q}_{R} \big| \, + \, \big| \wh{q} + \wh{q}_{L} \big|\; = \; \e{O}\Big( \tf{1}{L} \Big) \;. 
\label{ecriture deviation bord exacts par rapport a q}
\enq
A straightforward application of Watson's lemma to $\mf{R}_{N;1}\big[ \wh{\xi} \big]$ yields
\bem
\mf{R}_{N;1}\big[ \wh{\xi} \big](\la) \; = \; - \f{1}{4\pi^2 L^2} \bigg\{  \f{ R_{ I_{\wh{q}} }(\la,\wh{q}_R) }{ \wh{\xi}^{\prime}(\wh{q}_R) } \, - \,   \f{ R_{ I_{\wh{q}} }(\la,\wh{q}_L) }{ \wh{\xi}^{\prime}(\wh{q}_L) }   \bigg\}
\Int{\R}{} \ln \Big( 1 +  \ex{-|t|} \Big) \cdot \dd t \; + \; \e{O}\Big( \f{ 1 }{ L^{3} } \Big) \\
 =  -\f{1}{24 L^2 \xi^{\prime}_0(\wh{q}\mid \wh{q} \, ) } \Big( R_{ I_{\wh{q}} }( \la ,\wh{q} ) - R_{ I_{\wh{q}} }( \la,-\wh{q} )   \Big) \; + \; \e{O}\Big( \f{ 1 }{ L^{3} } \Big) \;. 
\end{multline}
Also, due to \eqref{ecriture deviation bord exacts par rapport a q}, it holds that 
\beq
\mf{R}_{N;2}\big[ \wh{\xi} \big](\la)  \, + \, \mf{R}_{N;3}\big[ \wh{\xi} \big](\la) \;=\;   R_{ I_{\wh{q}} }(\la,\wh{q}_R) \,  \wh{\xi}^{\prime}(\wh{q}_R) \f{ (\wh{q}_R-\wh{q})^2 }{ 2 } \, - \,  
R_{ I_{\wh{q}} }(\la,\wh{q}_L)  \, \wh{\xi}^{\prime}(\wh{q}_L) \f{ (\wh{q}+\wh{q}_L)^2 }{ 2 } \; + \; \e{O}\Big( \f{ 1 }{ L^{3} } \Big)\;. 
\enq
The above bounds already ensure that $\wh{\xi}$ admits the large-$L$ asymptotic expansion up to $\e{O}\big( L^{-3}  \big)$:
\beq
\wh{\xi}(\la) \; = \;  \xi_0(\la\mid \wh{q}) \; + \; \f{1}{L}\xi_1^{\, (s)}\big( \la \mid \{ \wh{x}_{p_a} \}_1^{n}  ;  \{ \wh{x}_{h_a} \}_1^n \big)
\; + \;  \f{1}{L^2}\wt{\xi}_2^{\,(s)}\big( \la \mid \ \{ \wh{x}_{p_a} \}_1^{n}  ;  \{ \wh{x}_{h_a} \}_1^n \big)
\; + \; \e{O}\big( L^{-3} \big)
\enq
where $\xi_0$ and $\xi_1^{\, (s)}$ are as defined in the statement of the proposition whereas the expression for $\wt{\xi}_2^{\,(s)}$ involves the $L$-dependent endpoints $\wh{q}_{R/L}$:
\beq
\wt{\xi}_2^{\,(s)}\big( \la \mid  \{ \wh{x}_{p_a} \}_1^{n}  ;  \{ \wh{x}_{h_a} \}_1^n \big) \; = \; - \sul{\eps = \pm }{} \eps \f{ R_{ I_{\wh{q}} }(\la, \eps \wh{q}\,) }{24 \, \xi^{\prime}_0(\wh{q}\mid \wh{q} \,) }
\; + \; \f{\xi^{\prime}_0(\wh{q}\mid \wh{q} )}{2} \Big\{   R_{ I_{\wh{q}} }(\la,\wh{q}\,)  ( \wh{q}_R - \wh{q} )^2  \, - \,  R_{ I_{\wh{q}} }(\la,-\wh{q} )  (\wh{q}+\wh{q}_L)^2  \Big\} \; . 
\label{expression pour xi2 avec dependenance au deviations du bord}
\enq
To conclude, it solely remains to obtain the asymptotic expansion of $\wh{q}_R - \wh{q}$ and $\wh{q}_L + \wh{q}$ up to $\e{O}(L^{-3})$. The latter can be obtained by expanding the defining relation
\beq
\f{1}{2L} \; = \; \wh{\xi}\big( \wh{q}_L \big) \qquad \e{and} \qquad \f{N+1/2}{L} \; = \; \wh{\xi}\big( \wh{q}_R \big) 
\enq
to the second order in $\wh{q}_R-\wh{q}$ and $\wh{q}_L+\wh{q}$ and using the \textit{a priori} estimates \eqref{ecriture deviation bord exacts par rapport a q} so as to separate slower and faster decaying terms. 
All-in-all one obtains the expansion \eqref{ecriture DA order 2 des endpoints} up to the second order, \textit{viz}. $r=2$, where $q_{\pm}^{(1)}$ is as defined in \eqref{ecriture coeff DA q L and R ordre 1}.  
However, at this stage of the analysis, the expression for $q_{\pm}^{(2)}$ has still not its final form in that it is given in terms of $\wt{\xi}_2^{\,(s)}$. To conclude, one should first insert 
the expansion \eqref{ecriture DA order 2 des endpoints} to the first order in $1/L$ into \eqref{expression pour xi2 avec dependenance au deviations du bord}
so as to get \eqref{ecriture DA xi a tout ordre} with $r=2$ along with the form of the coefficients \eqref{ecriture xi0 et xi1 dans DA xi hat}-\eqref{ecriture xi2 dans DA xi hat}. 
Starting from there, one readily obtains the claimed form of the second order coefficient $q_{\pm}^{(2)}$ as given by \eqref{ecriture coeff DA q L and R ordre 2}.

\subsubsection*{$\bullet$ Existence of the all order asymptotic expansion}

The existence of the all order 
asymptotic expansion is obtained by a bootstrap argument close in spirit to the one developed in Section 3.2 of \cite{KozProofexistenceAEYangYangEquation}. 
The main idea of the bootstrap is to use the fact that the action of $\mf{R}_{N}$ produces additional powers of $L^{-1}$ and the proof goes by induction. 
Hence, assume that the expansion \eqref{ecriture DA xi a tout ordre} and \eqref{ecriture DA order 2 des endpoints} hold up to a remainder $\e{O}\big( L^{-r-1} \big)$

The Taylor integral expansion of $\mf{R}_{N;2}[f]$ up to order $r+2$ and in respect to the endpoint of integration $\wh{q}$ takes the form
\bem
\mf{R}_{N;2}[f]\, = \, -\Int{ \wh{q} }{ \wh{q}_R[f]  } \hspace{-2mm} R_{I_{\wh{q}}}(\la,\mu) \Big( f(\mu)-f(\,  \wh{q}_R[f] ) \Big) \cdot \dd \mu  \\
\, = \,  \sul{k=0}{r} \f{ ( \, \wh{q}-\wh{q}_R[f]  \,   )^{k+2} }{ (k+2)!} \Dp{\mu}^{k+1}\Big\{  R_{I_{\wh{q}}}(\la,\mu) \Big( f(\mu)-f( \,\wh{q}_R[f] ) \Big)  \Big\}_{\mu = \wh{q}_R[f]}  \\
+ \f{ ( \, \wh{q}- \wh{q}_R[f]\, )^{r+2} }{ (r+1)!} \Int{0}{1}\dd t  (1-t)^{r+1}\cdot \Dp{\mu}^{r+1}\Big\{  R_{I_{\wh{q}}}(\la,\mu) \Big( f(\mu)-f( \, \wh{q}_R[f] ) \Big)  \Big\}_{\mu = (\, \wh{q}-\wh{q}_R[f]\, )t + \wh{q}_R[f]  }   
\end{multline}
The zeroth order term vanished because the integral vanishes when $\wh{q}=\wh{q}_{R}[f]$ while the first order vanished because the integrand vanishes at $\mu=\wh{q}_R[f]$. 
Owing to the assumed form of the asymptotic expansion of the counting function and of the endpoints, it follows that the integral remainder for $\mf{R}_{N;2}[\wh{\xi}]$
is already  a $\e{O}\big( L^{-r-2} \big)$ uniformly on $\ov{\mc{S}}_{\eta}(\R)$ for any $0<\eta<2\varkappa_{\De}$. The remainder is also holomorphic on $\mc{S}_{2\varkappa_{\De}}(\R)$. 
 owing the analytic structure of the resolvent kernel  $R_{I_{\wh{q}}}(\la,\mu)$. 
 
It  follows from the asymptotic expansion of $\wh{q}_R$ up to $\e{O}\big( L^{-r-1} \big)$ and of $\wh{\xi}$ up to $\e{O}\big( L^{-r-1} \big)$  that 

\beq
\la \mapsto \Dp{\mu}^{k+1}\Big\{  R_{I_{\wh{q}} }(\la,\mu) \Big( \wh{\xi}(\mu)\, - \, \wh{\xi}( \,\wh{q}_R  ) \Big)  \Big\}_{\mu = \wh{q}_R[f]}  \; = \; 
\sul{p=0}{r} \f{1}{L^p} \wt{\mf{r}}_{p}\big( \la \mid \{ \wh{x}_{p_a}\}_1^n ; \{ \wh{x}_{h_a}\}_1^n \big) \; + \;  \e{O}\Big(  \f{1}{L^{r+1}} \Big) \;. 
\enq
Again, it is easy to check that all building blocks have the desired analyticity and uniformness.

Finally, since $ ( \, \wh{q}-\wh{q}_R[f]  \,   )^{k+2}  $ admits an asymptotic expansion up to a remainder  $\e{O}\big( L^{-r-2-k} \big)$ and starting
with a $L^{-k-2}$ decay, the function
\beq
\la \mapsto ( \, \wh{q}-\wh{q}_R[f]  \,   )^{k+2}  \Dp{\mu}^{k+1}\Big\{  R_{I_{\wh{q}}}(\la,\mu) \Big( f(\mu)-f( \,\wh{q}_R[f] ) \Big)  \Big\}_{\mu = \wh{q}_R[f]}    
\enq
admits an asymptotic expansion, in the sense given in the statement of the proposition, up a remainder $ \e{O}\big( L^{-r-2-k} \big)$. Thus, all-in-all there exists analytic functions 
$\mf{r}_{p;2}$ on $\mc{S}_{2\varkappa_{\De}}(\R)$ and such that 
\beq
\mf{R}_{N;2}[f]\, = \, \sul{p=0}{r-1} \f{1}{L^{p+2}} \mf{r}_{p;2}\big( \la \mid \{ \wh{x}_{p_a}\}_1^n ; \{ \wh{x}_{h_a}\}_1^n \big) \; + \;  \e{O}\Big(  \f{1}{L^{r+2}} \Big)
\enq
with an analytic remainder on  $\mc{S}_{2\varkappa_{\De}}(\R)$ that is, furthermore, uniform on $\ov{\mc{S}}_{\eta}(\R)$ for any $0<\eta<2\varkappa_{\De}$.

One shows analogously that 
\beq
\mf{R}_{N;3}[f]\, = \, \sul{p=0}{r-1} \f{1}{L^{p+2}} \mf{r}_{p;3}\big( \la \mid \{ \wh{x}_{p_a}\}_1^n ; \{ \wh{x}_{h_a}\}_1^n \big) \; + \;  \e{O}\Big(  \f{1}{L^{r+2} } \Big)
\enq
for some functions $\mf{r}_{p;3}$ and remainder  enjoying the sought properties.

It thus remains to focus on $\mf{R}_{N;1}[\wh{\xi}]$. For later convenience, set 
\beq
G_{\la}(z ) \, = \,  \f{ R_{I_{\wh{q}}}\big( \la, \wh{\xi}^{-1}(z) \big)  }{ \wh{\xi}^{\prime}\circ  \wh{\xi}^{-1}(z)  } \;. 
\enq
Then, agreeing upon 
\beq
\wh{\Ga}^{(\e{ply})}_{\eps} \,  = \,  \wh{\Ga}_{\eps} \cap \mc{S}_{\tf{\a}{2}}(\R) \quad  \e{and} \quad  \wh{\Ga}^{(\e{exp})}_{\eps}  \, = \,  \wh{\Ga}_{\eps} \setminus \wh{\Ga}^{(\e{ply})}_{\eps}
\enq
with $\a>0$ as given in Figure \ref{Figure domaine G alpha hat} and small enough, $\mf{R}_{N;1}[\wh{\xi}]$ can be decomposed as 
\beq
\mf{R}_{N;1}[\wh{\xi}] \, = \, \mf{R}_{N;1}^{(\e{ply})}[\wh{\xi}] \, + \, \mf{R}_{N;1}^{(\e{exp})}[\wh{\xi}]
\enq
where 
\beq
 \mf{R}_{N;1}^{(\e{ply}/\e{exp})}[\wh{\xi}](\la) \; = \; -\sul{\eps=\pm}{}\; \Int{ \wh{\Ga}^{\, (\e{ply}/\e{exp})}_{\eps}  }{} \hspace{-3mm} G_{\la}(z) \ln \big[ 1-\ex{2\i\pi \eps L z} \big] \cdot \f{ \dd z}{2\i\pi L } \;. 
\enq
For given $\eta$, $0<\eta<2\varkappa_{\De}$, reducing $\a$ if necessary, one infers from the analytic structure of the resolvent that both   $\mf{R}_{N;1}^{(\e{ply}/\e{exp})}[\wh{\xi}]$ are holomorphic on $\mc{S}_{\eta}(\R)$. 
Furthermore, for such an appropriate value of $\a$, one has that 
\beq
 \Big| \mf{R}_{N;1}^{( \e{exp})}[\wh{\xi}](\la) \Big| \; \leq \; \norm{ G_{\la} }_{L^{\infty}(\mc{S}_{\eta}(\R)\times U_{\a})} 
\sul{\eps=\pm }{} \;  \Int{ \wh{\Ga}^{ \, ( \e{exp})}_{\eps}  }{}\Big| \ln \big[ 1-\ex{2\i\pi L z} \big] \Big| \cdot \f{ \dd z}{2 \pi L }  
\leq \; C  \cdot \f{1}{L}\sul{\eps=\pm}{} \Big| \wh{\Ga}_{\eps} \Big| \ex{-\pi \a L }
\enq
where $U_{\a}$ is such that $\wh{\Ga}_{+}\cup\wh{\Ga}_{-}\subset U_{\a}$ and, in the last bound, one uses that $R_{I_{\wh{q}}}$ is bounded on the relevant domain.  
The integrals arising in the definition of $\mf{R}_{N;1}^{( \e{ply})}[\wh{\xi}]$ can be recast as 
\beq
\Int{ \wh{\Ga}^{\,(\e{ply}) }_{\eps}  }{} G_{\la}(z) \ln \big[ 1-\ex{2\i\pi \eps L z} \big] \cdot \f{ \dd z}{2\i\pi L } \; = \; 
 \Int{ 0 }{ \tf{\a}{2} } \bigg\{ G_{\la}\Big( \i \eps s +\tfrac{N+\tf{1}{2}}{L} \Big) \, - \,  G_{\la}\Big( \i \eps s +\tfrac{1}{2L} \Big) \bigg\} \cdot \ln \big[ 1+\ex{-2 \pi L s} \big] \cdot \f{ \dd s }{ 2 \pi L } \;. 
\enq
so that  
\beq
 \mf{R}_{N;1}^{(\e{ply} }[\wh{\xi}](\la) \; = \; \Int{ 0 }{ \tf{\a}{2} }\msc{G}_{\la}(s) \cdot \ln \big[ 1+\ex{-2 \pi L s} \big] \cdot \f{ \dd s }{ 2 \pi L }
\enq
with
\beq
 \msc{G}_{\la}(s) \, = \, \sul{\eps=\pm}{} \bigg\{ G_{\la}\Big( \i \eps s +\tfrac{N+\tf{1}{2}}{L} \Big) \, - \,  G_{\la}\Big( \i \eps s +\tfrac{1}{2L} \Big) \bigg\} \;. 
\enq
Upon expanding $s \mapsto \msc{G}_{\la}(s)$ into a Taylor integral expansion of order $r$ around $s=0$, one gets
\bem
 \mf{R}_{N;1}^{(\e{ply}) }[\wh{\xi}](\la) \; = \;  \sul{p=0}{r} \f{ \msc{G}_{\la}^{(p)}(0) }{ p! } \bigg( \f{1}{2\pi L} \bigg)^{p+2} \Int{ 0 }{ \pi \a L  } s^{p}  \ln \big[ 1+\ex{- s} \big] \cdot  \dd s  \\
\; + \;
  \f{ 1 }{ (r+1)! } \bigg( \f{1}{2\pi L} \bigg)^{r+3} \Int{ 0 }{ \pi \a L  } \dd s  s^{r+1}  \ln \big[ 1+\ex{- s} \big]     \cdot  \Int{0}{1} \msc{G}^{(r+1)}_{\la}\Big( \f{ts}{2\pi L} \Big) (1-t)^{r+1} \dd t \;. 
\end{multline}
The second line produces a $\e{O}(L^{-r-3})$ remainder, uniform on $\ov{\mc{S}}_{\eta}(\R)$ for any $0<\eta <2\varkappa_{\De}$, and holomorphic on $ \mc{S}_{2\varkappa_{\De}}(\R)$ 
The integrals arising in the first line can be estimated as 
\beq
 \Int{ 0 }{ \pi \a L  } s^{p}  \ln \big[ 1+\ex{- s} \big] \cdot  \dd s  \; = \;  \Int{ 0 }{ +\infty } s^{p}  \ln \big[ 1+\ex{- s} \big] \cdot  \dd s  \; + \; \e{O}(L^{-\infty})
\; = \;  p! \zeta(p+2) (1-2^{-1-p})  \; + \; \e{O}(L^{-\infty})
\enq
where $\zeta(s)$ is the Riemann zeta function. All of this yields
\bem
\f{ \msc{G}_{\la}^{(p)}(0) }{ p! } \bigg( \f{1}{2\pi L} \bigg)^{p+2} \Int{ 0 }{ \pi \a L  } s^{p}  \ln \big[ 1+\ex{- s} \big] \cdot  \dd s \\
\; = \;  \bigg( \f{1}{2\pi L} \bigg)^{p+2}\zeta(p+2) (1-2^{-1-p}) 
 \sul{\eps=\pm}{}\Dp{s}^{p} \bigg\{ G_{\la}\Big( \i \eps s +\tfrac{N+\tf{1}{2}}{L} \Big) \, - \,  G_{\la}\Big( \i \eps s +\tfrac{1}{2L} \Big) \bigg\}_{\mid s=0} \; + \;    \e{O}\Big(  \f{1}{L^{r+2} } \Big) 
\end{multline}
with a remainder that is easily seen to enjoy the desired properties.

Since $\wh{\xi}^{-1}$ and $\wh{\xi}^{\prime}$ admit holomorphic asymptotic expansions up to  $\e{O}\Big(  L^{-r-1}  \Big)$, so does 
\beq
\msc{G}_{\la}^{(p)}(0) \, = \, \Dp{s}^{p} \bigg\{ G_{\la}\Big( \i \eps s +\tfrac{N+\tf{1}{2}}{L} \Big) \, - \,  G_{\la}\Big( \i \eps s +\tfrac{1}{2L} \Big) \bigg\}_{s=0} \, .
\enq
Thus, owing to the pre-factors $L^{-p-2}$, one gets
\beq
\mf{R}_{N;1}[f]\, = \, \sul{p=0}{r-1} \f{1}{L^{p+2}} \mf{r}_{p;1}\big( \la \mid \{ \wh{x}_{p_a}\}_1^n ; \{ \wh{x}_{h_a}\}_1^n \big) \; + \;  \e{O}\Big(  \f{1}{L^{r+2} } \Big) 
\enq
for holomorphic functions $ \mf{r}_{p;1}$ on $\mc{S}_{2\varkappa_{\De}}(\R)$ and a uniform and holomorphic remainder on  $\mc{S}_{\eta}(\R)$, $0<\eta<2\varkappa_{\De}$. 
Thus, by adding up the various contributions, the existence of the asymptotic expansion for $\wh{\xi}$ up to a $\e{O}\big(  L^{-r-2}  \big) $ remainder. 
 The latter implies the  existence of the asymptotic expansion for $\wh{\xi}^{-1}$ with the same precision, and thus, as well, for the two endpoints $\wh{q}_{R/L}$. \qed

\subsection{ The counting function when $D=\tf{1}{2}$ and $-1<\De \leq 1$}

\begin{theorem}
\label{Theorem Approximation effective de xi a densite un demi}

Let $-1 < \De \leq 1$ and $\tf{N}{L}=D=\tf{1}{2}$. Then, for any $\eps>0$ there exists $L_0$ such that the counting function $\wh{\xi}$ constructed from the 
Bethe roots associated with the choice of integers $\ell_a=a$ in the logarithmic Bethe equations admits the expansion  
\beq
\wh{\xi}(\la)\; = \; \xi_0(\la\mid +\infty) \; + \; \e{O}(\eps) \qquad with \qquad  \xi_0(\la\mid +\infty) \, = \, \Int{ 0 }{ \la } \rho_{\infty}(\mu) \cdot \dd \mu 
\enq
where the remainder in uniform in $L \geq L_0$.

\end{theorem}

\Proof

Let $\{\la_a\}$ be a solution to the logarithmic Bethe equations associated with the choice of integers $\ell_a=a$. 
In virtue of Proposition \ref{Proposition estimation fraction la GS qui sechappent a infini}, there exists $\La_{\eps}$ such that $\wh{c}_{\La_{\eps}}<\eps$, where 
$\wh{c}_{\La}$ has been defined in \eqref{borne et definition de la fraction c Lambda}.

The counting function  $\wh{\xi}$ associated with these roots is a sequence of holomorphic function on $\mc{S}_{ \varkappa_{\De} }(\R)$ and, as such, admits a converging subsequence to some $\xi$.
It remains to characterise the limit $\xi$ and hence the limit of $\wh{\xi}$ according to the reasoning of the proof of Theorem \ref{Theorem convergence et eqn NL pour ctg fct}. 

The first stage of the proof follows the analysis developed in the proof of  Theorem \ref{Theorem convergence et eqn NL pour ctg fct} relative to waiving-off the possibility that $\xi$ has 
several zeroes on $\R$. Taking the same definition of \eqref{definition X in et out} with $\La$ now being replaced by $\La_{\eps}$ one arrives to 
\eqref{ecriture hat xi prime dans l'hypothese de plusieurs intervalles de positivite pour la limite xi} with $\wh{\psi}_{\e{in}/\e{out}}$ defined exactly as in 
\eqref{definition hat psi in et out}. Again, by positivity of the integral operator and by invoking \eqref{ecriture borne inf sur moment habille intervalle J+}, one gets the lower bound  \eqref{ecriture borne inf sur xi hat}, \textit{viz}. 
\beq
 \wh{\xi}^{\prime} (\la) \; \geq  \; \rho_{\infty}(\la) \; + \;   \Big( \e{id} \;     +   \big[\msc{R}_{ J^{(+)} }\big]_{ J^{(-)} }  \Big) \Big[   \; - \, \wh{\psi}_{\e{out} }
\; + \; \sul{k=0}{r}\wt{\mf{r}}^{\, (k)}[\wh{\xi}] \Big](\la)   \;. 
\label{ecriture borne inf sur xi hat cas support infini des racines}
\enq
By continuity of  $ \e{id} \;     +   \big[\msc{R}_{ J^{(+)} }\big]_{ J^{(-)} }   $, it follows from $\wh{c}_{\La_{\eps}}<\eps$ and the bounds \eqref{bornage du cardinal de X out} on 
$ \# X^{(\e{out})}$ which allow one to control the cardinality of $ Y^{(\e{out})} \cap X^{(\e{out})}$  that 
\beq
 \limsup_{L\tend + \infty} \Big|   \Big( \e{id} \;     +   \big[\msc{R}_{ J^{(+)} }\big]_{ J^{(-)} }  \Big) \Big[   \; - \, \wh{\psi}_{\e{out} }\Big](\la) \Big| \; \leq \; 
	    C\cdot \Big\{ \de (r+1) \norm{\xi^{\prime}}_{L^{\infty}(I_{ \La_{\eps} } )}  +  \eps\Big\}\;. 
\enq
In the above formula, $r$ refers to the number of zeroes of $\xi^{\prime}$ on $\intoo{- \La_{\eps} }{ \La_{\eps} }$ and thus depends \textit{a priori} on $\eps$. 
One also obtains a similar bound on the second term, namely 
\beq
\limsup_{L\tend + \infty}\Big| \Big( \e{id} \;     +   \big[\msc{R}_{ J^{(+)} }\big]_{ J^{(-)} }  \Big) \Big[  \sul{k=0}{r}\wt{\mf{r}}^{\, (k)}[\wh{\xi}] \Big](\la) \Big| \; \leq  \; 
 C^{\prime} \cdot \de \cdot (r+1)  \cdot \norm{\xi^{\prime}}_{L^{\infty}(I_{ \La_{\eps} } )}  \;. 
\enq
Taking the pointwise in $\la$ $L\tend + \infty$ limit superior 
of \eqref{ecriture borne inf sur xi hat cas support infini des racines}, one finds the lower bound
\beq
 \xi^{\prime} (\la) \; \geq  \; \rho_{\infty}(\la)  \; - \; C^{\prime\prime} \Big(  \de \cdot (r+1) \cdot \norm{\xi^{\prime}}_{L^{\infty}(I_{ \La_{\eps} } )}  +  \eps \Big) \;. 
\enq
It then remains to send first $\de \tend 0^+$ and then $\eps \tend 0^+$ so as to get the strict positivity of $\xi$ on $\R$.

Now, taking for granted that $\xi^{\prime}>0$ on $\R$, one again picks $\eps>0$ and $\La_{\eps}$ such that $\wh{c}_{\La_{\eps}}<\eps$
and defines $\wh{q}_{R;\eps}$, resp. $\wh{q}_{L;\eps}$, to be the closed to $\La_{\eps}$, resp. $-\La_{\eps}$,  solution to $\exp\big\{ 2\i\pi L \wh{\xi}(\la)  \big\}=-1$  lying outside of $I_{\La_{\eps} }$. 
Repeating similar handlings to the ones carried out in the corresponding section of the proof of Theorem \ref{Theorem convergence et eqn NL pour ctg fct}, one gets 
\bem
\wh{\xi}_{\e{sym}}(\la) \; + \; \Int{ \wh{q}_{L;\eps} }{ \wh{q}_{R;\eps} }  K(\la-\mu) \wh{\xi}_{\e{sym}}(\mu)  \cdot \dd \mu \; = \; \f{ \mf{p}(\la) }{ 2\pi } \\
-\f{ 1  }{ 2\pi } \Big\{  \wh{\xi}_{\e{sym}}( \wh{q}_{R;\eps}) \cdot   \th\big( \la -  \wh{q}_{R;\eps} \big) \, - \,  \wh{\xi}_{\e{sym}}( \wh{q}_{L;\eps}) \cdot   \th\big( \la -  \wh{q}_{L;\eps} \big) \Big\} 
 \; + \; \mf{r}_1\big[ \wh{\xi} \big](\la) \, - \, \f{1}{2\pi L } \sul{ x \in \wt{Y}^{(\e{out})} }{ } \th(\la-x)  \; .
\end{multline}
Here $\wt{Y}^{(\e{out})}= \{ \la_a\; : \;  \la_a \not \in \intff{ \wh{q}_{L;\de} }{ \wh{q}_{R;\de} } \}$ and $\mf{r}_1\big[ \wh{\xi} \big](\la)$ is as given by \eqref{definition reste operateur integral r1} relatively to the above defined 
endpoints of integration $\wh{q}_{R/L;\eps}$. After a few handlings, one recasts the above equation as 

\bem
 \wh{\xi}_{\e{sym}} (\la)   \; = \;  p( \la \mid +\infty ) 
\,+\, Z(\la\mid +\infty) \f{ \pi-2\zeta }{2\pi} \cdot \Big\{  \wh{\xi}_{\e{sym}}( \wh{q}_{R;\eps}) +  \wh{\xi}_{\e{sym}}( \wh{q}_{L;\eps}) \Big\} \\
 \; + \; \mf{R}_{N;1}^{(\e{tot})}\big[ \wh{\xi} \big](\la)\, - \, \f{1}{ L } \sul{ x \in \wt{Y}^{(\e{out})} }{ } \vp(\la-x)  
\label{ecriture forme transformee pour xi s cas D Un et demi}
\end{multline}
where 
\bem
\mf{R}_{N;1}^{(\e{tot})}\big[ \wh{\xi} \big](\la) \; = \;   \mf{R}_{N;1}\big[ \wh{\xi} \big](\la) 
 \, +\, \Int{ \wh{q}_{R;\eps} }{+\infty}  R(\la-\mu) \big[ \wh{\xi}_{\e{sym}}(\mu) - \wh{\xi}_{\e{sym}}\big(  \wh{q}_{R;\eps}\big) \big]  \cdot \dd \mu  \\
 \, + \, \Int{-\infty}{ \wh{q}_{L;\eps} }  R(\la-\mu) \big[ \wh{\xi}_{\e{sym}}(\mu) - \wh{\xi}_{\e{sym}}\big(  \wh{q}_{L;\eps}\big) \big]  \cdot \dd \mu
\end{multline}
and 
\beq
\mf{R}_{N;1}\big[ \wh{\xi} \big](\la) \; = \; - \; \sul{\eps = \pm }{}\Int{ \wh{\Ga}_{\eps} }{} 
\f{  R\big( \la - \wh{\xi}^{-1}(z)  \big) }{ \wh{\xi}^{\prime}\big( \wh{\xi}^{-1}(z) \big) } \cdot \ln\Big[1- \ex{2\i \pi \eps L z} \Big] \cdot \f{ \dd z}{2\i \pi L }  \;. 
\enq
It remains to bound the various terms in \eqref{ecriture forme transformee pour xi s cas D Un et demi}. A counting of solution argument ensures that 
\beq
\Big[L \Big( \wh{\xi}_{\e{sym}}(-\La_{\eps} ) - \wh{\xi}_{\e{sym}}\big( -\infty \big)  \Big)  \Big] \, \leq \, L \wh{c}_{\La_{\eps}} \quad \e{and} \quad
\Big[L \Big(\wh{\xi}_{\e{sym}}\big( +\infty \big)- \wh{\xi}_{\e{sym}}(\La_{\eps} )   \Big)  \Big] \, \leq \, L \wh{c}_{\La_{\eps}} \;. 
\enq
Thus, 
\beq
\wh{\xi}_{\e{sym}}( \wh{q}_{L;\eps} ) - \wh{\xi}_{\e{sym}}\big( -\infty \big) \,  \leq  \, \wh{\xi}_{\e{sym}}(-\La_{\eps} ) - \wh{\xi}_{\e{sym}}\big( -\infty \big)  \, \leq \, \wh{c}_{\La_{\eps}}  \, + \, \tf{1}{L}  
\enq
and a similar bound holds for $\wh{\xi}_{\e{sym}}\big(\!+\infty\big)- \wh{\xi}_{\e{sym}}( \, \wh{q}_{R;\eps}  ) $. From there, since $\xi_{\e{sym}}(+\infty) + \xi_{\e{sym}}(-\infty)=0$, one readily 
bounds the second terms in \eqref{ecriture forme transformee pour xi s cas D Un et demi}. From the above bounds one also infers that 
\beq
\Big| \Int{ \wh{q}_{R;\eps} }{+\infty}  R(\la-\mu) \big[ \wh{\xi}_{\e{sym}}(\mu) - \wh{\xi}_{\e{sym}}\big(  \wh{q}_{R;\eps}\big) \big]  \cdot \dd \mu  \Big| \, \leq \, \norm{ R }_{L^{1}(\R) } \cdot \big( \, \wh{c}_{\La_{\eps}}  \, + \, \tf{1}{L}  \big) 
\enq
and analogously for its counterpart related to $\wh{q}_{R;\eps}$. Finally, one gets that 
\beq
\Big| \mf{R}_{N;1}\big[ \wh{\xi} \big](\la) \Big| \; \leq \; \f{ C }{ L^2 } \cdot \f{ \big(2\La_{\eps}+1\big)  \cdot \norm{ R }_{ L^{\infty}(\mc{S}_{\eta}(\R) ) }  }
{\inf \Big\{ \wh{\xi}^{\prime}(s) \, : \, s \in \mc{S}_{\eta}(I_{\La_{\eps}} ) \Big\}  } \;. 
\enq
Given $\eps>0$ fixed, the infimum is bounded from below due to the convergence of $\wh{\xi}$ on compact subsets. This entails the claim, upon taking $L$ large-enough. \qed

\section{Applications}
\label{Section Applications to comutation of limit}

In this section I establish several more or less direct applications of the existence of the large-$L$ asymptotic expansion of the counting function. 
The first of these corresponds to the proof of the existence of limits of the type \eqref{ecriture convergence somme sur racines de Bethe limit thermo}. 
The second application concerns the proof of the conformal behaviour of the spectrum of low-lying excitations above the ground state in the so-called massless regime of the model, 
namely when there exist zero energy excitations. This regime corresponds basically to $D\in \intoo{0}{\tf{1}{2}}$ for any $\De >-1$ and $D=\tf{1}{2}$ for 
$-1<\De \leq 1$. Below, I shall confine myself to the regime $0< D < \tf{1}{2}$ since the latter is technically much simpler to deal with.

\subsection{Densification of Bethe roots}

\begin{theorem}
\label{Theorem densification racines de Bethe}
Let $N,L\tend +\infty$ in such a way that $ \tf{N}{L} \tend D  \in \intof{0}{ \tf{1}{2} }$. Let $q$ be the unique solution to \eqref{ecriture probleme pour la bord de Fermi magnetique}
subordinate to $D$ and let the particle-hole integers \eqref{definition entiers ella en termes ha et pa} satisfy to the hypothesis of Theorem \ref{Theorem convergence et eqn NL pour ctg fct}.
Then, given any bounded Lipschitz function $f$ on $\R$ with Lipschitz constant $ \e{Lip}[f] $, it holds
\beq
\f{1}{L} \sul{a=1}{N} f(\la_a) \; \tend \; \Int{-q}{q} f( \la ) \rho( \la \mid q ) \cdot \dd \la
\enq
where $\{\la_a\}_1^N$ correspond to the unique solution to the logarithmic Bethe equations subordinate to the given choice of particle-hole integers.

\end{theorem}

\Proof

Due to the unbounded nature of the Bethe roots at $D=1/2$ and $-1< \De \leq 1$, I first deal with the simpler case of a uniformly bounded in $L$
distribution, namely when $0\leq D< \tf{1}{2}$ and $-1< \De \leq 1$ or $0\leq D \leq 1/2$ when $\De > 1$. 

For this range of $D$ and $\De$,  Proposition \ref{Proposition bornitude des racines de Bethe} ensures that 
the Bethe roots such that $\wh{\xi}(\la_a) = \tf{a}{L}$ all belong to the compact $\intff{-\La}{\La}$. 
Also, it has already been established that, provided $L$ is large enough
\beq
\norm{ \wh{\xi}^{-1} - \xi_0^{-1}(*\mid \wh{q}\,) }_{ L^{\infty}\big(  \xi_0\big( S_{\eta,\eps}(I_{\La}) \big) \big) } \; \leq \; \norm{ \wh{\xi} - \xi_0(*\mid \wh{q}\,) }_{ L^{\infty}\big(  S_{ \varkappa_{\De} }(I_{2\La}) \big) }
									  \; \leq \; \f{ C }{ L } 
\enq
for some $\eps, \eta>0$ small enough. Finally, for $L$ large enough, it holds $\tf{a}{L} \in \xi_0\big( I_{\La+\eps} \mid \wh{q} \, \big)$ for any $a\in \intn{1}{N}$. The finite sum one starts with can be decomposed as 
\beq
\f{1}{L} \sul{a=1}{N}f(\la_a) \; = \; \msc{S}_1 \, + \, \msc{S}_2\, + \, \msc{S}_3 \, + \, \msc{S}_4 
\qquad \e{with} \qquad \msc{S}_1\;=\; \f{1}{L} \sul{a=1}{N}\Big[ f\circ\wh{\xi}^{-1}\Big(\f{a}{L} \Big) \, - \, f\circ\xi^{-1}_0\Big(\f{a}{L} \mid \wh{q}\, \Big)  \Big]
\enq
while 
\beq
\msc{S}_2 \; = \; \f{1}{L}\sul{a=1}{n} \Big[ f\big( \wh{x}_{p_a} \big) -  f\big( \wh{x}_{h_a} \big) \Big] \quad , \qquad 
\msc{S}_3 \; = \; \f{1}{L}\sul{a=1}{N} \Big[ f\circ\xi^{-1}_0\Big(\f{a}{L} \mid \wh{q} \, \Big) -  f\circ\xi^{-1}_0\Big(\f{a}{L} \mid q \Big) \Big]  
\enq
and 
\beq
\msc{S}_4 \; = \; \f{1}{L}\sul{a=1}{N} f\circ\xi^{-1}_0\Big(\f{a}{L} \mid q \Big) \;. 
\enq
$\msc{S}_4$ is a Riemann sum and, as such, converges
\beq
\msc{S}_4 \tend \Int{0}{D} \hspace{-1mm} f\circ\xi^{-1}_0(s\mid q) \cdot \dd s \;= \;  \Int{-q}{q} \hspace{-1mm} f( \la ) \, \rho( \la \mid q ) \cdot \dd \la
\enq
where the last equality follows from $\xi_0\big(\intff{-q}{q} \mid q \big)=\intff{0}{D}$ and $\xi_0^{\prime}(\la\mid q) = \rho( \la \mid q ) $. 
Further, $R_{I_{\a}}(\la,\mu)$ is readily seen to be smooth in $\a$ and satisfy $ \norm{ \Dp{\a} R_{I_{\a}} }_{L^{\infty}(\R^2)} \leq C$ for $\a$ bounded. The latter leads to 
\beq
\msc{S}_{3}\, \leq \, C^{\prime} \cdot \e{Lip}[f] \cdot |\wh{q}-q| \cdot \f{N}{L }  \limit{L}{+\infty} 0 \;. 
\enq
Clearly, $\msc{S}_2 \; = \; \e{O}(L^{-1})$ since $f$ is bounded while
\beq
\big| \msc{S}_1 \big| \;\leq \; \f{N}{L} \cdot  \e{Lip}[f] \cdot \norm{ \wh{\xi}^{-1} - \xi_0^{-1}(*\mid \wh{q}\,) }_{ L^{\infty}\big(  \xi_0\big( I_{\La+\eps} \mid q \big) \big) } 
 \; = \; \e{O}\big( L^{-1} \big)\;. 
\enq

\vspace{2mm}

I now focus on the remaining cases, \textit{viz}. $D=1/2$ and $-1<\De \leq 1$ and picking some $M>0$ decompose the sum of interest as 
\beq
\f{1}{L} \sul{a=1}{N}f(\la_a) \; = \; \f{1}{L} \sul{ |\la_a|>M  }{ }f(\la_a)   \; + \; \f{1}{L} \sul{ a\, : \,  |\la_a| \leq M }{}\Big[ f\circ\wh{\xi}^{-1}\Big(\f{a}{L} \Big) \, - \, f\circ\xi^{-1}_0\Big(\f{a}{L} \mid +\infty\Big)  \Big]
\; +\; \f{1}{L} \sul{ |\la_a| \leq M }{} f\circ\xi^{-1}_0\Big(\f{a}{L} \mid +\infty \Big)   \;. 
\enq
The last term converges as a Riemann sum 
\beq
\f{1}{L} \sul{ |\la_a| \leq M }{} f\circ\xi^{-1}_0\Big(\f{a}{L} \mid +\infty \Big)   \; \tend \; \Int{ - M }{ M } f(\la) \rho_{\infty}(\la) \dd \la \;. 
\enq
The estimates of Theorem \ref{Theorem Approximation effective de xi a densite un demi} ensure that, provided $\eps$ is small enough, one has 
\beq
 \limsup_{L\tend +\infty} \Big| \f{1}{L} \sul{ |\la_a| \leq M }{}\Big[ f\circ\wh{\xi}^{-1}\Big(\f{a}{L} \Big) \, - \, f\circ\xi^{-1}_0\Big(\f{a}{L} \mid +\infty \Big)  \Big] \Big| \; \leq \; 
C \cdot \e{Lip}[f]\cdot  \eps \;. 
\enq
 Finally, 
\beq
 \limsup_{L\tend + \infty}\Big| \f{1}{L} \sul{ |\la_a|>M  }{ }f(\la_a) \Big|  \; \leq \; 
\norm{f}_{ L^{\infty}(\R) } \cdot  \limsup_{L\tend + \infty} \Big(  \f{1}{L} \# \big\{ a \in \intn{1}{N} \;: \; |\la_a| > M \big\} \Big) \;. 
\enq
Thus, upon sending $L\tend +\infty$ and then $\eps \tend 0^+$ one gets 
\beq
\f{1}{L} \sul{a=1}{L}f(\la_a) \; \tend \; \Int{-M}{M} f(\la) \rho_{\infty}(\la) \dd \la  \; + \; \e{O}\bigg(    \limsup_{L\tend + \infty} \Big(  \f{1}{L} \# \big\{ a \in \intn{1}{N} \;: \; |\la_a| > M \big\} \Big) \bigg)  \;. 
\enq
Finally, it remains to relax $M\tend +\infty$. The remainder term, in virtue of Proposition \ref{Proposition estimation fraction la GS qui sechappent a infini}, tends to 0. \qed

\subsection{First corrections to the densification of Bethe roots}

The present sub-section  will improve the results of the last one by providing the first two sub-dominant corrections to the sums $\sum_{ a = 1 }^{ N^{\prime} } f(\la_a)$
in which $\{\la_a\}_1^{N^{\prime}}$ is a set of particle-hole Bethe roots and $f$ is holomorphic in an open neighbourhood of 
the interval $I_q$ where the Bethe roots form a dense distribution.

In order to state the result it is convenient to introduce the function
\beq
F(\la) \; = \; f(\la)- \Int{ - \wh{q} }{ \wh{q} } R_{I_{\wh{q}}}(\la,\mu) f(\mu) \cdot \dd \mu \;,
\label{definition fct F}
\enq
\textit{i.e.} the solution to $\Big(\e{id}+K_{I_{\wh{q}}}\Big)[F]=f$. The below linear combination 
\beq
F_{\e{eff}}(\la) \, = \, F(\la) \, + \, F(\, \wh{q}\,) \,\vp(\wh{q},\la) \, - \,  F(-\wh{q}\,) \,\vp(-\wh{q},\la)
\label{definition fct Feff}
\enq
will also be of use.

\begin{prop}
\label{Proposition DA ordre 2 avec fct holomorphe}
Let $0\leq D < \tf{1}{2}$,  $\De >-1$ and let $q$ be the associated value of the magnetic Fermi boundary. Let $f$ be holomorphic in an open 
neighbourhood $U_q$ of $I_q$ and let $\{\la_a\}_1^{N+s}$ be a solution to the Bethe Ansatz equation as discussed in Proposition \ref{Proposition DA counting fct}. Then, under the assumptions of Theorem \ref{Theorem convergence et eqn NL pour ctg fct},  it holds
\beq
\sul{a=1}{N+s} f(\la_a) \; = \;  L \cdot \mc{S}_{0}[f]( \, \wh{q} \,  ) 
\; + \; \sul{k=1}{2} \f{1}{L^{k-1}} \mc{S}_{k}[f] \Big(   \{ \wh{x}_{p_a^{\prime}} \}_1^{ n_{p}^{(s)} }  \, ; \, \{ \wh{x}_{ h_a^{\prime} } \}_1^{ n_{h}^{(s)} }  \Big) \; + \; \e{O}\Big( L^{-2} \Big)
\enq
where 
\beq
 \mc{S}_{0}( \, \wh{q} \, )\; = \;   \Int{ - \wh{q} }{ \wh{q} } \!\! f(\mu)  \, \rho(\mu \mid \wh{q} ) \cdot \dd \mu
\enq
\beq
\mc{S}_{1}[f] \Big( \{ \wh{x}_{p_a^{\prime}} \}_1^{ n_{p}^{(s)} }  \, ; \, \{ \wh{x}_{ h_a^{\prime} } \}_1^{ n_{h}^{(s)} } \Big) \; = \; 
\sul{a=1}{ n_{p}^{(s)} } F_{\e{eff}}\big(\,  \wh{x}_{p_a^{\prime} } \big)  \, - \,  \sul{a=1}{ n_{h}^{(s)} }F_{\e{eff}}\big( \, \wh{x}_{ h_a^{\prime} } \big)
-\f{s}{2} Z(\, \wh{q}\mid \wh{q}) \sul{\eps=\pm}{} F(\eps \wh{q}\, )
\enq
 with $F$ and $F_{\e{eff}}$ as defined through \eqref{definition fct F}-\eqref{definition fct Feff}, $p^{\prime}_a$, $h^{\prime}_a$ as introduced in \eqref{definition des racines p prime et h prime}.  Furthermore, one has 
\beq
\mc{S}_{2}[f] \Big(  \{ \wh{x}_{p_a^{\prime}} \}_1^{ n_{p}^{(s)} }  \, ; \, \{ \wh{x}_{ h_a^{\prime} } \}_1^{ n_{h}^{(s)} }  \Big) \; = \; 
 \f{1}{2}\sul{\eps=\pm}{} F_{\e{eff}}^{\, \prime}( \eps \wh{q} \, ) \cdot \bigg\{  \big( q_{\eps}^{(1)} \big)^2   \xi_{0}^{\prime}(\,  \wh{q} \mid \wh{q} \, )  \, - \,    \f{  1   }{12 \,  \xi_{0}^{\prime}( \, \wh{q}\mid \wh{q}\,  )  } \bigg\} \, , 
\enq
with $q^{(1)}_{\pm}$  defined by \eqref{ecriture coeff DA q L and R ordre 1}. 

\end{prop}

\Proof

Since $f$ is holomorphic in a neighbourhood of the real axis, one has the representation
\bem
\sul{a=1}{N+s} f(\la_a)  \; = \;  L \Int{ \wh{q}_L }{ \wh{q}_R } f(\mu) \cdot \wh{\xi}^{\, \prime}(\mu) \cdot \dd \mu 
\, + \, \sul{a=1}{ n_{p}^{(s)} } f(\, \wh{x}_{p_a^{\prime}} ) \, - \,  \sul{a=1}{ n_{h}^{(s)} } f(\, \wh{x}_{h_a^{\prime}} )  \\
\, -\, \sul{\eps= \pm }{} \Int{ \wh{\Ga}_{\eps} }{} \f{ f^{\prime}\big( \wh{\xi}^{-1}(z) \big) }{ \wh{\xi}^{\prime}\big( \wh{\xi}^{-1}(z) \big) } \ln \big[ 1-\ex{2\i \pi \eps L z}  \big] \cdot \f{ \dd z }{ 2 \i \pi } \;, 
\nonumber
\end{multline}
where $\wh{\Ga}_{\eps}$ is as defined on Figure \ref{Figure domaine G alpha hat} and $\wh{\xi}(\,\wh{q}_R)=\tf{(2N+1)}{2L}$ and $\wh{\xi}(\,\wh{q}_L)=\tf{1}{2L}$. 
A straightforward application of Watson's lemma leads to 
\beq
\sul{\eps= \pm }{} \Int{ \wh{\Ga}_{\eps} }{} \f{f^{\prime}\big( \, \wh{\xi}^{-1}(z) \,  \big) }{ \wh{\xi}^{\prime}\big( \, \wh{\xi}^{-1}(z) \, \big) } \ln \big[ 1-\ex{2\i \pi \eps L z}  \big] \cdot \f{ \dd z }{ 2 \i \pi } 
\; = \;- \f{ 1  }{24 L  }\sul{\eps=\pm}{}  \eps  \f{ f^{\prime} ( \, \wh{q}\, )}{ \xi_{0}^{\prime}( \eps \wh{q}\mid \wh{q} \, )}  \; + \; \e{O}\Big( \f{1}{L^2}\Big) \;. 
\enq
In its turn, by Proposition \ref{Proposition DA counting fct}, the one-fold integral admits the expansion 
\beq
\Int{ \wh{q}_L }{ \wh{q}_R } f(\mu) \cdot  \wh{\xi}^{\, \prime} (\mu) \cdot \dd \mu \; = \; \sul{k=0}{2} \f{1}{L^k} \Int{ \wh{q}_L }{ \wh{q}_R } f(\mu) \cdot \big(\xi^{(s)}_k\big)^{\prime}(\mu) \cdot \dd \mu  \; + \; \e{O}\Big( \f{1}{L^3}\Big)
\enq
where, so as to lighten the notation, I have dropped the dependence of the function $\xi_{k}^{(s)}$ on the roots $\wh{x}_{p_a /h_a}$. 
It then remains to use the Taylor expansion valid for a sufficiently regular function g
\beq
\Int{ \wh{q}_L }{ \wh{q}_R } g(\mu) \cdot \dd \mu  \; \simeq  \; \Int{ -\wh{q} }{ \wh{q} } f(\mu) \cdot \dd \mu  \; + \; 
\sul{k \geq 0}{}\bigg\{ \f{ g^{(k)}(\, \wh{q} \, ) }{ (k+1)! } \big( \, \wh{q}_R -\wh{q} \, \big)^{k+1}   \; - \; \f{ g^{(k)}(-\wh{q}\,) }{ (k+1)! } \big( \, \wh{q}_L + \wh{q} \, \big)^{k+1}   \bigg\}\; ,
\enq
and the differential identities satisfied by the dressed phase \eqref{ecriture derivee lambda phase habillee} so as to get the claim after some longish but straightforward algebra. \qed

The above proposition deals with the case of sums involving holomorphic functions. It is possible to apply similar techniques so as to obtain the expansions in a less regular case, typically 
when the summand is holomorphic with the exception of a finite number of points where it has power-law or logarithmic singularities. However, this is a case-by-case study which depends on the 
problem of interest and the details of obtaining an asymptotic expansion can be readily figured out.

\subsection{The conformal structure of the low-lying excitations}

The XXZ chain embedded in an external magnetic field $h$ refers to the Hamiltonian $\op{H}_{\De;h}=\op{H}_{\De}-\op{S}^z\frac{h}{2}$ on $\mf{h}_{XXZ}$. Its eigenvalues and eigenvectors are readily deduced from those of $\op{H}_{\De}$ owing to 
$\big[ \op{H}_{\De}, \op{S}^{z} \big]=0$. The eigenvalue of $\op{H}_{\De;h}$ associated with the eigenvector $\Psi( \{\la_a\}_1^N )$ of $\op{H}_{\De}$ that is parametrised by the Bethe roots 
$\{\la_a\}_1^N$ takes the from 
\beq
\mc{E}\Big( \{\la_a\}_1^N \Big) \; = \; \Big(J \De \, - \, \f{h}{2} \Big)L \; + \; \sul{a=1}{N} \mf{e}(\la_a) 
\enq
with $\mf{e}$ as defined in \eqref{definition dressed and bare energies}. The large-$L$ expansion of the above eigenvalue involves the effective dressed energy defined as 
\beq
\veps_{\e{eff}}\big( \la \mid Q \big) \; = \; \veps\big( \la \mid Q \big) \, + \, \veps( Q \mid Q ) \Big( \vp(Q,\la\mid Q ) - \vp(-Q ,\la \mid Q ) \Big)  \;. 
\label{definition energie habille modifiee}
\enq
and is a direct consequence of Proposition \eqref{Proposition DA ordre 2 avec fct holomorphe}.

\begin{cor}
Under the assumptions of Theorem \ref{Theorem convergence et eqn NL pour ctg fct}, for any $0\leq D < \tf{1}{2}$ and  $\De >-1$ it holds
\beq
\mc{E}\Big( \{\la_a\}_1^{N+s} \Big) \; = \;  L \cdot \mc{E}_{0}( \, \wh{q} \,  ) 
\; + \; \sul{k=1}{2} \f{1}{L^{k-1}} \mc{E}_{k}\Big( \wh{q} \mid  \{ \wh{x}_{p_a^{\prime}} \}_1^{ n_{p}^{(s)} }  \, ; \, \{ \wh{x}_{ h_a^{\prime} } \}_1^{ n_{h}^{(s)} }  \Big) \; + \; \e{O}\Big( L^{-2} \Big)
\enq
where 
\beq
 \mc{E}_{0}( \, \wh{q} \, )\; = \; J \De \, - \, \f{h}{2} \, + \, \Int{ - \wh{q} }{ \wh{q} } \!\! \mf{e}(\mu)  \, \rho(\mu \mid \wh{q} ) \cdot \dd \mu
\enq
\beq
\mc{E}_{1}\Big( \wh{q} \mid \{ \wh{x}_{p_a^{\prime}} \}_1^{ n_{p}^{(s)} }  \, ; \, \{ \wh{x}_{ h_a^{\prime} } \}_1^{ n_{h}^{(s)} } \Big) \; = \; 
\sul{a=1}{ n_{p}^{(s)} } \veps_{\e{eff}}\big( \wh{x}_{p_a^{\prime} } \mid \wh{q} \big)  \, - \,  \sul{a=1}{ n_{h}^{(s)} }\veps_{\e{eff}}\big( \wh{x}_{ h_a^{\prime} } \mid q \big)
\enq
 with $\veps_{\e{eff}}$ as defined in \eqref{definition energie habille modifiee} and, finally, 
\beq
\mc{E}_{2}\Big( \wh{q} \mid \{ \wh{x}_{p_a^{\prime}} \}_1^{ n_{p}^{(s)} }  \, ; \, \{ \wh{x}_{ h_a^{\prime} } \}_1^{ n_{h}^{(s)} }  \Big) \; = \; - \f{    \veps_{\e{eff}}^{\, \prime}( \wh{q}\mid \wh{q} )  }{12 \,  \xi_{0}^{\prime}( \wh{q}\mid \wh{q} ) }
\; + \;  \f{ 1}{2}     \veps_{\e{eff}}^{\, \prime}( \wh{q}\mid \wh{q} ) \,  \xi_{0}^{\prime}( \wh{q}\mid \wh{q} ) \cdot \Big\{  \big( q_+^{(1)} \big)^2  +  \big( q_-^{(1)} \big)^2   \Big\}
%
%
%
%
%
%
\enq
with $q^{(1)}_{\pm}$ as defined by \eqref{ecriture coeff DA q L and R ordre 1}. 

\end{cor}

\begin{lemme}
Let  $ h_{c} > h > 0$ for $-1<\De \leq 1$ and $ h_{c} > h \geq  h_{c}^{(L)}$ for $1<\De $, then the function $Q \mapsto \mc{E}_0(Q)$ attains a unique minimum 
at $Q_{\e{F}} \in I_{\iota} \cap \R^{+}$, the so-called Fermi boundary of the model. Furthermore, $Q_{\e{F}}$ is such that 
\beq
\veps(\la\mid Q_{\e{F}})_{\mid I_{ Q_{\e{F}} } }  \, < \,  0 \qquad , \quad 
\veps(\la\mid Q_{\e{F}})_{\mid I_{ Q_{\e{F}} }^{ \e{c} } } \, > \,  0 \qquad and \quad \veps( Q_{\e{F}} \mid Q_{ \e{F} } )=0\;. 
\enq

\end{lemme}

\Proof

It follows from straightforward algebra that 
\beq
\Big( \mc{E}_{h}^{(0)}\Big)^{\prime}(Q) \, = \, \veps(Q\mid Q) \cdot \rho(Q\mid Q)\;. 
\enq
Since, $\rho(Q\mid Q) >0$, the derivative has the sign of $\veps(Q\mid Q)$. It follows form Proposition \ref{Proposition propriete de dressed energy and fermi boundary}
that $\veps(Q\mid Q)<0$ on $\intfo{ 0 }{ Q_{\e{F}} }$ and $\veps(Q\mid Q)>0$ on $I^{\e{c}}_{  Q_{\e{F}} }$. This entails that $ Q\mapsto \mc{E}_{0}(Q)$ admits a unique minimum
at $Q_{\e{F}} $. \qed

Define the Fermi density $D_F$ by $D_F=p(Q_F\mid Q_F)$. Given $L$ large enough it seems reasonable to expect that the ground state of $\op{H}_{\De}^{(N)}$ with $N$ such that 
$|D_F-\tf{N}{L}|$ is minimal will give rise to a\symbolfootnote[2]{There could, in principle, more than one ground state since $|D_F-\tf{N}{L}|$ could attain two minima for some fixed value of $L$.} ground state of $H_{\De;h}$. 
In the present state of the art, I am however not in position to prove the statement due to the lack of a sufficient uniform in $L,N$ control close to $\tf{N}{L}=\tf{1}{2}$
of $\mc{E}\big( \{\la_a\}_1^N\big)$.

\begin{theorem}

Let $h$ be such that $D_F\in \intoo{0}{\tf{1}{2}}$ and assume that $N,L$ are such that $ D_F - \tf{N}{L}=\e{O}(L^{-3})$. Let $s$ be fixed and consider the solution to 
the logarithmic Bethe equations  $\{\la_a\}_1^{N+s}$ subordinate to the choice of particle-hole integers $\{p_a\}_1^n$ and $\{ h_a \}_1^n$ 
\beq
p_a\,=\, 1-p_a^- \qquad for \quad a=1,\dots, n_p^- \qquad and \qquad p_{n_p^-+a}\,=\,N+s+ p_a^+ \qquad for \quad a=1,\dots, n_p^+ \;, 
\enq
with $p_a^{\pm} \in \mathbb{N}^*$ fixed in $N,L$,
\beq
h_a\,=\, h_a^- \qquad for \quad a=1,\dots, n_h^- \qquad and \qquad h_{n_h^-+a}\,=\,N+s+1-h_a^+ \qquad for \quad a=1,\dots, n_h^+ \;, 
\enq
with $h_a^{\pm} \in \mathbb{N}^*$ fixed in $N,L$, and $n_h^++n_h^-=n=n_p^++n_p^-$. 

 Then, it holds
\bem
\mc{E}\Big( \{\la_a\}_1^{N+s} \Big)  \, - \, L \mc{E}_0( Q_F ) \; = \; \f{ v_F }{ L } \bigg\{  -\f{1}{12} + \ell^2 \big[ Z^2( Q_F \mid Q_F )-1 \big]   + \f{s^2}{ 4 Z^2( Q_F \mid Q_F ) } \\
\, + \, \sul{\eps = \pm} {}\Big(   \sul{a=1}{n_p^{\eps} } [p_a^{\eps}-\tf{1}{2}] \, + \, \sul{a=1}{n_h^{\eps} } [h_a^{\eps}-\tf{1}{2}]   \Big)  \bigg\} 
\; + \; \e{O}\Big( \f{ 1 }{ L^2 } \Big) 
\end{multline}
where 
\beq
v_F \, = \, \f{  \veps^{\prime}( Q_F \mid Q_F )  }{ p^{\prime}( Q_F \mid Q_F )  }
\enq
refers to the velocity of the excitation lying on Fermi boundary.

\end{theorem}

\Proof

Observe that the choice of integers given in the statement of the theorem always gives rise to a solution of the logarithmic Bethe equations 
since the conditions stated in Proposition \ref{Proposition existence solution Log BAE} are always verified provided that $L$ is large enough.  
Further, since $D \mapsto q(D)$ is smooth, it holds that $Q_F-\wh{q}=\e{O}(L^{-3})$ where $\wh{q}$ is the magnetic Fermi boundary associated with $ \wh{D}=\tf{N}{L}$. Also, since for any fixed $a \in \mathbb{Z}$
\beq
\lim_{L\tend +\infty}  \f{ N + a}{L}  \; = \; D_F \qquad \e{and} \qquad \lim_{L\tend +\infty}  \f{ a }{L}  \; = \; 0\;
\enq
one has
\beq
\lim_{L\tend +\infty} \wh{x}_{ N + a }  \; = \; Q_F \; \qquad  \e{and} \qquad \lim_{L\tend +\infty}  \wh{x}_{ a  } \; = \; -Q_F \;. 
\enq
These pieces of information are enough so as to obtain the large-$L$ expansion of the particle-hole roots. One gets 
\beq
\wh{x}_{ N + a } \, = \, Q_F \, + \, \f{ a  - \xi_1^{(s)}(Q_F) }{ L \xi_0^{\prime}(Q_F\mid Q_F) } \; + \; \e{O}\Big(  \f{1}{L^{2}} \Big) \qquad \qquad 
\wh{x}_{  a  } \, = \, -Q_F \, + \, \f{ a-\xi_1^{(s)}(-Q_F) }{ L \xi_0^{\prime}(-Q_F\mid Q_F) } \; + \; \e{O}\Big(  \f{1}{L^{2}} \Big) 
%
%
\enq
%
%
%
%
%
%
%
%
%
%
%
%
Note that, above, the dependence of $\xi_1^{(s)}$ on the position of the particle-hole roots is kept implicit. Inserting the expansion of the roots up to $ \e{O}(L^{-1})$ one gets that 
\beq
\xi_1^{(s)}\big(\la \mid \{ \wh{x}_{ p_a } \}_1^{n} \, ; \, \{ \wh{x}_{h_a} \}_1^n \big) \; = \; \f{1}{2} \, - \, \ell \Big\{ Z(\la\mid Q_F) - 1 \Big\} 
\, + \, \f{s}{2} \Big\{ 1 -\vp(\la,Q_F\mid Q_F) - \vp(\la,-Q_F\mid Q_F)  \Big\} \, + \, \e{O}\Big(  \f{1}{L} \Big)\;. 
\enq
Straightforward algebra based on these expansions leads to  the claim. \qed

\section*{Conclusion}

This paper develops tools allowing one to prove the existence and form of the large-$L$ asymptotic expansion of the counting function associated with the XXZ spin-$1/2$ chain. 
The method is robust in that it does not rely on details of the model such as the sign or the magnitude of the $L^{\infty}$ norm of the derivative of the bare phase. 
As such it allows one to step out of the setting where the model's Yang-Yang action is strictly convex or where the model is a small perturbation of a non-interacting model. 
For these reasons it appears plausible that the method of large-$L$ analysis proposed in this paper is applicable to many other quantum integrable models,
this independently of the value of their coupling constant, \textit{viz}. the sign of the Lieb kernels. 

The method should also allow one to study the case of excited states above the ground state described by Bethe roots such that a fixed number thereof is 
complex valued. I plan to investigate this issue in a forthcoming publication.

\section*{Acknowledgements}

K.K.K. is supported by CNRS. This work has been partly done within the financing of the  2014 FABER grant "Structures et asymptotiques d'int\'{e}grales multiples". 
K.K.K. would like to thank the Cathedral of mathematical physics of the University of Warsaw for its warm hospitality during the period were a substantial part of this work was done.

\appendix

\section{Auxiliary results}
\label{Appendix Section Auxiliary results}

Below, by  $f$ holomorphic on $F$ with $F$ not open, it is meant that $f$ is a holomorphic function on some open neighbourhood of $F$. Also, given a segment $I$
the box-open neighbourhood $\mc{S}_{\eta,\eps}(I)$ is as defined in \eqref{definition voisinages boites autour de intervalle} while $\Dp{ }\mc{S}_{\eta,\eps}(I)$ stands for the canonically oriented boundary of 
$\mc{S}_{\eta,\eps}(I)$.

\begin{prop}
\label{Proposition invertibilite de fL en fct de f}

Let   $I=\intff{a}{b}$ be a segment and $X$ be a compact in $\Cx$ such that $I \subset \e{Int}(X)$. 
 Let $f_L$ be a sequence of holomorphic functions on $X$ converging to a holomorphic function $f$ on $X$.  Assume that $f$ is such that $\e{sgn}\big(f^{\prime}_{\mid \intff{a}{b} }  \big)$ is constant. 
Then, there  exists  $\eta, \eps>0 $ such  that $ \mc{S}_{2\eta,2\eps}(I) \subset X$ and 
\beq
 \qquad   f \; : \; \mc{S}_{2\eta,2\eps}(I) \; \tend \; f\big( \mc{S}_{2\eta,2\eps}(I) \big)\quad 
is \; a  \; biholomorphism. 
\label{ecriture domaine biholomorphie de f}
\enq
Furthermore, there exists $L_0$ such that for any $L\geq L_0$  one has the inclusions 
\beq
f_L(I) \subset f\big( \mc{S}_{\eta,\eps}(I) \big)  \subset f_L\big( \mc{S}_{2\eta,2\eps}(I) \big)  
\enq
 and 
\beq
f_L \; : \;  \mc{S}_{2\eta,2\eps}(I) \cap f_L^{-1}\Big( f\big( \mc{S}_{\eta,\eps}(I) \big)  \Big) \tend f\big( \mc{S}_{\eta,\eps}(I) \big) 
\enq
is a biholomorphism. Also, there exists a constant $u_{\eta,\eps}[f]$ depending on $f$ such that, for any $L\geq L_0$, it holds 
\beq
 \inf_{ \la \in  \Dp{} \mc{S}_{2\eta,2\eps}(I) }  \inf_{ z \in f\big( \mc{S}_{\eta,\eps}(I) \big)  } \big| f_L(\la) - z \big| \geq  u_{\eta,\eps}[f]>0 \;.
\label{ecriture borne inf sur distance entre image par f de deux ensembles S eta eps}
\enq
Finally, for any $z \in f\big( \mc{S}_{\eta,\eps}(I) \big) $, it holds
\beq
f_L^{-1} (z) \; = \; \Oint{ \Dp{} \mc{S}_{2\eta,2\eps}(I) }{}  \hspace{-3mm} \f{ \la  \cdot f^{\prime}_L(\la) }{ f_L(\la) - z  } \cdot \f{ \dd \la  }{2\i \pi } \qquad and \qquad  
\norm{ f_L^{-1} - f^{-1}  }_{ L^{\infty}\big( f\big( \mc{S}_{\eta,\eps}(I) \big)  \big) } \; \leq \; C[f] \cdot 
\norm{ f_L - f  }_{ L^{\infty}( X ) }
\label{ecriture rep int pour fL}
\enq
for some constant $C[f]>0$ only depending on $f$, $\eps, \eta$ and the segment $I$. 
 
\end{prop}

\Proof

I first establish the statement relative to $f$. It is enough to consider the case where $f^{\prime}_{\mid \intff{a}{b} } >0 $. 
Then, by continuity, there exists $\wt{\eta}, \eps>0$ such that $\mc{S}_{3\wt{\eta},3\eps}(I)  \subset X$ and $f^{\prime}(\la) \not= 0 $ for any $ \la \in \ov{ \mc{S}_{3\wt{\eta},3\eps}(I) } $.
Hence, $f$ is a local biholomorphism on $\mc{S}_{3\wt{\eta},3\eps}(I) $. Thus it remains to show that there exists $\eta$, $\wt{\eta} > \eta > 0$ such that 
$f_{\mid \mc{S}_{3 \eta,3 \eps}(I) }$ is injective. If not, then there exists two sequences 
\beq
z_1^{(n)} \not= z_2^{(n)} \;  , \quad a-3\eps < \Re\big( z_k^{(n)} \big) < b+3\eps  \quad \e{such} \, \e{that} \quad 
\Im\big( z_k^{(n)} \big)\limit{n}{+\infty} 0 \quad \e{and} \quad  f(z_1^{(n)} ) \, = \,  f(z_2^{(n)}) \; .
\enq
By compactness of $\ov{\mc{S}}_{2\wt{\eta},2\eps}(I)$  one  may build converging subsequences $z_a^{(p_n)}\tend x_a \in \intff{a-3\eps}{b+3\eps}$. Yet
since $f^{\prime}_{\mid \intff{a-3\eps}{b+3\eps} } >0 $, $f$ is bijective on $\intff{a-3\eps}{b+3\eps}$ and thus $f(x_1)=f(x_2)$ implies that 
one  has $x_1=x_2=x$. There exists an open neighbourhood $U_x$ of $x$ such that $f_{\mid U_x}$ is a biholomorphism. 
For $n$ large enough, one has that $z_a^{(p_n)} \in U_{x}$, $a=1,2$. Thence, by injectivity of $f$ on $U_x$, 
$z_1^{(p_n)}=z_2^{(p_n)}$ for $n$ large enough, a contradiction. 
Thus $f$ is a biholomorphism on $\mc{S}_{3 \eta,3 \eps}(I)$ for some $\eta >0$ and, \textit{a fortiori} \eqref{ecriture domaine biholomorphie de f} is established.

It remains to establish the statements relative to $f_L$. Since $f$ is injective on  $\mc{S}_{3 \eta,3 \eps}(I)$, 
it follows that \newline $f\big( \Dp{} \mc{S}_{2 \eta,2 \eps}(I) \big) \cap \ov{ f\big( \mc{S}_{  \eta, \eps}(I) \big) } = \emptyset$. 
Thus, by compactness of these sets,  one has 
\beq
2 u_{\eta,\eps}[f] \; = \; \inf_{z \in \mc{S}_{\eta,  \eps}(I) } \inf_{s \in \Dp{} \mc{S}_{2\eta, 2\eps}(I) } \big|f(z)-f(s) \big| \, > \, 0 \; . 
\label{estimation distance dans preuve invertibilite fL}
\enq
Furthermore, the invertibility of $f$ on $\mc{S}_{2 \eta,2 \eps}(I)$ also ensures that  
\beq
2 \iota_{\eta,\eps}[f] \; = \; \inf\Big\{  |f^{\prime}(\la)| \; : \; \la \in  \mc{S}_{2\eta, 2\eps}(I) \Big\} >0 \;. 
\enq
Since $f_L \tend f$ on $X$, one has that there exists $L_0$ such that 
\beq
\norm{ f_L - f  }_{ L^{\infty}\big( \Dp{} \mc{S}_{3\eta,3\eps}(I) \big) } \, < \, \e{min}\Big\{  u_{\eta,\eps}[f]  , \tf{ \iota_{\eta,\eps}[f] }{ C_{1} } \Big\}  \qquad \e{provided} \; \e{that} \; L \geq L_0 \; . 
\label{ecriture borne sup sur distance de fL a f}
\enq
Here, $C_{1}$ only depends on $\eta$ and $\eps$ and is such that for any holomorphic functions $g$ on $ \ov{\mc{S}}_{2\eta,2\eps}(I) $
\beq
\norm{ g^{\prime} }_{ L^{\infty}\big(  \mc{S}_{2\eta,2\eps}(I) \big) } \, \leq  \, C_1 \cdot \norm{ g^{\prime} }_{ L^{\infty}\big(  \Dp{} \mc{S}_{3\eta,3\eps}(I) \big) } \;. 
\label{borne derivee via fct}
\enq

This bound implies \eqref{ecriture borne inf sur distance entre image par f de deux ensembles S eta eps} and ensures that $2\iota_{\eta,\eps}[f_L]> \iota_{\eta,\eps}[f]>0$ so that $f_L$ is a local biholomorphism on $\mc{S}_{2\eta, 2\eps}(I)$. 
Furthermore,  in virtue of Rouch\'{e}'s theorem, 
\beq
\e{for} \; \e{any} \quad  z \in f\big( \mc{S}_{\eta, \eps} (I) \big)   \qquad s \mapsto f_L(s) - z 
\enq
has exactly one simple zero in $  \mc{S}_{2\eta, 2\eps}(I)   $, and is thus injective on this set. Therefore, it follows that 
\beq
f_L \, : \, \mc{S}_{2\eta, 2\eps}(I) \cap f_L^{-1}\Big(   f\big( \mc{S}_{\eta, \eps}(I) \big)   \Big)  \; \tend  \; f\big( \mc{S}_{\eta, \eps}(I) \big)
\enq
is a biholomorphism. The above also implies that $f\big( \mc{S}_{\eta, \eps}(I) \big) \subset f_L\big( \mc{S}_{2\eta, 2\eps}(I) \big)$. Also, since $f$ is strictly increasing on 
$\intff{a-\eps}{b+\eps}$ it is clear that for $L$ large enough $f\big( \intff{a-\eps}{b+\eps} \big) \supset f_L(I)$. From there the statement about inclusion follows.

Finally, the bounds \eqref{estimation distance dans preuve invertibilite fL} and \eqref{ecriture borne sup sur distance de fL a f}  ensure that the 
integral representation \eqref{ecriture rep int pour fL} for $f_L$ holds. The same representation holds with $f \leftrightarrow f_L$ so that 
\beq
f_L^{-1}(z)-f^{-1}(z) \; = \; \Oint{ \Dp{} \mc{S}_{2\eta,2\eps}(I) }{}  \hspace{-3mm} \la \cdot \f{    f_L^{\prime}(\la) \big(f(\la) - z \big) -f^{\prime}(\la)  \big(f_L(\la)-z\big)   }
{ \big( f_L(\la) - z \big) \big(f(\la)-z\big)  } \cdot \f{ \dd \la  }{2\i \pi } \;. 
\enq
The difference is then bounded thanks to  \eqref{ecriture borne inf sur distance entre image par f de deux ensembles S eta eps}, \eqref{estimation distance dans preuve invertibilite fL}, 
\eqref{ecriture borne sup sur distance de fL a f}
and the bound  \eqref{borne derivee via fct}. \qed

\end{document}